\documentclass[journal]{IEEEtran}

\usepackage[noadjust]{cite}
\usepackage{multirow} 
\usepackage{algpseudocode}
\usepackage{algorithm}
\usepackage{rotating}
\usepackage{kantlipsum} 
\usepackage{commath}
\allowdisplaybreaks
\usepackage{mathtools}  
\usepackage{verbatim}
\usepackage{xr-hyper} 
\usepackage{enumitem}

\usepackage{makecell}
\usepackage{epstopdf}
\usepackage{wrapfig}
\usepackage{latexsym}
\usepackage{amssymb}
\usepackage{amsthm}
\usepackage{amsfonts}
\usepackage{amsmath} 
\usepackage{graphicx}
\usepackage{latexsym}
\usepackage{booktabs}
\usepackage{support-caption} 
\usepackage{subcaption} 
\usepackage{xspace}
\usepackage{tabularx}
\usepackage[normalem]{ulem} 

\usepackage{array}

\usepackage[T1]{fontenc}

\usepackage{url}

\usepackage{listings}

\usepackage{blindtext}

\graphicspath{{./fig/}}
\DeclareGraphicsExtensions{.png}

\ifCLASSINFOpdf
\else
\fi

\usepackage{placeins}

\newcommand{\dq}[1]{``#1''}

\usepackage[dvipsnames]{xcolor}

\newcommand{\commentBy}[3]{\textcolor{#1}{\textbf{#2:} #3}}

\newif\ifcommentson

\newcommand{\ste}[1]{\ifcommentson \commentBy{blue}{SS}{#1} \fi}
\newcommand{\fl}[1]{\ifcommentson \commentBy{RoyalPurple}{FL}{#1} \fi}

\newif\ifthesis
\thesisfalse

\newcommand{\thesis}[1]{\ifthesis #1\else \fi}

\newif\ifpaper
\papertrue

\newcommand{\paper}[1]{\ifpaper #1\else \fi}

\newif\ifextended
\newif\ifshortver

\shortvertrue

\newcommand{\extended}[1]{\ifextended \ifshortver \textcolor{purple}{#1} \else \textcolor{black}{#1} \fi  \fi}

\newif\ifrevision

\begin{document}

\bstctlcite{IEEEexample:BSTcontrol}



\title{Extending Kubernetes Networking to make use of\\
Segment Routing over IPv6 (SRv6)}

\author{Francesco~Lombardo,
        Stefano~Salsano,
        Ahmed~Abdelsalam,
        Daniel~Bernier,
        Clarence~Filsfils

\IEEEcompsocitemizethanks{\protect
\IEEEcompsocthanksitem Stefano Salsano and Francesco Lombardo are with the Department of Electronic Engineering at the University of Rome \dq{Tor Vergata} and the Consorzio Nazionale Interuniversitario per le Telecomunicazioni (CNIT) - Rome, Italy E-mail: \{stefano.salsano, francesco.lombardo\}@uniroma2.it;\\
Ahmed Abdelsalam and Clarence Filfils are with Cisco System - USA E-mail: \{ahabdels, cfilsfil\}@cisco.com;\\
Daniel Bernier is with Bell Canada, Canada E-mail: daniel.bernier@bell.ca.
}
}

\markboth{Submitted to a journal}%
{Shell \MakeLowercase{\textit{et al.}}: Extending Kubernetes Networking}



\maketitle

\begin{abstract}
Kubernetes is the leading platform for orchestrating containerized applications. In this paper, we extend Kubernetes networking to make use of SRv6, a feature-rich overlay networking mechanism. Integration with SRv6 can be very beneficial when Kubernetes is used in large-scale and distributed multi-datacenter scenarios. We have focused on the Calico CNI plugin, one of the most used Kubernetes networking plugins. In particular, we consider Calico-VPP, a version of the Calico plugin based on the VPP (Vector Packet Processing) data plane, which provides support for SRv6 operations with very high performance. The proposed SRv6 overlay networking solution for Kubernetes offers several advantages compared to a traditional overlay (e.g. IP in IP), in particular the possibility to use Traffic Engineering for the overlay tunnels. In the paper, we provide the architecture and the detailed design of the SRv6 based overlay and describe our open source implementation. We consider the research and technological question on how to extend Kubernetes networking to support large-scale and distributed multi-datacenter scenarios, which is an important goal for Cloud and Network providers. In this respect, we compare two different solutions for the control plane architecture of the SRv6 capable Kubernetes networking plugin, one based on the BGP routing protocol and another one based on extending the Kubernetes control plane. Finally, we report a performance evaluation of the data plane of the proposed SRv6 overlay networking, showing that it has comparable performance to existing overlay solutions (e.g. IP in IP), while offering a richer set of features. 
\end{abstract}



%
\IEEEpeerreviewmaketitle

\paper{\section{Introduction}}

\paper{\IEEEPARstart{K}{ubernetes}} \thesis{Kubernetes} is the leading system for automating deployment, scaling, and management of containerized applications. The network communications in Kubernetes rely on software components called CNI (Container Networking Interface) plugins, which interact with the IP networking infrastructure supporting the Kubernetes clusters. 

With the current industry race towards cloud-native 5G core deployments and the growing cloudification of Telco software stacks, Cloud Service Providers face a challenge. Kubernetes was initially not designed for complexities of operator environments with non typical protocols, massive network segmentation and large scale multi-tenancy needs. If we add the growing complexities added with 5G slicing, MEC (Multi-access Edge Computing) applications deployment, latency sensitive workloads and the Kubernetes massive consumption of IPv4 addressing, a new approach needs to be looked at. A highly scalable and highly flexible technology for Telco operations is needed, at the same time simple enough not to break the basic networking model of Kubernetes and its APIs.

Segment Routing over IPv6 (SRv6) is a networking architecture that can be used in IP backbones and data centers. SRv6 technology is gaining a lot of traction with several large-scale deployments that have been recently made public. With SRv6, operators can implement services like overlay networking, VPNs, traffic engineering, protection/restoration in a scalable and effective way.


When Kubernetes is used in large-scale and/or distributed multi-datacenter scenarios, the integration with a feature-rich overlay networking solution like SRv6 would be very beneficial for service providers to address the above identified challenges. Unfortunately, no Kubernetes networking plugin (CNI plugin) currently supports SRv6. The extension of networking plugins to support this enhanced overlay networking solution is not a trivial task, for a number of reasons: 

\begin{itemize}
    \item Generally speaking, IPv6 support in Kubernetes networking plugins is not fully mature. Our target is to have support of IPv6 in the infrastructure/underlay (as we want to use SRv6 for transport) and support of both IPv4 and IPv6 in the pods because Kubernetes cluster should be able to support workloads based on IPv4, IPv6 or both (dual stack).
    \item For the extension to be successfully deployed in the real world, it needs to smoothly integrate into an existing plugin without losing the existing features or breaking compatibility
    \item The networking model of Kubernetes has been designed to be general and it makes difficult to introduce specific networking features for a CNI plugin in a clean way, without breaking compatibility with other plugins.
\end{itemize}

Considering these issues, we can identify a number of research and technological questions related to the extension of Kubernetes networking to support advanced networking features. The first set of questions concerns the interaction with the existing CNI API and Kubernetes configuration mechanisms. Is it possible to introduce the support for the new features in an optional way? Is it possible to have a smooth coexistence of legacy configurations and new advanced scenarios? Should the advanced features be completely \emph{hidden} to the regular Kubernetes configuration, or is it useful and possible to expose them? 

A second set of research questions concerns the control and configuration mechanisms to be used in Kubernetes when dealing with the new advanced networking features. In fact, such features require the dynamic control and coordination of a potentially large number of nodes which could also be distributed in a large geographical area across multiple datacenters. In this scenario, how to simplify the advanced networking configuration of a large and distributed Kubernetes cluster, minimizing the manual configuration operations? Should we use routing protocols (e.g. BGP) or native Kubernetes control plane mechanism for the dynamic configuration of the cluster nodes related to the advanced networking aspects?

The main contributions of this \paper{work}\thesis{thesis} are the following:

\begin{itemize}
    \item the design of an overlay networking solution based on SRv6 for Kubernetes, offering additional features and advantages compared to a traditional overlay (e.g. IP in IP), in particular the possibility to use Traffic Engineering for the overlay tunnels
    \item the implementation of the proposed SRv6 overlay by extending the existing Calico-VPP Kubernetes networking plugin; our implementation has been released and merged in the Calico-VPP open source project
    \item the design and implementation of two mechanisms for the control and coordination of the nodes of a Kubernetes cluster to support advanced networking features (Traffic Engineering), one based on extending the BGP routing protocol and one based on Kubernetes control plane
    \item the validation of the proposed SRv6 overlay solution in a replicable virtual testbed
    \thesis{
    \item the design and implementation of a tool capable of automatically evaluating the performance of the Kubernetes infrastructure. 
    }
\end{itemize}
\paper{
The rest of the paper is organized as follows. 
In Section~\ref{sec:k8s-model}, we provide an overview of the Kubernetes networking model. Section~\ref{sec:calico-vpp} describes Calico-VPP, the networking plugin that we have extended. We illustrate the overlay networking models that are already supported by Calico-VPP and point out the missing features that we want to support. Section~\ref{sec:ip-ip-calico} goes into more detail on the IP in IP overlay networking approach implemented in Calico-VPP, because it is the one that we have extended to support SRv6 overlay networking.
In Section~\ref{sec:srv6-overlay-calico-vpp} we describe how we have introduced the new overlay networking model based on SRv6 in the Calico-VPP networking plugin. We also include here a short introduction to the SRv6 network programming model, needed to understand its features and in particular how an SRv6 overlay can be enhanced with Traffic Engineering.
Section~\ref{sec:testbed} describes the testbeds that have been used for development and testing, pointing also to the instructions for replicating the environment. We describe and discuss the results of the performance evaluation experiments in section~\ref{sec:perf-eval} and finally we draw conclusions in Section~\ref{sec:conclusions}.


}
\thesis{
The rest of the thesis is organized as follows.
In section~\ref{sec:kubernetes} I provide an overview of the core concept of Kubernets, in particular in section~\ref{sec:k8s-model} an overview of the Kubernets networking model and CNI (Container Networking Interface). Section~\ref{sec:srv6stateart} provides an overview of SRv6, specifically the SRv6 data plane first, followed by the SRv6 Networking model and the SRv6 control plane.
In section~\ref{sec:vppstateart} I introduce the VPP platform and I give a brief description of the technology used by Calico-VPP.  Section~\ref{sec:calico-vpp} describes Calico-VPP, the networking plugin that I have extended. From here, I walk through the overlay networking models that are already supported by Calico-VPP and report the missing features that I want to support.
Section~\ref{sec:ip-ip-calico} goes deeper in the details of the IP in IP overlay networking approach implemented in Calico-VPP, because it is the one that I have extended to support SRv6 overlay networking.
Section~\ref{sec:srv6-overlay-calico-vpp} describes how I integrated this new overlay networking model based on SRv6 into the Calico-VPP networking plug-in. In this section, I also demonstrate how an SRv6 overlay can be enhanced with traffic engineering. Section~\ref{sec:testbed} describes the testbeds that have been used for the development and for the testing. In section  \ref{sec:srv6basciscenario} I describe how to configure the base testbed and how to configure the CNI to have it deployed without using Traffic Engineering. Section \ref{sec:srv6testbedte} describes how to set up the complete test bed and how to use the CNI with the mechanisms implemented for Traffic Engineering. In in section~\ref{sec:perf-eval} I describe and discuss the results of the performance evaluation experiments. In Chapter~\ref{chap:kites} I explain the need for a tool capable of automatically evaluating the performance of the Kubernetes infrastructure, in particular in section~\ref{sec:kitesintro} and ~\ref{sec:kitesintro} I introduce the tool "KITES" of which I contributed to the design and development. Section~\ref{sec:kitesperformanceindicator} describes the performance indicators that Kites supports and  Section~\ref{sec:kitescomparisonstateart} provides a brief comparison with the state of the art. 

}
\paper{\section{Kubernetes networking model (CNI plugins)}}
\thesis{\subsection{Kubernetes networking model (CNI plugins)}}
\label{sec:k8s-model}

\paper{\begin{figure*}[t!]
    \centering
    \includegraphics[width=\linewidth]{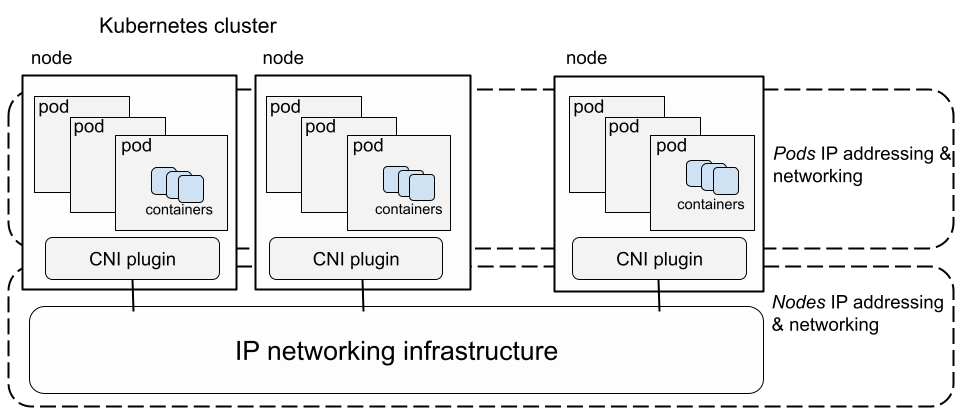}
    \caption{Networking view of a Kubernetes cluster: nodes, CNI plugins, pods, containers}
    \label{fig:overall-kubernetes-cluster}
\end{figure*}}

\thesis{\begin{figure}[ht]
    \centering
    \includegraphics[width=\linewidth]{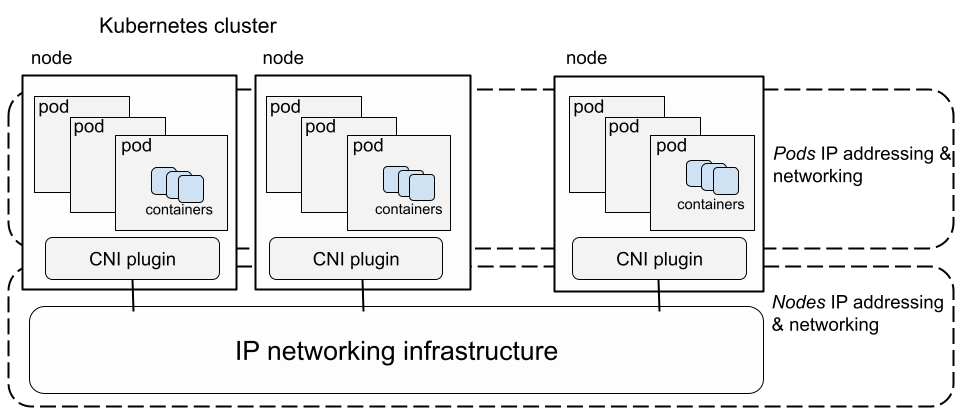}
    \caption{Networking view of a Kubernetes cluster: nodes, CNI plugins, pods, containers}
    \label{fig:overall-kubernetes-cluster}
\end{figure}}

To introduce the fundamental concepts of Kubernetes networking, we refer to Fig.~\ref{fig:overall-kubernetes-cluster}. A Kubernetes cluster consists of a set of \emph{nodes} that can host the containerized applications. In particular, within each node, one or more \emph{pods} can run the \emph{containers} that constitute the \emph{workload} of the cluster. 




From the point of view of networking, each pod in the cluster gets its own IP address (IPv4 or IPv6) which is used by the pod to communicate with all other pods in the cluster, both in the same node and in other nodes. Fig.~\ref{fig:overall-kubernetes-cluster} illustrates that the pods communicate using a \dq{Pods IP addressing}. The nodes in the cluster also need to have their IP addresses, which are used by the Kubernetes agents residing on the nodes to communicate with each other (\dq{Nodes IP addressing} in Fig.~\ref{fig:overall-kubernetes-cluster}). 

The Kubernetes network model~\cite{kubernetes-network-model} imposes some requirements on the implementation of the communication among the entities (pods and nodes). In particular, all pods need to be able to communicate with each other using \dq{pod IP addressing} \emph{without using Network Address Translation (\paper{NAT}\thesis{\acrshort{NATLabel}})}. Moreover, a Kubernetes agent in a node must be able to communicate with all pods in the same node \emph{without using (NAT)}. It is possible to meet these requirements in many different ways, and the Kubernetes architecture does not prescribe a specific way to implement the networking, also because how the nodes communicate depends on the environment in which the Kubernetes cluster is deployed. For example: i) nodes can be bare metal servers or Virtual Machines, or even a combination of the two cases; ii) all nodes can be on the same layer 2 subnet, or they can belong to multiple IP subnets; iii) the nodes can be located in the same datacenter or in multiple datacenters. 

\thesis{
Based on these considerations, it is possible to define the different configurations with which Kubernetes components are able to communicate with each other.
\subsubsection{Intra-pod communications}
One of the principles on which Kubernetes is based is the fact that each Pod has its own unique IP address. Additionally, each newly created Pod is placed inside a network namespace within which the containers can share resources.
For this reason, the containers started inside this Pod share the same network space, therefore they are able to communicate with each other via the localhost address. Obviously, two containers within the same Pod cannot use the same network port, just like two processes running on the same host. However, it is possible to use the same port for two containers located on different Pods, even if they belong to the same node.
This is possible since each Pod has its own IP address and therefore the port refers to its network. The only problem could occur when the ports are exposed to the host and therefore the affinity between Pod and node should be taken into consideration.
\begin{figure}[ht]
    \centering
    \includegraphics[width=\linewidth]{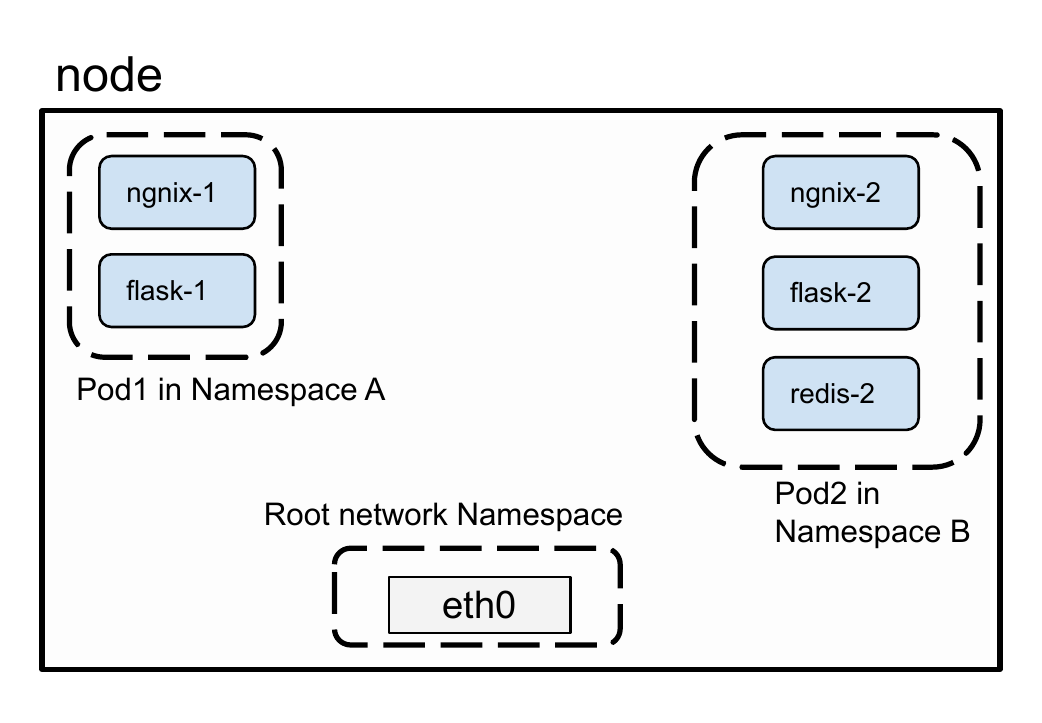}
    \caption{Intra pod communications}
    \label{fig:comm-pods-same-pod}
\end{figure}
Therefore, taking as an example the node in Figure \ref{fig:comm-pods-same-pod}, we can say that the container nginx-1 inside Pod 1 and nginx-2 inside Pod 2 can both be connected to port 8080, while this is not possible for the other containers within the same Pod.

\subsubsection{Inter-Pod Communications}
Pods within a Kubernetes cluster have a unique IP address which allows them to communicate directly with other pods without the need for NAT, tunnel, proxy or any other intermediate layer. It is important to consider the case in which communication between Pods occurs between Pods allocated within the same physical node or through different nodes within the cluster.

\subsubsection{Communications between Pods within the same node}
The mechanism typically used is to attach one interface of the pair to the host namespace and the other interface to the Pod namespace in question.
By convention, the interface on the host namespace will be named veth <xxx>, while the interface on the Pod will be named eth0. To make communication between the Pods possible, a bridge is implemented that connects the veth interfaces in the host namespace.
The eth0 interfaces in the Pod namespaces will then be assigned an IP address that is in the network address range of the aforementioned bridge.
Since the bridge within the host namespace connects the different Pod namespaces attached to it, it is now possible to send traffic between different Pods within the same node.
\begin{figure}[ht]
    \centering
    \includegraphics[width=\linewidth]{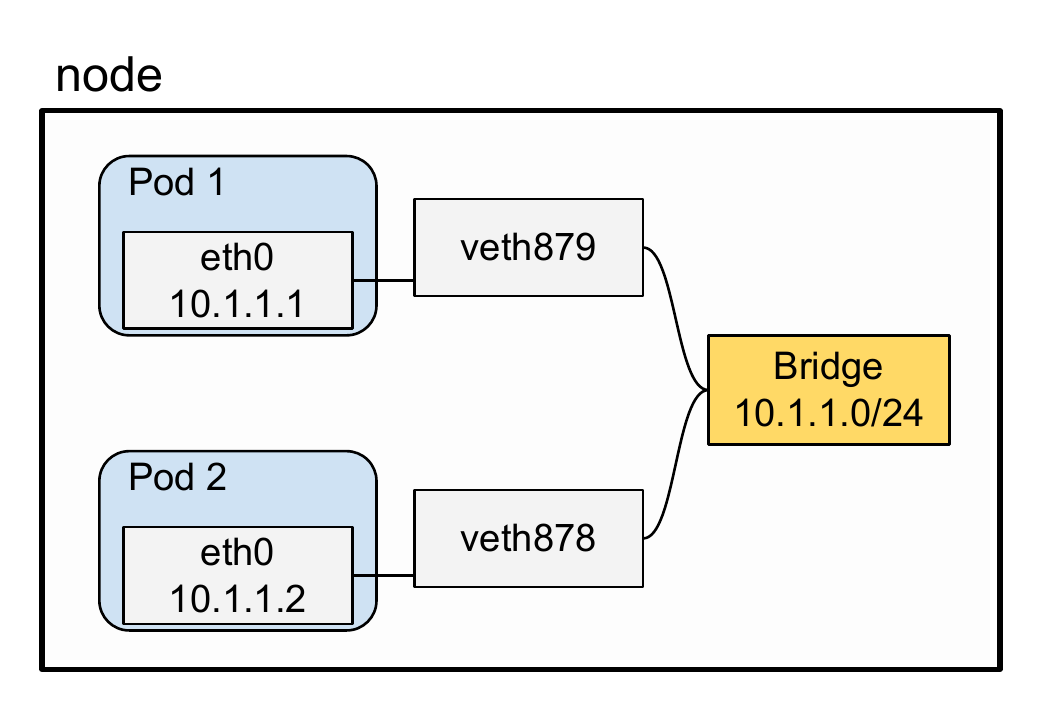}
    \caption{Communications between Pods within the same node}
    \label{fig:comm-pods-same-node}
\end{figure}
\subsubsection{Communications between Pods belonging to different nodes}
There are several ways to allow connection between bridges on different nodes: via overlay networks (which we will analyze later) or via a normal Layer 3 routing solution, which we will analyze below. Therefore, the standard layer 3 networking procedure used by Kubernetes ensures that the physical interface of the node is connected to the various bridges, in order to support communication between Pods located on different nodes.
Furthermore, the routing tables on the different nodes must be configured to ensure that traffic is redirected to the correct node.
Finally, note that this solution only works if the nodes are connected to the same network switch, without router in between.
\begin{figure}[ht]
    \centering
    \includegraphics[width=\linewidth]{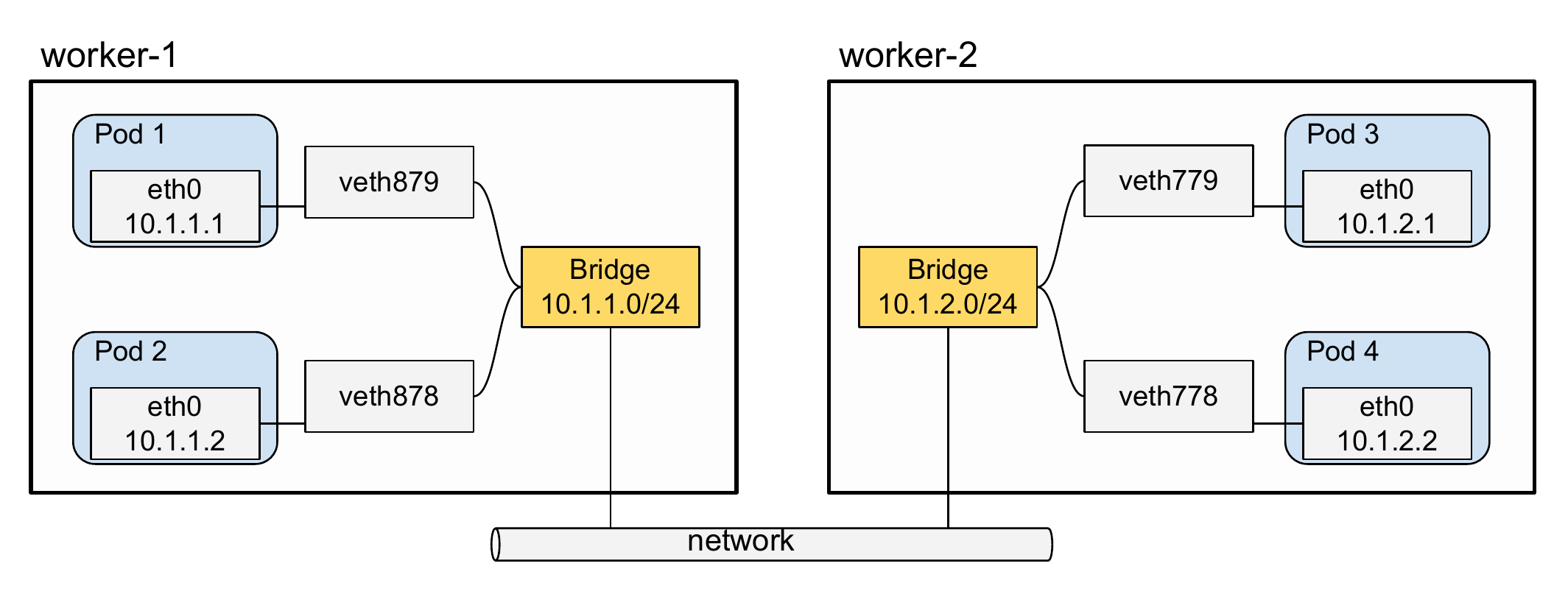}
    \caption{Communications between Pods belonging to different nodes}
    \label{fig:comm-pods-different-node}
\end{figure}
}

To cope with these different scenarios, the Kubernetes architecture introduces the concept of the \emph{Container Network Interface} (CNI)~\cite{github-containernetworking-cni} and of the \emph{CNI plugins}. As shown in Fig.~\ref{fig:overall-kubernetes-cluster}, the CNI plugin inside each node interconnects the Pods with the underlying IP networking infrastructure. The role of the CNI plugin is to allow the pods to communicate transparently using the \dq{pod IP addressing/networking}, adapting it to the \dq{node IP networking} environment in which the cluster nodes are deployed. Several different CNI plugins are currently available; a non-exhaustive list can be found in~\cite{k8s-addons-networkingl}. 

A CNI plugin can operate in two ways, depending on how the \dq{pod IP networking} interacts with the underlying networking environment at node level: i) flat networking, in which the IP packets of the pods can be routed through the node IP networking layer without modification; ii) overlay networking, in which the pods' IP packets need to be encapsulated to cross the node IP networking layer. Some CNI plugins only support one of the two modes (flat or overlay); other ones can be configured to operate with flat or overlay networking, or even with a combination of the two modes. 

The flat networking model has the advantage of better performance because packets do not need to be encapsulated/decapsulated. The disadvantage of the flat networking model is that it cannot be applied in several deployment scenarios. In some cases it is simply not feasible, in other cases it does not support all the requirements (e.g. support of multiple tenants, scalable operations, simplicity in configuration). On the other hand, the overlay networking model is very flexible and it can be applied in all circumstances. The overlay networking model supports multiple tenants and can scale well in complex deployment scenarios, distributed over multiple geographical locations.

A Kubernetes CNI plugin is logically decomposed into data plane and control plane functions. Data plane functions concern the forwarding of the packets (from pod to pod, from pods to the external world and vice versa). Control plane functions concern the dynamic configuration of IP routing at the \dq{pod IP networking} level and the configuration of forwarding operations in the nodes of the cluster, to ensure the proper operation of the data plane. In turn, the control plane operations are configured and managed by Kubernetes with configuration files provided by the system administrator and/or with commands entered manually by the system administrator using the Kubernetes CLI (Command Line Interface).


Let us consider complex and large-scale deployment scenarios that require the use of overlay networking. In these scenarios, having a powerful overlay networking mechanism is beneficial. Segment Routing over IPv6 (SRv6 in short) offers a feature-rich overlay networking mechanism. It can support traffic engineering in the underlay, encapsulate both IPv6 and IPv4 packets (and also layer 2 frames), and offer transit to multiple tenants. The integration of these advanced features into Kubernetes is not easy, because the Kubernetes model is meant to be general and to avoid relying on specific features of the networking CNI plugin.

\paper{In this work, we }\thesis{In this thesis, I }have considered the open source Calico CNI plugin (in particular, Calico with VPP dataplane, or Calico-VPP in short) and extended it to support SRv6. \paper{We }\thesis{I } have designed the configuration and control mechanisms that are needed to integrate SRv6 in Kubernetes and take advantage of its powerful networking features.

\section{Calico and Calico-VPP}
\label{sec:calico-vpp}

Calico~\cite{web-about-calico} is an open-source networking solution to interconnect entities (e.g. Containers, Virtual Machines, and bare-metal Servers) in Cloud Computing scenarios. It supports complex interconnection policies and it can enforce security. Thanks to its flexibility, the use of Calico is not limited to Kubernetes, but it can be used in other orchestrator platforms (e.g. OpenShift, OpenStack). Here, we only consider the use of Calico as a \paper{CNI}\thesis{\acrshort{CNILabel}} networking plugin for Kubernetes. Calico Networking is documented in \cite{web-calico-networking}, see also an introduction in~\cite{mackrory_website_2020}.

\thesis{ As described in ~\cite{mackrory_website_2020} Calico uses IP networking and BGP for more efficient routing and handling of traffic within the cluster. Indeed the use of BGP allows the network to be easily scaled, and it is highly effective at managing the systems with distributed architectures. The use of IP networking also makes troubleshooting much more straightforward. Kubernetes relies on NetworkPolicy~\cite{kubernetesio-netpolicy} as its primary tool to secure the network, in fact Kubernetes provides a basic definition of it. But the network plug-in is responsible for enforcing the policy rule. As described in \cite{web-calico-networking} Calico also allows administrators to define network policies that manage the flow of traffic between specific pods, providing granular control over the flow of network traffic.}

With respect to the two operating modes of a CNI plugin described in Section~\ref{sec:k8s-model} (flat networking and overlay networking), the Calico CNI plugin can be configured to work in both modes. We are interested here in the overlay networking approach, as it is the most useful and widely used in large and complex cloud computing scenarios involving multiple data centers and multiple customer sites.

Overlay networking in Calico relies on two types of encapsulation: VXLAN and IP in IP. We will call these solutions \emph{VXLAN overlay} and \emph{IP in IP overlay}, which means that VXLAN encapsulation and IP in IP encapsulation are used, respectively.
As anticipated in the Introduction, we are interested in adding a third type of overlay/encapsulation: the \emph{SRv6 overlay}, i.e. based on Segment Routing over IPv6. The reason for adding the \paper{SRv6}\thesis{\acrshort{SRv6Label}} overlay solution is that it can support very powerful features and gives the possibility to benefit from advanced services in the underlay transport network, such as traffic engineering, fault protection/restoration, support of VPN addresses. 

\begin{figure}[ht]
    \centering
    \includegraphics[width=\columnwidth]{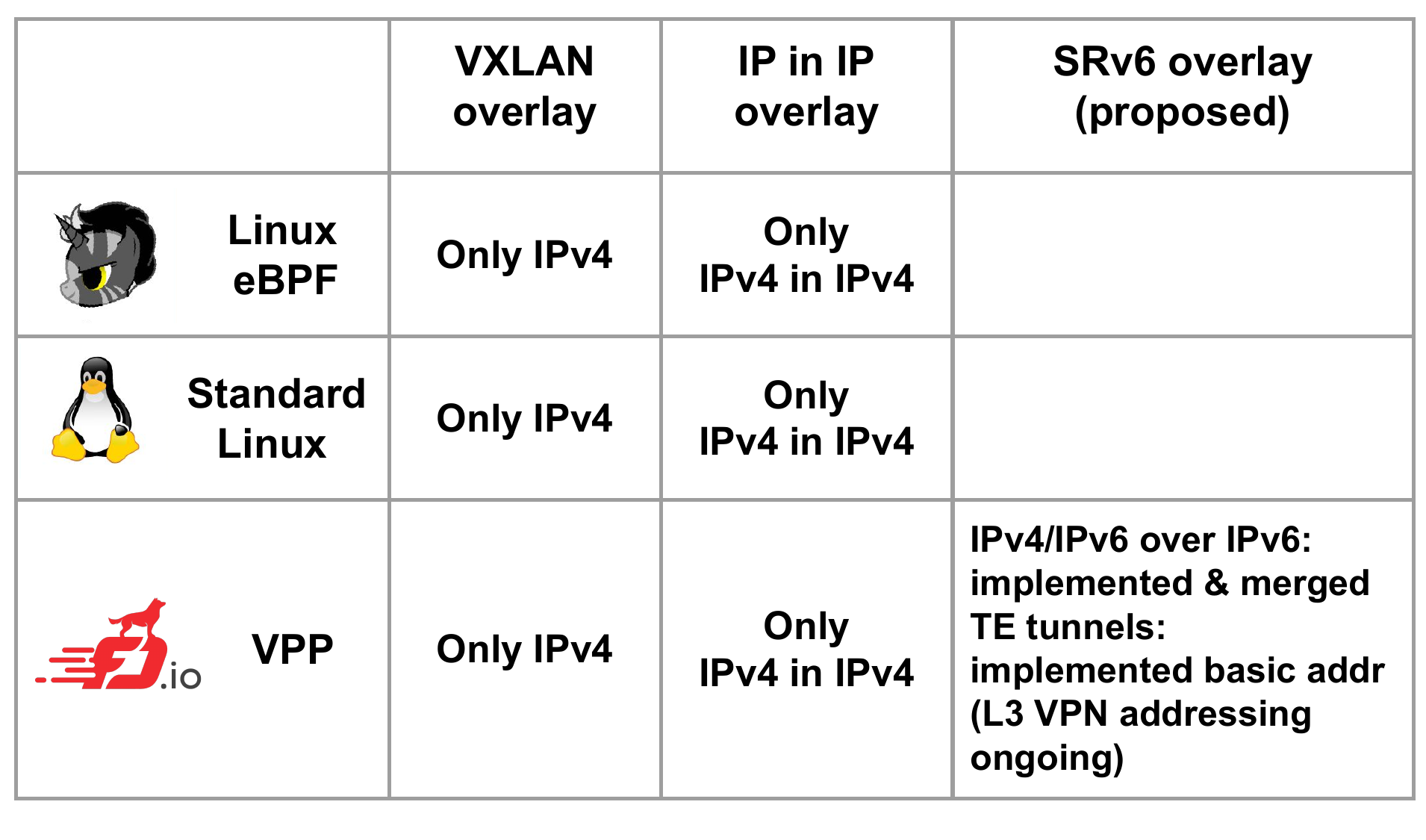}
    \caption{Calico Dataplanes and Overlay types}
    \label{fig:dataplanes-and-overlays}
\end{figure}

The Calico project offers the possibility to choose among a set of different \emph{packet forwarding engines} or \emph{dataplanes} as they are called in~\cite{web-about-calico}: a Linux \paper{eBPF}\thesis{\acrshort{eBPFLabel}} dataplane, a Standard Linux dataplane, and a Windows \paper{HNS}\thesis{\acrshort{HNSLabel}} (Host Networking Service) dataplane. In addition, there is a fourth dataplane, currently in \dq{tech preview} status, called Calico-VPP and based on the Vector Packet Processing (VPP) technology~\cite{whatisvpp}. VPP is a high performance packet processing stack for Linux. It can boost the packet processing performance of Linux based nodes~\cite{fd-io-terabit}, especially when coupled with the Data Plane Development Kit (\paper{DPDK}\thesis{\acrshort{DPDKLabel}}) technology~\cite{dpdk-wikipedia}.  

In the design of our solution, we have decided to extend the Calico-VPP dataplane, because VPP already provides a high-performance support for SRv6. The advantages of the VPP dataplane over the standard Linux networking dataplane are discussed in~\cite{web-calico-vpp-get-started}. In particular, it scales to higher throughput, especially when encryption services are enabled. Moreover, the VPP dataplane supports the Kubernetes \emph{Service} concept~\cite{kubernetesio-concepts-services} in a very efficient way, by using a VPP native NAT service instead of relying on the \emph{kube-proxy} component described in~\cite{kubernetesio-concepts-services} (we refer the interested reader to~\cite{calico-vpp-implem-details} for further details on the Calico-VPP dataplane).

In Fig.~\ref{fig:dataplanes-and-overlays}, we show a table that compares the support of overlay networking of the Linux eBPF dataplane, the Standard Linux dataplane and the Calico VPP dataplane. The three dataplanes support VXLAN and IP in IP overlays only for IPv4 addresses. Our proposed SRv6 overlay for the Calico-VPP data plane supports the encapsulation of both IPv4 and IPv6 pods addresses. Moreover, our solution is the only one that supports Traffic Engineering.  

\section{IP in IP overlay in Calico-VPP}
\label{sec:ip-ip-calico}



\begin{figure}[ht]
    \centering
    \includegraphics[width=\columnwidth]{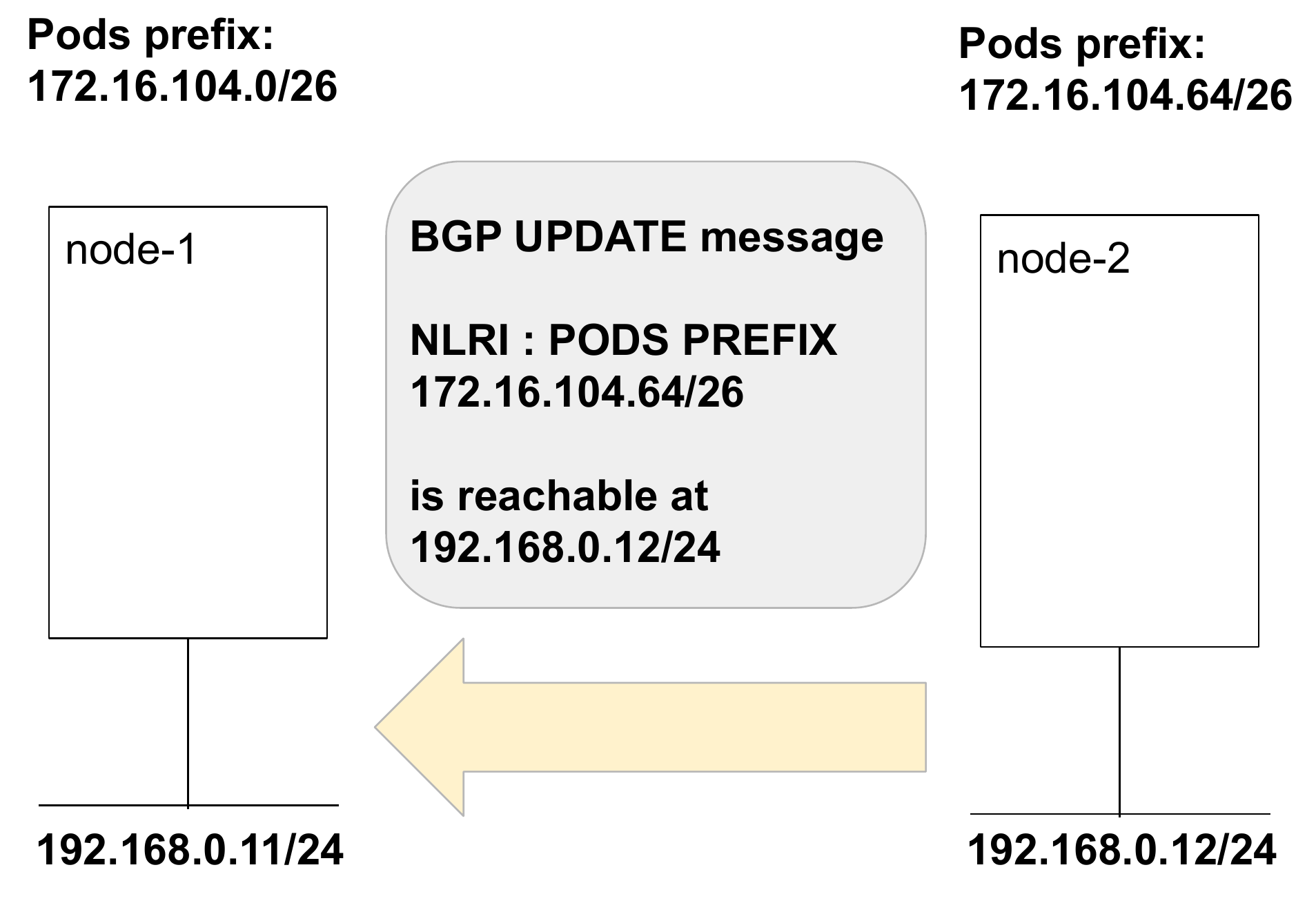}
    \caption{BGP mechanism for IP-in-IP tunnels}
    \label{fig:bgp-mechanism-for-ip-in-ip}
\end{figure}

Let us illustrate the operations of the IP in IP overlay in Calico-VPP considering an example a scenario with two cluster nodes (node-1 and node-2), as shown in Fig.~\ref{fig:bgp-mechanism-for-ip-in-ip}. When Kubernetes assigns the pods to the nodes (e.g. node-1 and node-2 in the figure), these pods need to receive their IP addresses at the pod networking level. For example, when the first pod is assigned to node-2, a dedicated subnet is assigned to all pods that will be hosted by node-2. This means that a portion of the pod IP networking address space of the cluster is dedicated to node-2 in question. In Fig.~\ref{fig:bgp-mechanism-for-ip-in-ip}, this portion assigned to node-2 is indicated as Pods prefix 172.16.104.64/26. Using an IP in IP overlay, this Pods prefix allocated to node-2 will be reachable through the infrastructure level IP address of node-2 (192.168.0.12 in Fig.~\ref{fig:bgp-mechanism-for-ip-in-ip}). To ensure that IP in IP overlay is established between all cluster nodes, the association between the pods prefix allocated to node-2 and the node-2 infrastructure address must be communicated to all nodes.
As mentioned in the previous sections, the Calico IP in IP overlay networking relies on the BGP protocol to distribute the routing information about which pods prefixes are present in which nodes. In particular, the BGP protocol is used to advertise an \paper{NLRI (Network Layer Reachability Information)}\thesis{\acrshort{NLRILabel} (\acrlong{NLRILabel})} that contains the IPaddressPrefix. In BGP jargon, the NLRIs are the prefixes that can be reached through an advertising BGP neighbor. As seen in Fig.~\ref{fig:bgp-mechanism-for-ip-in-ip}, node-2 sends a BGP UPDATE message that contains the pod prefix (IPv4 type) along with the node IPv4 infrastructure address.
Likewise, node-1 advertises the same information for the reachability of its Pods prefix. In this way, the CNI agents present on the nodes can configure the routing rules, which VPP can use to encapsulate and allow the pods to communicate in a completely transparent way (the software architecture is described in the next subsection).

We highlight that the above described procedure is \emph{dynamic} and \emph{automatic}. The network administrator of the Kubernetes cluster just needs to initially configure the networking plugin, then the software components of the plugin are able to react to the events like the allocation of pods to the nodes and to exchange the needed information with remote nodes using the BGP signalling.

\subsection{Calico-VPP software architecture}

In this section, we describe the software architecture of Calico-VPP, as needed to understand our work on the integration of the SRv6 overlay. In the original Calico CNI plugin an instance of the BIRD BGP agent is running inside all Calico nodes in the cluster as a separate container, to implement the BGP based interactions. Each BIRD agent in a node distributes the Pods prefixes (i.e. the pods subnets addresses) to all other BIRD agents using iBGP sessions when the subnets are activated on the node. By default this is done using a full mesh of BGP sessions among all nodes in a Calico based cluster, but a more scalable configuration using a centralized BGP reflector is also possible. The Calico BIRD BGP agent is based on the open source BIRD Internet Routing daemon~\cite{bird-daemon}.

Another software component running in a separate container is called Felix~\cite{web-calico-felix}. Felix programs the routes and the policies/\paper{ACLs}\thesis{\acrshort{ACLsLabel}} (Admission Control Lists) on the node, as required to provide connectivity to the pods running in the node. The detailed description of the software architecture of the original Calico CNI plugin can be found in~\cite{web-calico-comp-arch}.

\begin{figure}[ht]
    \centering
    \includegraphics[width=\paper{0.5}\thesis{1}\textwidth]{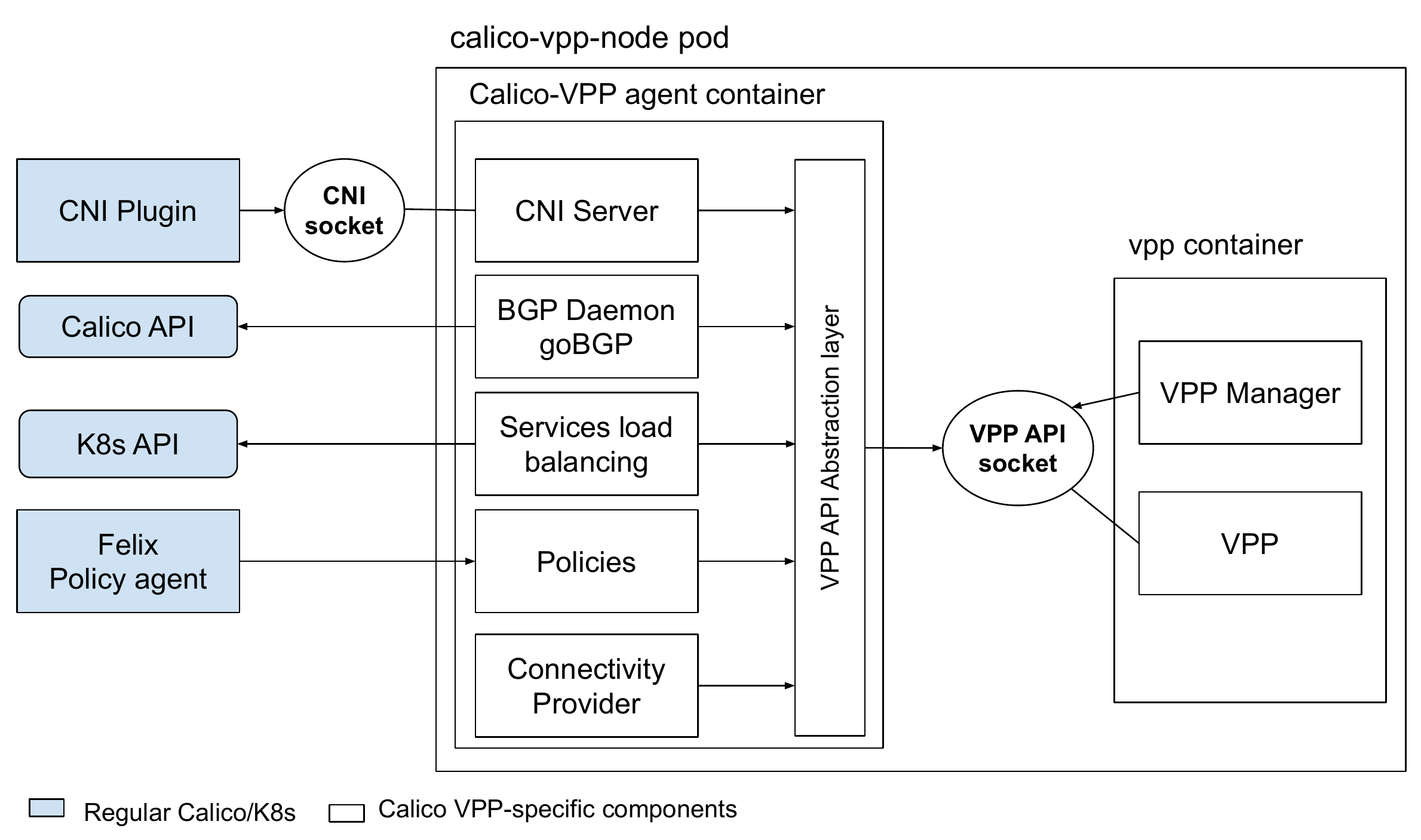}
    \caption{The Calico VPP dataplane Software Architecture }
    \label{fig:vpp-soft-arch}
\end{figure}

Calico-VPP (i.e. Calico using the VPP dataplane) uses a slightly modified control architecture with respect to the original Calico CNI plugin, as shown in Fig.~\ref{fig:vpp-soft-arch}. The control components are deployed inside a pod called calico-vpp-node. The Calico-VPP agent is a container running inside this pod that implements all the control functions. Calico-VPP agent components are implemented using the Go language. The BIRD BGP agent is replaced by a \emph{GoBGP Daemon}~\cite{go-bgp} running in the Calico-VPP agent container. A subset of the operations performed by Felix is directly performed by a new component called \emph{Connectivity Provider}, while for the operations related to the Policies, the \emph{Policies} component interacts with the regular Felix Policy agent. 

The extensions to the original Calico CNI plugin are implemented through the concept of \dq{CNI chaining} (see~\cite{cni-spec, cni-chaining}). In particular, a component in the Calico-VPP agent (called CNI Server) implements a server that receives gRPC requests from the Calico CNI (configured with a gRPC dataplane) through a Unix socket mounted on the \paper{k8s}\thesis{\acrshort{K8sLabel}} node (CNI socket in Fig.~\ref{fig:vpp-soft-arch}). These requests concern the configuration of the networking for the containers that are hosted in a node (e.g. add a container to a network). The CNI Server adds an interface to the container, assigns the proper IP address (from the pods prefix) and the gateway for the default route. Then it interacts with the VPP data plane to provide the proper networking configuration so that the container can send and receive IP packets.

\subsection{Calico-VPP networking configuration}

\begin{figure}[ht]
    \centering
    \includegraphics[width=\paper{0.5}\thesis{0.5}\textwidth]{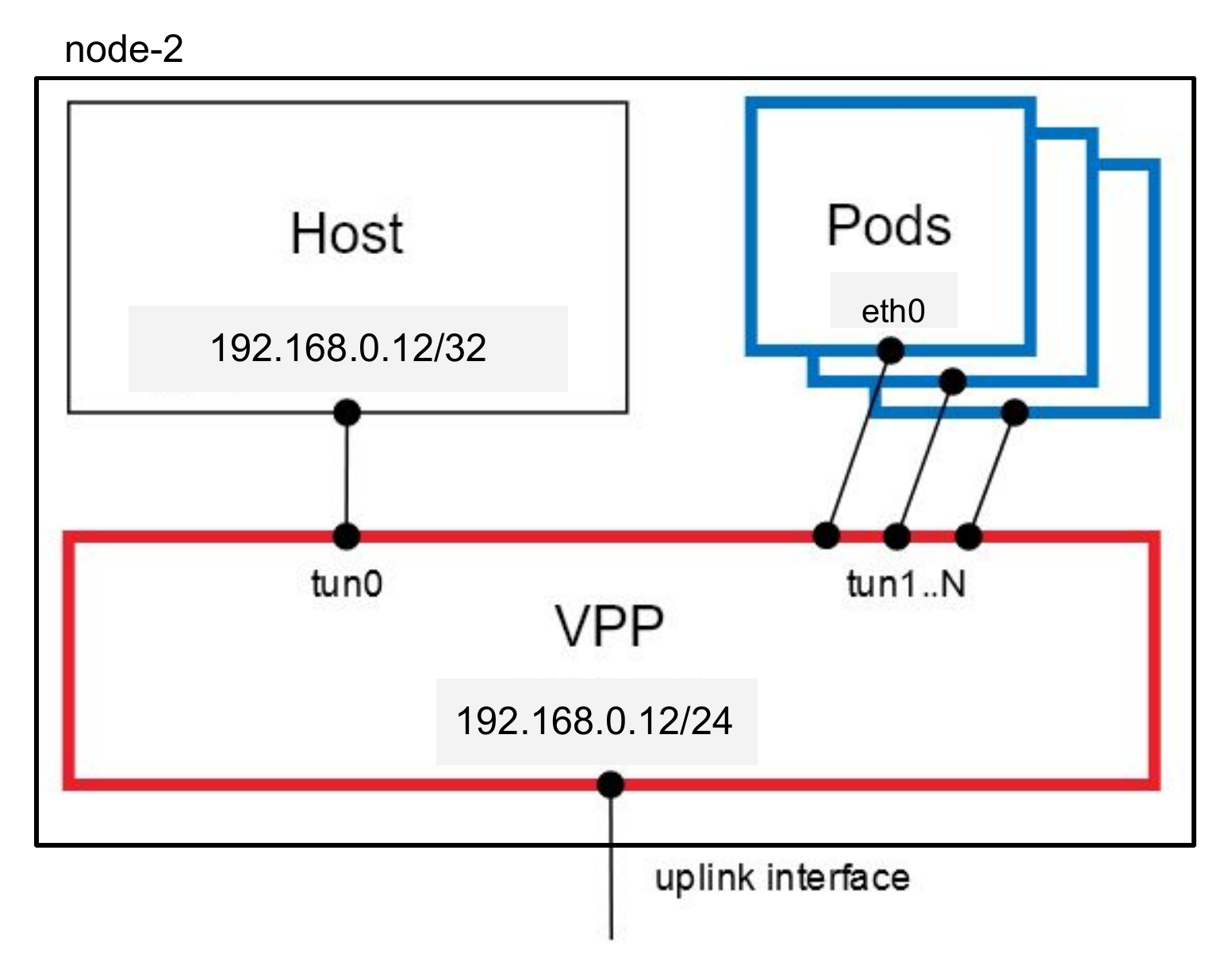}
    \caption{Calico VPP networking configuration in a node}
    \label{fig:vpp-network-conf}
\end{figure}

The networking configuration is represented in Fig.~\ref{fig:vpp-network-conf}, considering as an example node-2 in Fig.~\ref{fig:bgp-mechanism-for-ip-in-ip}. VPP takes full control of the external (\dq{uplink}) interface of the node and creates a set of tun interfaces. One tun interface is connected with the Host (which runs the Kubernetes control components), the other tun interfaces are used to connect the pods. Inside the host and the pods there is a virtual interface which is connected to the tun interfaces. In particular, the virtual interface in the Host is configured with the external infrastructure IP address, so that the Kubernetes control components can transparently use the infrastructure IP addresses to communicate with other Kubernetes control components. On the other hand, the virtual interfaces in the pods (eth0 in the figure) are configured with the pod address belonging to the Pods prefix and VPP performs the encapsulation and decapsulation operations needed to transmit and receive the packets on the infrastructure network.

\section{SRv6 overlay for Calico VPP}
\label{sec:srv6-overlay-calico-vpp}

\subsection{SRv6 basics}
\label{sec:srv6}

Segment Routing for IPv6 (SRv6 for short) is the instantiation of the Segment Routing concept \cite{filsfils2015segment, rfc8402} over the IPv6 dataplane. SRv6 introduces the concept of \emph{network programming} \cite{rfc8986srv6netprog}: the source node provides a list of \emph{segments} which represents a \emph{network program}. Each segment can represent a waypoint and/or an operation to be performed on the packet by a node. The operations that can be performed are also called \emph{behaviors}. A large set of well-known behaviors have been standardized by IETF \cite{rfc8986srv6netprog} and work is ongoing to further extend this set. A complete technical tutorial on SRv6 can be found in the survey \cite{ventre2021survey}, hereafter we provide a basic explanation to help understand our solution.

In SRv6, each segment is identified by an IPv6 address, which is referred to as \emph{Segment IDentifier} (\paper{SID}\thesis{\acrshort{SIDLabel}}). A segment list (i.e. a sequence of SIDs) is inserted by the source node in an Extension Header of the IPv6 header, called \emph{Segment Routing Header} (\paper{SRH}\thesis{\acrshort{SRHLabel}}) \cite{rfc8754srh}. When the SRv6 packet is forwarded, the IPv6 Destination Address is set to the current (or \emph{active}) segment (so the source node will copy the first SID into the Destination Address). In this way, the packet can be simply forwarded considering the IPv6 Destination Address, until the node associated to the active segment is reached. When this node is reached, the SRv6 operation (behavior) associated with the SID will be executed. The simplest operation is denoted as the \emph{End} behavior and it consists in considering the next segment in the segment list carried in the SRH: the active segment becomes the next one, its address is copied in the Destination Address and the packet is forwarded considering the new destination. In this way, a number of waypoints can be added to a packet in order to implement a traffic engineering goal (e.g. avoiding a congested link) or a restoration goal (i.e. avoiding a failure on a node or a link), each waypoint is implemented with an \emph{End} behavior in the node to be crossed. An example of a more complex operation is the \emph{End.X} behavior, where X stands for cross-connect. This behavior forces the forwarding of the packet towards a specific next-hop of the crossed node. This behavior can be used to force the forwarding of packets over some interfaces that would otherwise not be selected by the regular routing and this is again needed in several traffic engineering or restoration scenarios.

A set of operations of our interest are the \emph{encap} and \emph{decap} behaviors, which can be used for VPN services based on SRv6. In these VPN services, the packets of the VPN users can be IPv4 and/or IPv6 and they are encapsulated in IPv6 packets with the SRH header. In particular, the \emph{H.Encaps} behavior is defined as \dq{SR Headend with Encapsulation in an SR Policy}. This operation is executed by the \emph{SR Headend} node, that encapsulates a packet into an outer IPv6 packet with its Segment Routing Header carrying the segment list. Note that the segment list is denoted here \emph{SR Policy}, this notation will be often used in the \paper{paper}\thesis{thesis} from now on. In other words, the SR policy is the list of instructions that the source node adds to the SRv6 packet. Several decapsulation operations are specified in \cite{rfc8986srv6netprog}, we only describe here the two behaviors used in our solution: \emph{End.DT4} and \emph{End.DT6}. \emph{End.DT4} is defined as \dq{Endpoint with decapsulation and specific IPv4 table lookup}. It is expected that an \emph{End.DT4} behavior is the last segment of a segment list. The receiving node that executes the \emph{End.DT4} behavior extracts (decapsulates) the internal packet, which needs to be an IPv4 packet, and then uses a specific routing table to take the forwarding decision for the extracted packet. In this way, it is possible to run a \dq{multi-tenant} VPN and the different tenants can have overlapping address spaces for the \dq{internal} IPv4 address without any problem. The routing table to be used is associated to the SID that identifies of the \emph{End.DT4} behavior. In other words, if a node supports multiple tenants, there will be multiple instances of the \emph{End.DT4} behavior, each one identified with a different SID (IPv6 address). The \emph{End.DT6} behavior works in the same way, but it supports IPv6 user packets.

A basic multi-tenant VPN service can be realized with segment lists containing only one segment: the SID of the \emph{End.DT4} or \emph{End.DT6} is used at the same time to: i) forward the packet up to the destination edge node; ii) trigger the decapsulation operation in the destination node; iii) identify the tenant, i.e. the specific user IPv4 or IPv6 routing table to be considered. Another possibility is to use two segments: the first one to forward the packet to the destination node, the second one to trigger the decapsulation and to identify the tenant. The fist possibility (single- segment) is more efficient as it saves 16 bytes in the encapsulating header, while the second one (double-segment) can be simpler to implement and operate. 

This basic VPN service can be easily extended by combining it with a Traffic Engineering service thanks to the features of SRv6 network programming. All that is needed is to extend the segment list (SR policy) inserted in the source edge node, by prepending a sequence of SIDs representing the needed waypoints (\emph{End} behavior).

In general, the great advantage of using SRv6 network programming with respect to other approaches is that the state information to be configured in the internal nodes is reduced to the minimum because the instructions are carried inside each packet. For example, the Traffic Engineering instructions can be configured (and updated when needed) only in the source edge nodes, while the internal nodes are \dq{stateless} in this respect.


\paper{In our proposed solution, we }\thesis{In this work of thesis we }first show how Kubernetes can use a basic SRv6 based VPN service (subsections~\ref{sec:srv6-vpp-impl} and~\ref{sec:design-of-srv6-overlay-vpp}). In particular, we will use the End.DT6 and End.DT4 behaviors thanks to the support offered by the VPP dataplane.  Then we show how this solution can be extended to a VPN that offers Traffic Engineering capabilities (subsections~\ref{sec:srv6-te-overlay-bgp} and~\ref{sec:srv6-te-overlay-configmap}).  

\subsection{VPP implementation of SRv6 based VPN}
\label{sec:srv6-vpp-impl}

\begin{figure*}[ht]
    \centering
    \includegraphics[width=\textwidth]{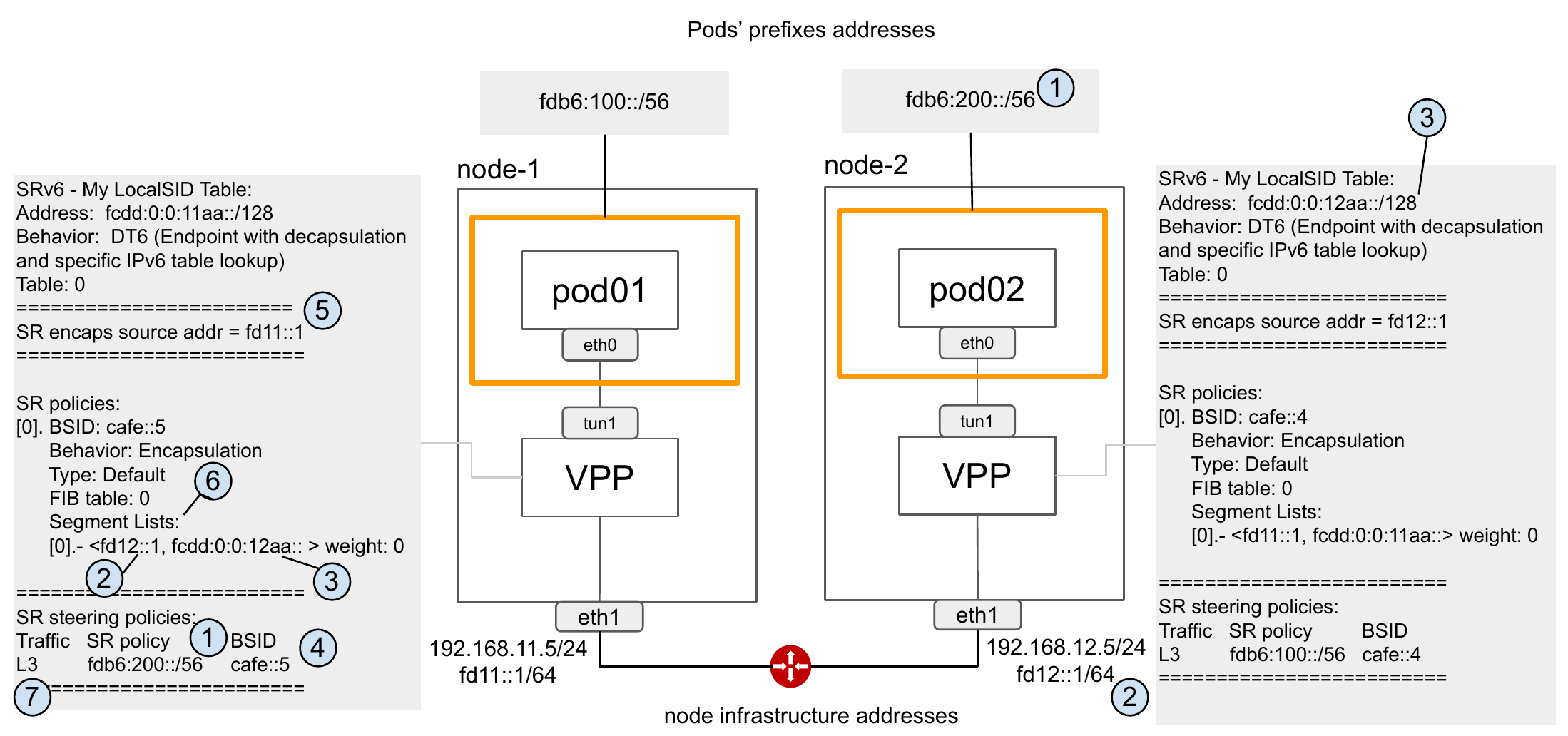}
    \caption{Two nodes of a cluster in different remote subnets with their SRv6 configuration}
    \label{fig:two-nodes-srv6-conf}
\end{figure*}

Let us describe how an SRv6 tunnel can be established using the VPP dataplane between two nodes to encapsulate (and decapsulate) packets belonging to the Pods networks. We refer to Fig.~\ref{fig:two-nodes-srv6-conf}, assuming that node-1 is the \textit{source} node that encapsulates the packets and node-2 is the \textit{destination} node that receives the packets destined to the pod network and executes the decapsulation function (End.DT6 or End.DT4). The source and destination nodes respectively play the role of \emph{ingress} and \emph{egress}  nodes of an SRv6 domain. The VPP configuration of node-1 and of node-2 are reported in Fig.~\ref{fig:two-nodes-srv6-conf} respectively on the left and on the right of the figure.

The \emph{SR localSID} is the Segment Identifier (i.e. an IPv6 address) locally associated with the decapsulation function to be executed by the \textit{destination} node (3 in the right of Fig.~\ref{fig:two-nodes-srv6-conf}). The SR localSID is added as last element of the Segment List inserted by the encapsulating (\textit{source}) node (3 in the left of Fig.~\ref{fig:two-nodes-srv6-conf}). Therefore, the source node that performs the encapsulation (node-1 in our example) needs to know the SR localSID used in the destination node (node-2). When the destination node processes this localSID in the Segment List, it understands that that packet needs to be decapsulated and then delivered towards the destination Pod (we are considering here a single tenant solution).

The \emph{SR policy} defines the Segment List (6 in  Fig.~\ref{fig:two-nodes-srv6-conf}) that will be applied to a packet by the source node. A packet is \textit{steered} into an SR policy with a classification based on its IP destination address. The packet is encapsulated in an outer packet, and the Segment List corresponding to the policy is written in the Segment Routing Header (SRH) of the outer packet.

In VPP, an SR policy is identified by a \emph{Binding SID} (4 in  Fig.~\ref{fig:two-nodes-srv6-conf}). This binding SID is used to configure the classification and encapsulation procedures in the source node. In particular, the classification is defined by adding a \emph{steering rule}  (7 in  Fig.~\ref{fig:two-nodes-srv6-conf}) that associates a destination prefix with a \thesis{\acrlong{BSIDLabel}} \paper{Binding SID}. In turn, the Binding SID is associated with the Segment List, and this enforces the proper encapsulation of the packet by the VPP dataplane.

In the source node, VPP also requires the configuration of the source address of the outer packet to be used in the SRv6 tunnel (5 in  Fig.~\ref{fig:two-nodes-srv6-conf}). 

The BGP based communication mechanism that we have described in Section~\ref{sec:ip-ip-calico} (Fig.~\ref{fig:bgp-mechanism-for-ip-in-ip}) cannot be used to communicate the information needed to configure VPP as explained above, therefore we needed to extend this mechanism. The following subsection describes the approach we have designed and implemented.

\subsection{Design of the SRv6 overlay for Calico-VPP}
\label{sec:design-of-srv6-overlay-vpp}

In this subsection, we present the design of our solution that extends Calico-VPP enabling the support of SRv6 overlays. 

In a generic large-scale and multi-data center scenario, our goal is to have a cluster of Kubernetes nodes interconnected by an SRv6 overlay where the pods of each node can communicate with each other. The obvious preliminary requirement to create the SRv6 overlay is that the infrastructure supports IPv6 connectivity among the nodes. Note that the pods do not necessarily need to use IPv6, they can either use IPv4 or IPv6, as both IPv4 and IPv6 networking at the pod level are supported.

As highlighted in Section \ref{sec:ip-ip-calico} the operations of the existing IP-in-IP overlay of Calico-VPP are \emph{dynamic} and \emph{automatic} and rely on the initial configuration of the networking plugin by the network administrator. In this section, we will show how the operations of the SRv6 overlay are more complex as they require the correct configuration of several parameters that need to be aligned among the different nodes. The research challenge, as anticipated in the Introduction, is to design the mechanisms for the dynamic and automatic operations, keeping at a minimum the configuration complexity.

Let us consider the same scenario discussed in the previous subsection, with two nodes, denoted as node-1 and node-2, which respectively host a pod denoted pod01 and a pod denoted as pod02, as shown in Fig.~\ref{fig:two-nodes-srv6-conf}.
The two nodes are SRv6 enabled with the SRv6 implementation based on VPP. As explained in Section \ref{sec:srv6-vpp-impl}, the SR localSIDs in node-1 and node-2 associated with the decapsulation functions of the SRv6 overlay must be allocated. In particular, on each node two localSIDs are needed, one for the DT4 behavior and one for the DT6 behavior. The DT4 behavior is needed to extract the encapsulated IPv4 packets when pods use IPv4 addressing, while the DT6 behavior is needed when pods use IPv6 addresses. Both DT4 and DT6 can be used at the same time if needed.

Let us start by considering the basic scenario without Traffic Engineering. In order to create an SRv6 overlay between the two nodes using the VPP dataplane, the following information is needed in node-1 for the tunnel from node-1 to node-2, as explained in Subsection \ref{sec:srv6-vpp-impl} (the number refers to Fig.~\ref{fig:two-nodes-srv6-conf}):
\begin{itemize}
    \item (1) The Pods prefixes addresses (IPv4 and/or IPv6) assigned in node-2
    \item (2) The IPv6 infrastructure address of node-2
    \item (3) The SR localSID(s) assigned in node-2
    \item (4) The Binding SID (BSID) to be used in node-1
    \item (5) The source IPv6 addresses for the outer packets 
\end{itemize}

With this information, node-1 is able to create the two SR policies (for IPv4 and IPv6 traffic) and the two Layer 3 steering rules that match the pods addresses in node-2. The information needed in node-2 to set up the tunnel from node-2 to node-1 is obviously symmetric.

As explained in Section~\ref{sec:ip-ip-calico}, in Calico-VPP for each Pods prefix assigned to a node, a BGP UPDATE message is sent by the Calico-VPP agent toward all the remote nodes. We have re-used this mechanism, as shown in Fig.~\ref{fig:bgp-mechanism-for-ip-in-ip} (step 1). In particular, this message only contains the Pods prefix address (1) and the infrastructure IPv6 address (2). There is an important difference with respect to the mechanism currently implemented for the IP-in-IP overlay in Calico-VPP: for the SRv6 overlay we always have to communicate the IPv6 infrastructure address of the node, as the SRv6 tunnels are based on (external) IPv6 addresses and can support (internal) IPv4 or IPv6 addresses for the pods network. The existing IP-in-IP overlay uses the IPv4 infrastructure address and can support only IPv4 pods addresses.

\begin{figure}[ht]
    \centering
    \includegraphics[width=\columnwidth]{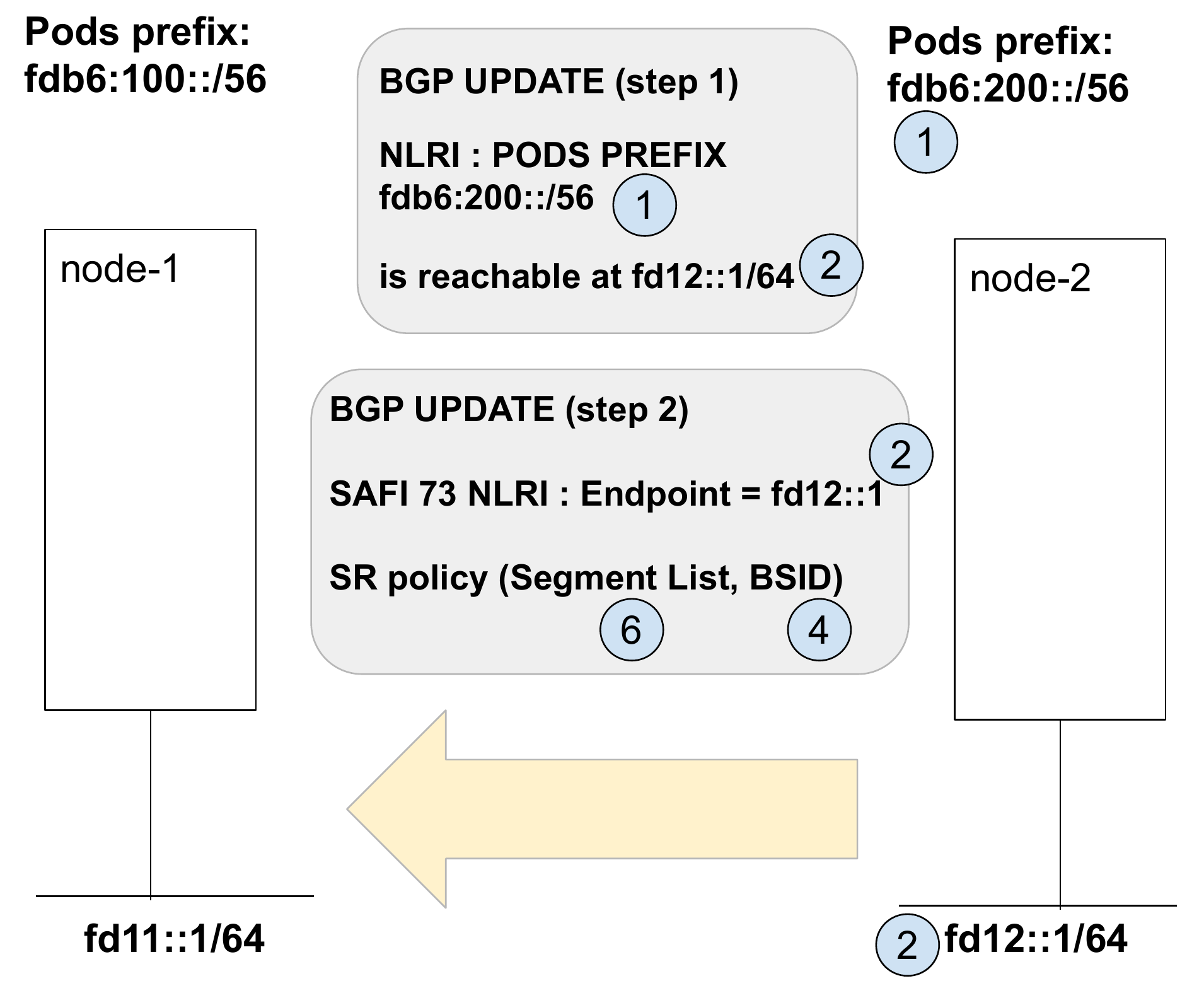}
    \caption{BGP mechanism for SRv6 tunnels}
    \label{fig:bgp-mechanism-for-srv6}
\end{figure}

In order to communicate the local SIDs and the Binding SIDs we decided to add a second phase with a separate BGP UDPATE message. The modified Calico-VPP agent will create this second UPDATE message for each SR localSID supported by the node (e.g. End.DT4 and/or End.DT6), carrying the SR policy (6) and the BSID (4), as shown in Fig.~\ref{fig:bgp-mechanism-for-srv6} (step 2). This UPDATE message must be sent to all remote nodes (only two nodes are represented in Fig.~\ref{fig:bgp-mechanism-for-srv6} for simplicity).

\begin{figure}[ht]
    \centering
    \includegraphics[width=\columnwidth]{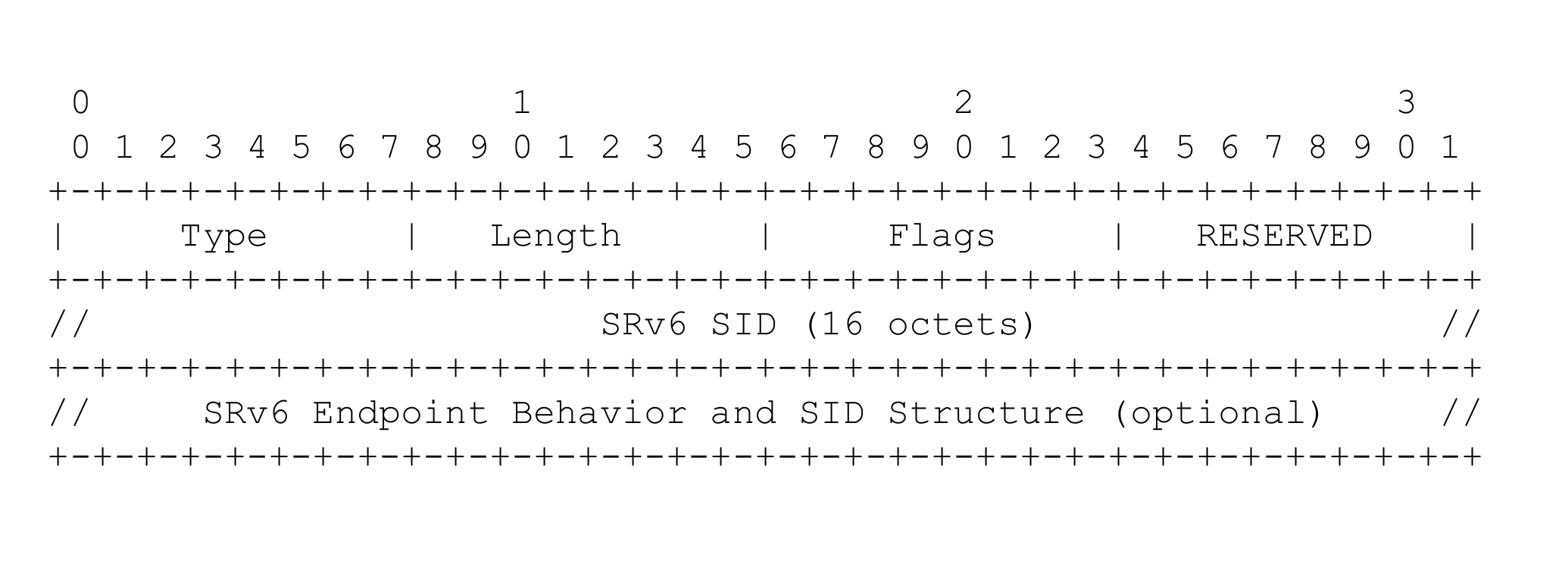}
    \caption{SRv6-sid-tlv}
    \label{fig:SRv6-sid-tlv}
\end{figure}

For this second message, we decided to leverage the advertising of candidate path of a Segment Routing (SR) Policy using the BGP \paper{SAFI}\thesis{\acrshort{SAFILabel}} (Subsequent Address Family Identifiers) defined in \cite{draft-ietf-idr-segment-routing-te-policy} (BGP SAFI 73). The structure of this SR Policy SAFI is shown in Listing~\ref{lst:sr-policy-encoding}. The SR Policy SAFI encoding structure contains all the information needed to advertise a generic SR Policy, in particular the Binding SID (BSID) and the Segment List. Regarding the Segment List it can support several types of Segments as defined in~\cite{draft-ietf-idr-segment-routing-te-policy}, for our purposes we used the so-called Type B that represents a SRv6 SID (see Fig.~\ref{fig:SRv6-sid-tlv}).

\begin{lstlisting}[float,floatplacement=h,caption={SRv6 Policy Encoding},label={lst:sr-policy-encoding}, basicstyle=\footnotesize,stepnumber=1,showstringspaces=false,tabsize=4,breaklines=true,breakatwhitespace=false,xleftmargin=0.8cm, belowskip=-0.8 \baselineskip]
SR Policy SAFI NLRI: 
    <Distinguisher, Policy-Color, Endpoint>
   Attributes:
      Tunnel Encaps Attribute (23)
        Tunnel Type: SR Policy
            Binding SID
            SRv6 Binding SID
            Preference
            Priority
            Policy Name
            Policy Candidate Path Name
            Explicit NULL Label Policy (ENLP)
            Segment List
                Weight
                Segment
                Segment
                ...
            ...
\end{lstlisting}

The BGP messages received by the nodes are processed by the GoBGP daemon inside the Calico-VPP agent. Upon receiving the first BGP UPDATE (step 1 in Fig.~\ref{fig:bgp-mechanism-for-srv6}), no VPP configuration operations are executed. The received information (infrastructure address of the sending node, Pods prefix) is stored in a data structure in the memory (RAM). When the second BGP UPDATE is received (step 2 in Fig.~\ref{fig:bgp-mechanism-for-srv6}) the \emph{endpoint} of the NLRI (Network Layer Reachability Information) is again the infrastructure address of the sending node. After receiving this, the agent of the receiver node can lookup the infrastructure address in the data structure in memory, retrieving the Pods prefix associated to it. By combining the content of the two messages, the agent has the information needed to create an SR policy with a related SR steering rule using the VPP dataplane, with the exception of the source IPv6 address to be used for the SRv6 tunnels. In our solution, we simply use the node infrastructure address as source IPv6 address for the VPP configuration. Specifically, our Connectivity Provider retrieves the node infrastructure address during its startup. The retrieved address is used to set the source encapsulation using the VPP configuration interface.

We have explained how the information is communicated by each destination node toward all the other nodes that can properly configure the encapsulation operations in the VPP dataplane. Let us now discuss how the destination nodes allocate the addresses and choose the SR policy that is distributed using the second BGP update.

As mentioned in Section \ref{sec:srv6} we need to assign a localSID for each SRv6 function, in our case End.DT4 and End.DT6. We used the IP Address Management (IPAM) system of Kubernetes to manage the allocation of the localSIDs. Using the \paper{IPAM}\thesis{\acrshort{IPAMLabel}} it is possible to configure a pool of IP addresses (called \emph{IPPool}) for a given purpose, so that a node can request the IPAM to allocate an address from a specific IPPool. Using the IPAM system, each node will automatically select a suitable address without the need for manual allocation. Of course, a proper configuration of the IPAM system at cluster level is needed. 

As discussed earlier, there are two options for an SRv6 basic VPN: single-segment and double-segment. In the double-segment case, the localSIDs do not need to be routable, we configure a single IPPool with an arbitrary prefix and all nodes can ask the IPAM to assign a localSID from this common pool. In the single-segment case the localSIDs for End.DT4 and/or End.DT6 need to be routable up to the destination node. For this reason, we would need to configure a different IPPool specific for each node, using a prefix that will be routed in the infrastructure network up to the node itself. In our prototype, we have used the double-segment approach, i.e. the destination nodes prepare the step 2 BGP UPDATE messages using SR policies with two segments. Note that for the source nodes that receive the UPDATE messages, there is no difference in the procedure as they use the received SR policy without caring for the number of segments inside the policy.





As for the binding SIDs, they are chosen by the destination nodes and communicated to the source nodes, hence we need to avoid that two destination nodes send the same binding SIDs for two different SR policies to the same node. We solve this issue by using a single common IPPool for which all nodes allocate the binding SID. In this way, we can guarantee that the binding SIDs are unique across all the networks, while it would suffice to have them unique for each single node. Considering the huge address space of IPv6, this is not an issue and it avoids configuring a different IPPool for each node for the binding SIDs.


\begin{figure}[ht]
    \centering
    \includegraphics[width=0.6\columnwidth]{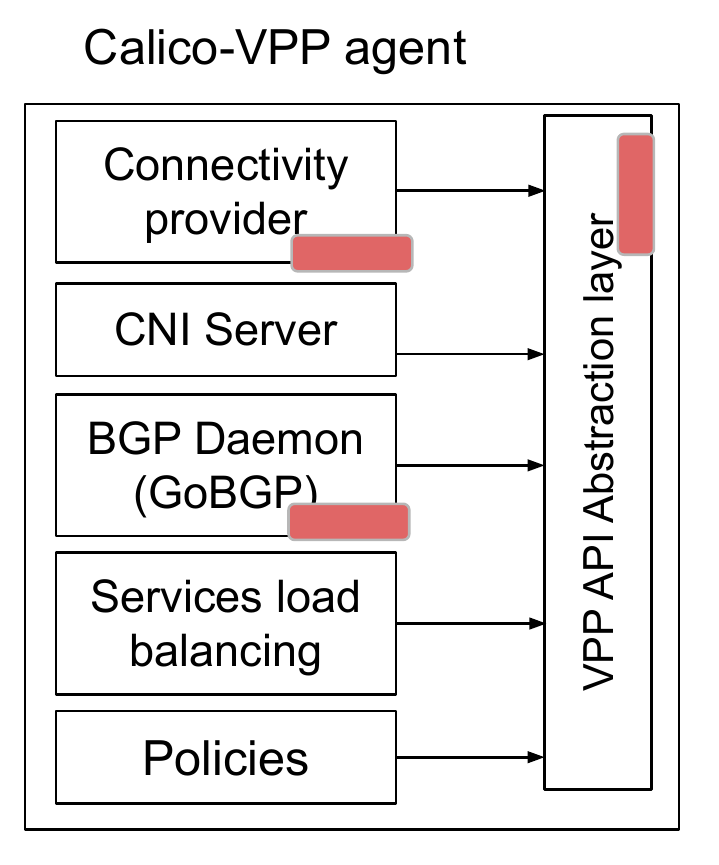}
    \caption{Modifications to Calico-VPP compoments}
    \label{fig:contributions-to-calico-vpp}
\end{figure}

Taking into account the software architecture, the components that have been modified to implement the proposed design are shown in Fig.~\ref{fig:contributions-to-calico-vpp}. In the GoBGP daemon, we have extended the existing support for the SR Policy BGP SAFI (73), in particular we implemented the support of the segment Type B. This modification has been merged upstream in the GoBGP distribution and it is now part of the official release. 


We have added a new version of the Connectivity Provider that is SRv6 capable and can send the SRv6 configuration commands to the VPP manager using the RPC API provided by goVPP. For the new Connectivity provider to work, we had to extend the VPP API abstraction layer so that it can deal with SRv6. The new Connectivity Provider and the extensions to the VPP API Abstraction Layer have been merged upstream in the Calico-VPP project and are now part of the official release. This means that it is possible to configure a Kubernetes cluster to use the SRv6 overlay just by properly modifying the configuration files. In this respect, we have achieved the \dq{feature parity} of the proposed SRv6 overlay with the existing IP-in-IP overlay. Note that we support IPv6 traffic in the overlay and in the transport infrastructure, while the existing IP-in-IP overlay only supports IPv4.

\subsection{SRv6-TE overlay with BGP}
\label{sec:srv6-te-overlay-bgp}



We have shown in the previous section how to implement an SRv6 overlay that has the same features of IP in IP overlay (i.e. it properly encapsulates the packets, forwards them to the destination node and decapsulates them). In this section, we show how we can take advantage of the SRv6 overlay for supporting more advanced features, in particular the capability of specifying the path to be followed by a tunnel in the infrastructure network taking into account Traffic Engineering aspects.

In the basic SRv6 overlay approach described so far, assuming that N nodes belong to the cluster and runs the pods, there will be a full mesh of tunnels between all the N nodes to interconnect all pods network with each other. In the infrastructure network, the tunnels will just follow the default (best-effort) paths and some links may become congested while other links remain under-utilized. This can especially happen when Kubernetes is used in large-scale and/or distributed multi-datacenter scenarios (such as for edge computing). In these scenarios, it can be useful to allocate some tunnels to specific Traffic Engineered Paths (TE Paths) to distribute load and avoid congestion. We refer to this approach as \emph{SRv6-TE overlay}.

As discussed in Subsection~\ref{sec:srv6}, SR-TE overlay is done by adding waypoints to the SR policy associated with a tunnel. The fundamental difference with respect to the basic overlay (i.e. without TE) is that the SR policies used to reach a given destination node may be different, while in the basic case all the SR policies that define tunnels used to reach a destination are identical (using either a single-segment or a double-segment approach). This means that in the TE enabled overlay the destination node (the tunnel decapsulation node) should communicate to each source node a potentially different SR policy. 

\begin{figure}[ht]
    \centering
    \includegraphics[width=\columnwidth]{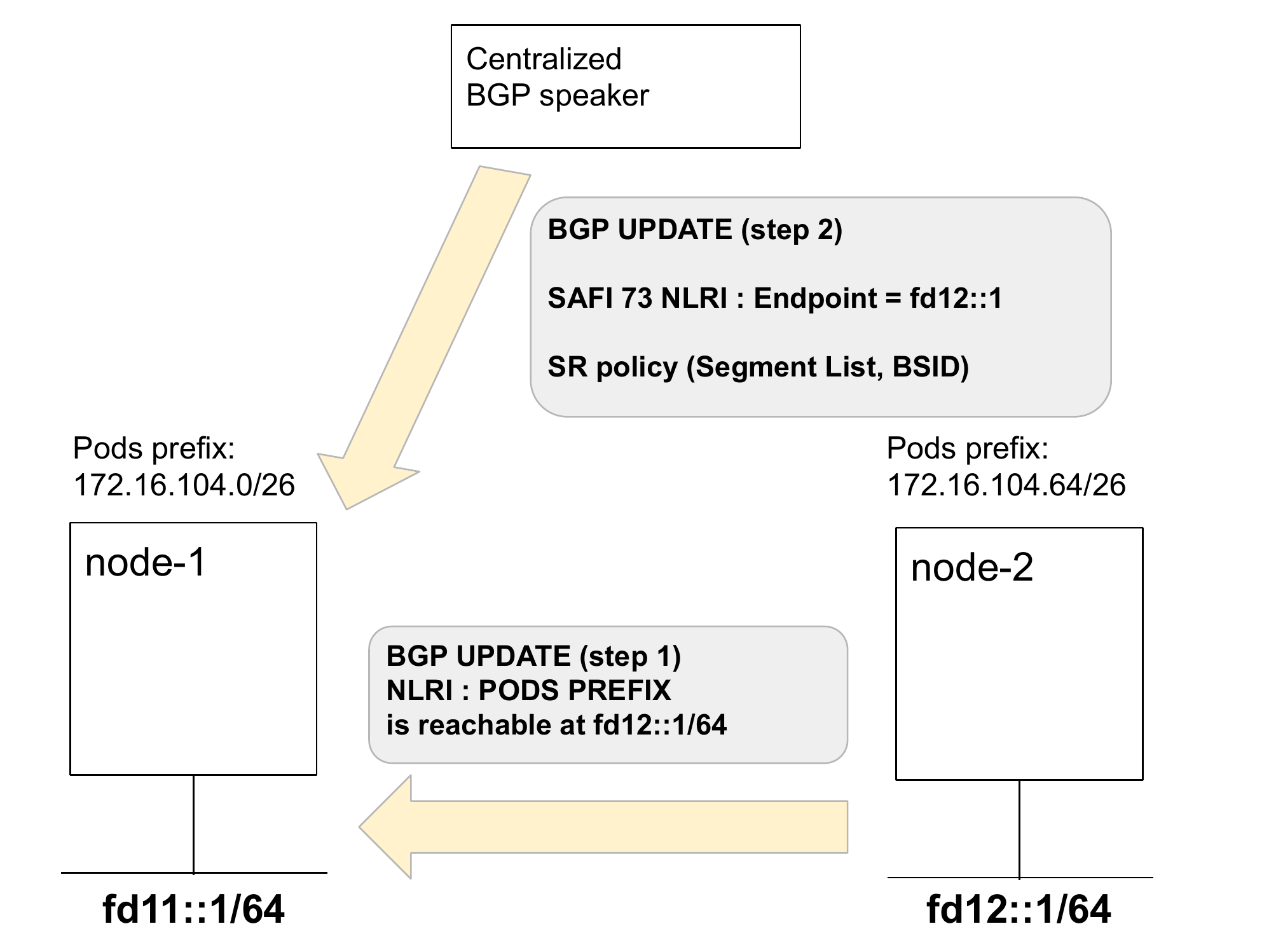}
    \caption{BGP mechanism for SRv6-TE overlay}
    \label{fig:bgp-te-mechanism-for-srv6}
\end{figure}

We think it is not a good approach, considering that typically the TE paths are not autonomously evaluated by the destination nodes. A centralized entity (we refer to it as \emph{TE engine}) is responsible for selecting the TE paths. If we keep the same BGP signalling approach illustrated in Fig.~\ref{fig:bgp-mechanism-for-srv6}, the centralized TE engine should send the appropriate SR policies to all destination nodes, which can then send the step 2 BGP updates to the source nodes. A much better solution in our opinion is that the centralized TE engine directly injects the SR policies into the source nodes. The BGP protocol natively supports this approach, as any BGP speaker (with the proper authorization) can interact with a BGP node and send the required BGP UPDATE messages. The proposed approach is visualized in \ref{fig:bgp-te-mechanism-for-srv6}. The centralized TE engine obviously needs to have a vision of the nodes and of the infrastructure network topology, according to an SDN-based approach. 

Note that if the TE path of a tunnel needs to be updated, the centralized TE engine can send a new BGP UPDATE message (step 2) with the updated SR policy. 

\begin{figure}[ht]
    \centering
    \includegraphics[width=\columnwidth]{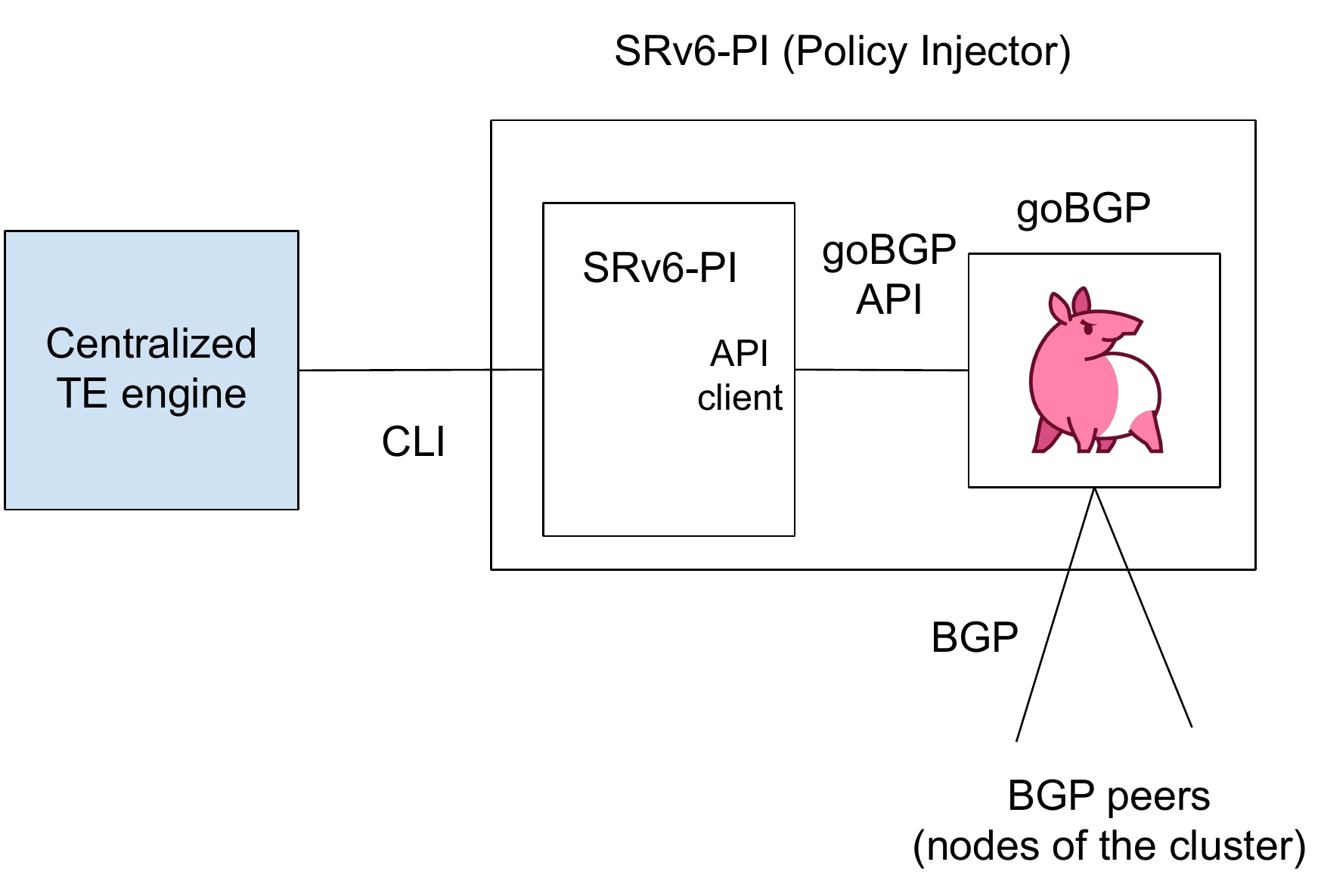}
    \caption{Architecture of SRv6-PI (Policy Injector)}
    \label{fig:srv6-pi}
\end{figure}

To demonstrate the feasibility of our solution, we have implemented a BGP peer capable of injecting the policies to dynamically modify the paths between the various nodes involved. We have called this entity SRv6-PI (SRv6 Policy Injector). Its role is similar to that of a BGP Route Reflector, as it is a logically centralized peer that interacts with all other peers and avoids the needs of a full mesh of BGP connections. The SRv6-PI is implemented in go and is based on a go BGP client that uses the API offered by goBGP, as shown in Fig.~\ref{fig:srv6-pi}. SRv6-PI offers a CLI to insert the SR policies, and it is meant to be a generic re-usable component. A centralized TE engine that selects the TE paths can use the services offered by SRv6-PI to inject the policies into all BGP peers. The SRv6-PI implementation is open source and available at~\cite{rose-k8s-srv6}.

In order to integrate the proposed solution with a Kubernetes cluster using the Calico-VPP plugin, the cluster nodes must be able to accept BGP policies from any BGP-Peer that is not necessarily part of the cluster. Calico already has a configuration option to configure external BGP peers other than cluster nodes, so we simply had to enable this option in the cluster configuration (see \cite{web-calico-conf-bgp-peering}). Apart from this configuration, it is important to note that the SRv6-TE overlay solution based on BGP with a centralized speaker does not need any changes to the Calico-VPP agent source code with respect to the basic SRv6 overlay. In fact, the destination node can perform step 1 and step 2 as in the basic overlay, so that the source nodes will setup the full mesh of SRv6 tunnels using best-effort paths. Afterwards, the centralized entity can decide to select TE paths for all or for some of the tunnels, and consequently will send the BGP updates (step 2) to the relevant source nodes. Note also that from the point of view of the source node, there is no difference if a received SR policy includes waypoints and hence it is a TE path or if it only includes the SID(s) to define the basic tunnel with no TE. This is an interesting properties of the SRv6 overlay solution, that offers a seamless coexistence of best effort and TE paths (and a smooth transition from the basic best-effort solution and a \paper{TE}\thesis{\acrshort{TELabel}} enabled overlay).


\subsection{SRv6-TE overlay with Kubernetes control plane}
\label{sec:srv6-te-overlay-configmap}


The solution described in the previous subsection is based on the BGP protocol. BGP messages are used to dynamically configure the Calico-VPP agents running in the cluster nodes with the TE paths associated with the overlay tunnels. Considering that the cluster nodes are also participating in the Kubernetes control plane, we have considered also an alternative design. The idea is that the Calico-VPP agents can be triggered and (re)configured using Kubernetes control APIs instead of the BGP protocol. With this approach, the centralized TE engine that selects the SR policies does not need to use a BGP speaker to interact with the Calico-VPP agents in the cluster nodes, but it can use Kubernetes native communication mechanisms in the control plane.

In particular, a Kubernetes ConfigMap \cite{config-map} is an API object that can be used to inject containers with configuration data separately from application code. By writing information in a ConfigMap, the administrator of the cluster can distribute the configuration over the nodes of the cluster. The ConfigMap is logically a key/value store. The information written in a ConfigMap can be consumed in different ways, in our case the Calico-VPP agents subscribe to get updates whenever the ConfigMap changes, and can react when that happens (more details later).  

Compared to SRv6-TE with BGP (see Fig.~\ref{fig:bgp-te-mechanism-for-srv6}), we have decided to keep using BGP for step 1, that is, to distribute the prefixes of the Pods subnets and associate them to the infrastructure address of the node. We use the ConfigMap based mechanism for step 2, i.e. to distribute the SR policies with the TE paths for the tunnels from each ingress node to each egress node. The main reason is to minimize the differences in the implementation of the Calico-VPP agent for the different overlays. In fact, step 1 is performed in the same way in the IP-in-IP overlay, in the basic SRv6 overlay and in the SRv6-TE overlay and no changes to the code base are needed. 

The information that needs to be provided to a given source node (i.e. the node that encapsulates the packet) for all the destination nodes is: the infrastructure node address of the destination node, the SR policy and the BSID to be used for the configuration of the tunnel towards the destination node. This information is encoded in YAML format as shown in Listing~\ref{lst:example-config map}. The example refers to a single source node and includes two destination nodes (\emph{egress\_node} in the YAML), the details for the second destination node are omitted. In addition to the information mentioned above, we have also added the localSIDs to be used by the node when it acts as the destination node. Using this information we can avoid using the IPAM and configuring the IPPool for the allocation of localSIDs (the approach described in Section~\ref{sec:design-of-srv6-overlay-vpp}), as the node can directly read its localSIDs from the ConfigMap.

\begin{lstlisting}[float,floatplacement=h,caption={Definition of SR-TE policies with ConfigMap},label={lst:example-config map}, basicstyle=\footnotesize,stepnumber=1,showstringspaces=false,tabsize=4,breaklines=true,breakatwhitespace=false,xleftmargin=0.8cm, belowskip=-0.8 \baselineskip]
localsids:
  DT4: "fcdd::aa:34b8:247c:36da:db44"
  DT6: "fcdd::aa:34b8:247c:36da:db45"
node: master
policies:
  - egress_node: "fd11::1000"
    bsid: "cafe::1c3"
    segment_list:
    - "fcff:3::1"
    - "fcdd::11aa:c11:b42f:f17e:a683"
    traffic: IPv6
  - egress_node: "fd15::1000"
    [...]
\end{lstlisting}

In the first version of our prototype, we use a single ConfigMap for all source nodes. The key used to store the information is the node ID of the source node. The value associated with a node ID is a YAML object, as shown in Listing~\ref{lst:example-config map}. The Calico-VPP agent in each node registers for changes to this ConfigMap. When it is notified of a change, the agent makes a query using its node ID as key. Note that this is a query to the \emph{kube-API-server} in the Kubernetes control plane. The API server returns the YAML object, which includes the sequence of policies for all the destination nodes. The source node compares each received policy with the one that is currently associated to the destination node, if there is a change it will update the tunnel. 

The drawback of this solution is that each source node will be notified of all changes, also when these changes only affect other source nodes. This creates a number of unneeded queries to the kube-API-server (increasing the control traffic in the network and the load on the kube-API-server). It also wastes CPU resources in the source nodes that will have to scan through the sequence of policies and check one by one if a policy toward a destination node has changed. Hence, we have designed a more efficient solution, with a separate ConfigMap for each source node. We used a naming convention for this set of ConfigMaps, based on the node ID of the source node. Using this convention, the centralized TE engine that needs to update a tunnel belonging to a given source node can identify the ConfigMap to be used. At the same time, the source node will subscribe to the updates related to the ConfigMap of the node itself and it will only be triggered by the changes of its interest. This second solution may reduce query traffic and related CPU load in the server and in the agent by up to a factor N, where N is the number of nodes in the cluster. 

Note that the under the hood there is no notification message coming from the control plane when the ConfigMap is changed. Rather, the agent performs a periodic poll to the kube-API-server checking if there is a new version of the ConfigMap. This polling traffic (and the related processing load on kube-api-server and on the agents) is constant in the first and in the second solution. The polling load on the agent does not depend on the number of nodes, while it grows linearly with the number of nodes in the kube-API-server. 

In a production system, the configuration of the SR policies in the ConfigMaps would be performed by the centralized TE engine. In our prototype, we simply use the kubectl command line to write the content of a local file into the ConfigMaps.   

In our proposed solution, it is possible to configure a cluster to use the BGP based solution or the Kubernetes control plane based solution using a configuration option (which is stored in the internal ConfigMap used by the Calico-VPP agent). We call them \emph{BGP mode} and \emph{ConfigMap mode} respectively. When the ConfigMap mode is activated: 1) destination nodes do not perform step 2 BGP update; 2) source nodes ignore the received step 2 BGP updates; 3) the source nodes subscribe to changes to the ConfigMap associated to their node ID and (re)configure the tunnels using the policies contained in the ConfigMaps. In particular, the Calico-VPP agents in the source node watch trough the Kubernetes API the changes that occur to the ConfigMap and react configuring the VPP dataplane with the received SR policies (TE paths).



Let us compare the solution based on BGP and the one based on Kubernetes control plane (using the ConfigMaps) for implementing the SRv6-TE overlay. The BGP-based solution has the advantage of already being compatible with the Calico-VPP mainstream release. It leverages BGP signalling, therefore it can also be used to interact with remote nodes that do not belong to the Kubernetes cluster; this may be needed in specific scenarios as will be discussed later. On the other hand, when all nodes belong to the Kubernetes cluster and there is no specific need of using BGP, the use of the Kubernetes control plane has the main advantages of being conceptually simpler and hence it can facilitate the development of new features. For this reason we believe that it is worth adding the support of the ConfigMap based approach as an option in the Calico-VPP mainstream release and the related work is ongoing. 

The solution based on Kubernetes control plane (ConfigMap) is available in a public repository forked by the Calico-VPP as it has to be discussed with the Calico-VPP community. It represents a departure from the current model based on BGP; therefore, the potential advantages should be compared with the disadvantages of adding a second solution which increases the code base to be maintained and with the needed work for code review. 



\section{Experimental Testbeds}

\label{sec:testbed}


We have deployed the proposed solutions in two replicable virtual testbeds, which we refer to as \emph{basic testbed} and \emph{full testbed}. All the steps to setup the testbeds and replicate our experiments are reported in a detailed walk-through available in~\cite{rose-k8s-srv6}.

\begin{figure}[ht]
    \centering
    \includegraphics[width=\paper{0.48}\thesis{1}\textwidth]{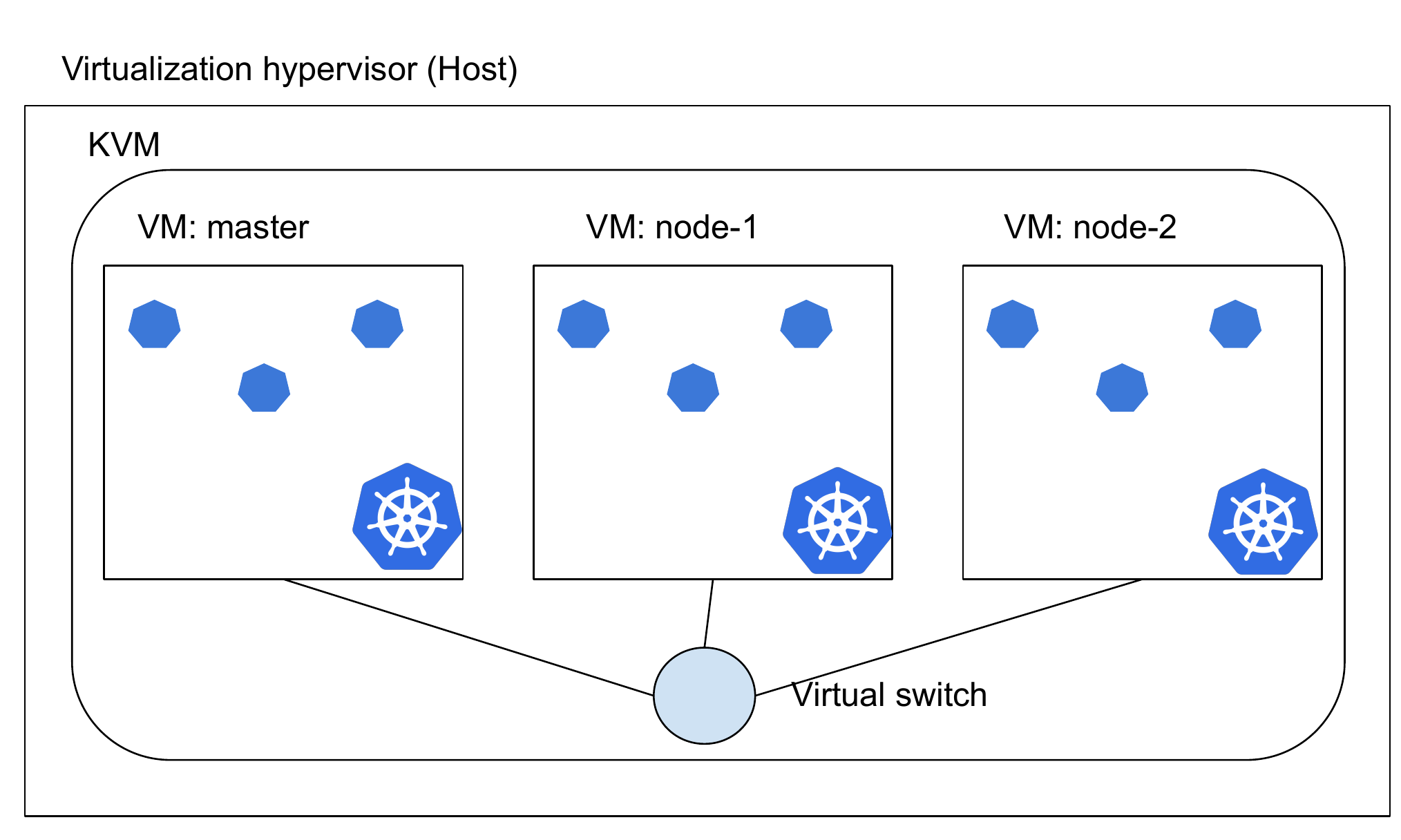}
    \caption{Basic testbed (VMs on KVM)}
    \label{fig:basic-testbed}
\end{figure}

The \emph{basic testbed} is used mostly for development. It is based on a simple switched network with three nodes: one master and two workers. The nodes are Virtual Machines \paper{(VMs)}\thesis{(\acrshort{VMsLabel})}running in a \paper{KVM}\thesis{\acrshort{KVMLabel}} hypervisor inside a Linux Host. The three VMs are connected via a virtual switch provided by KVM, as shown in Fig.~\ref{fig:basic-testbed}. This configuration is derived from the Calico-VPP development testbed, the setup of the testbed is facilitated by the Vagrant tool~\cite{vagrant}.

\paper{
\begin{figure*}[ht!]
    \centering
    \includegraphics[width=0.8\textwidth]{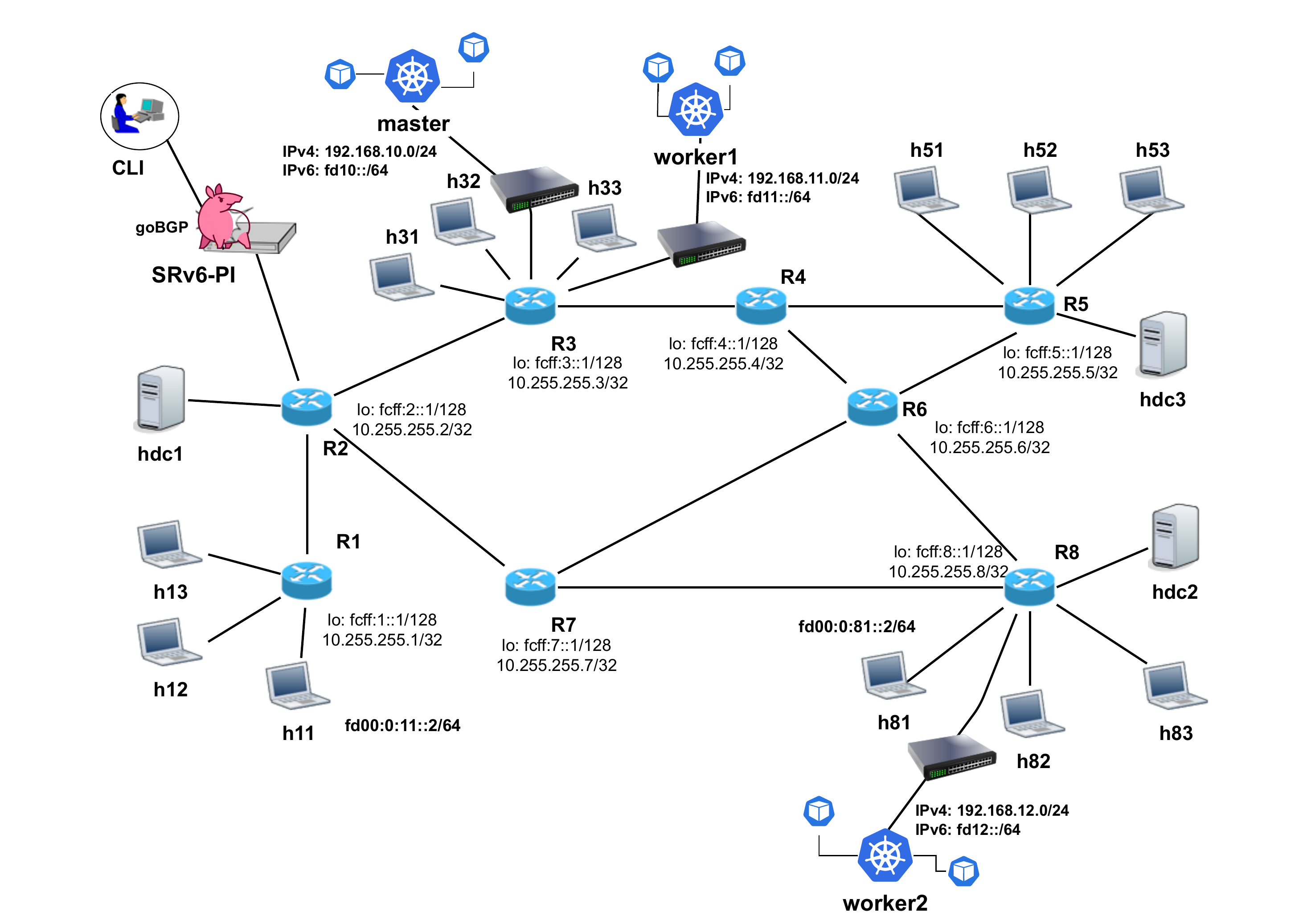}
    \caption{Full testbed with emulated network backbone}
    \label{fig:srv6-te-with-bgp}
\end{figure*}}

\thesis{
\begin{figure}[ht]
    \centering
    \includegraphics[width=1\textwidth]{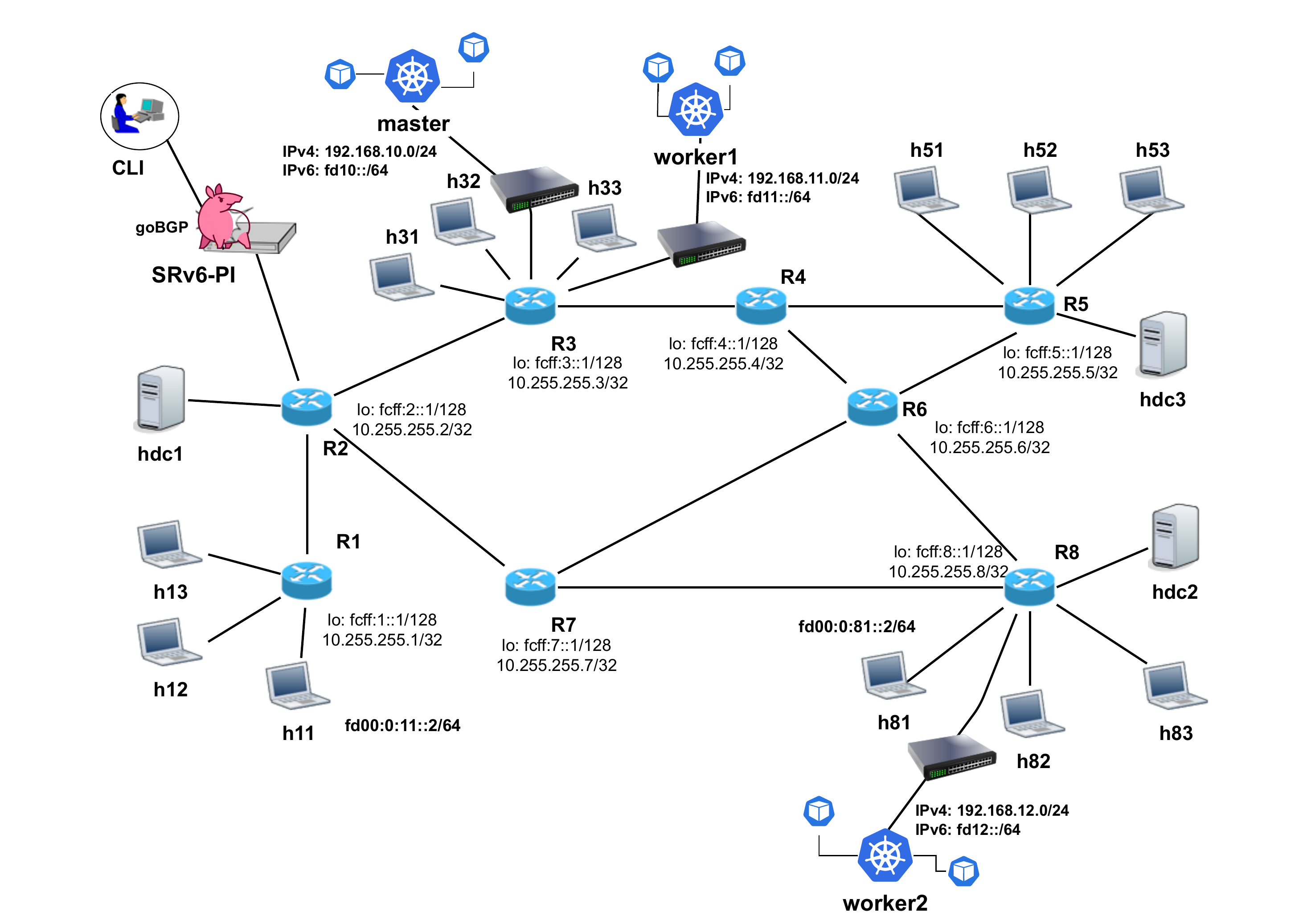}
    \caption{Full testbed with emulated network backbone}
    \label{fig:srv6-te-with-bgp}
\end{figure}}

\begin{figure}[ht]
    \centering
    \includegraphics[width=\paper{0.48}\thesis{1}\textwidth]{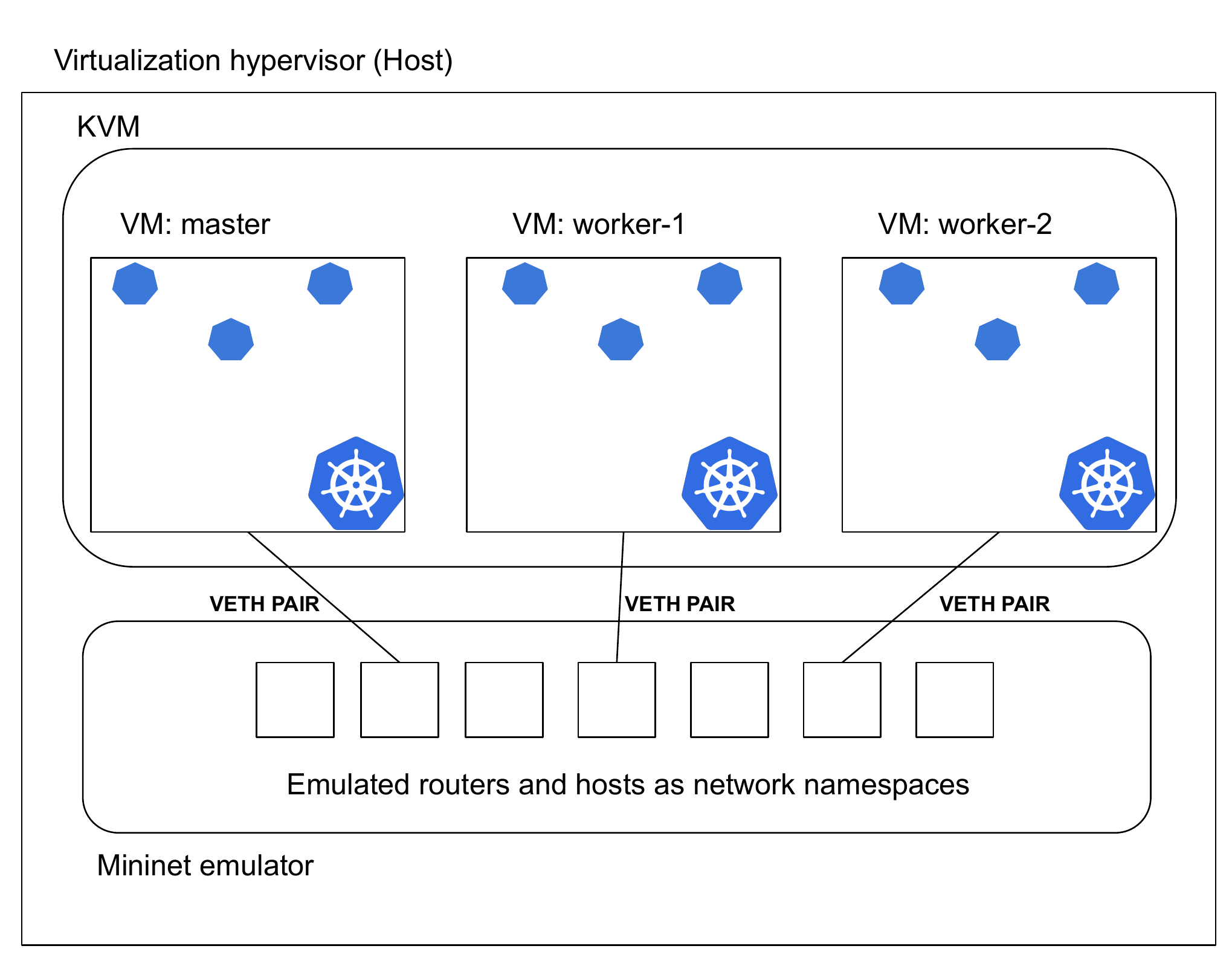}
    \caption{Full testbed: interconnection of VMs through the emulated network}
    \label{fig:full-testbed}
\end{figure}

The \emph{full testbed} is used to execute the functional tests and the experiments with a Kubernetes cluster empowered by SRv6 Traffic Engineering functionality. The two solutions proposed in Section~\ref{sec:srv6-overlay-calico-vpp} to dynamically configure the SRv6-TE overlay (respectively based on BGP and on the Kubernets control plane) can be deployed in the full testbed. This testbed provides a virtualized network topology as shown in Fig.~\ref{fig:srv6-te-with-bgp}. From the point of view of the Kubernetes cluster, we still have three VMs as shown in Fig.~\ref{fig:full-testbed}. Unlike the basic testbed, the VMs are not interconnected by a virtual switch, but through an emulated network backbone composed of eight routers. The emulated networking scenario is implemented using the Mininet tool~\cite{mininet,lantz2010network}. In the Mininet network, we have configured dynamic routing using the \paper{ISIS}\thesis{\acrshort{ISISLabel}} intra-domain routing protocol, to achieve a realistic emulation of a real backbone network of an Internet Service Provider. The three VMs of the Kubernetes cluster are connected using virtual Ethernet pairs to the eight routers emulated with Mininet. \paper{In the detailed walk-through of the experiments available in~\cite{rose-k8s-srv6} we show how to enforce paths for the overlay tunnels that follow an arbitrary route across the routers of the topology of Fig.~\ref{fig:srv6-te-with-bgp}.}


\thesis{
In this walkthrough we describe how to set up and configure a Calico VPP networking plugin for a Kubernetes cluster using a new type of overlay based on SRv6.

First we present the SRv6 baseline scenario (without traffic engineering). Subsequently we present a more complex distribution where we can demonstrate an SRv6 overlay with Traffic Engineering (SRv6-TE). More in detail, two different solutions for SRv6-TE are demonstrated: one based on BGP and another based on the Kubernetes control plane.

\section{SRv6 basic scenario (no Traffic Engineering)}
\label{sec:srv6basciscenario}

The basic testbed scenario is based on a simple switched network with three nodes: one
master and two workers as shown in the following figure.

In this basic scenario we will deploy a cluster using the SRv6 overlay (i.e. the pods packets will be tunneled over SRv6 tunnels). We are not using any traffic engineering in the SRv6 underlay, therefore the segments of the SRv6 tunnels only identify the egress node of the tunnel and the decapsulation operation to be performed in the egress node. In particular, a single segment will be used, which at the same time identifies the egress node and the decap operation. 

\subsection{Environment Setup}

In order to deploy this basic scenario we rely on the development testbed provided by the vpp-dataplane repository which can be cloned from here \cite{fork-vpp-dataplane-repo}.
After installing the requirements as documented in \cite{fork-vpp-dataplane-repo}, we can proceed with the deployment of the scenario, by entering the repository directory from the command line and running:
\begin{center}
\begin{verbatim}
    $ make start-test-cluster
\end{verbatim}
\end{center}

The command takes care of setting up the virtualization environment using Vagrant.

\subsection{Configure and install Calico VPP}

At this point we have the Kubernetes cluster that needs to be configured with our CNI. In order to do this we execute the command that takes care of the complete deployment and configuration of our Calico VPP.
\begin{center}
\begin{verbatim}
    $ make test-install-calicovpp-srv6
\end{verbatim}
\end{center}

The above command basically runs a series of kubectl apply commands that allow the creation of all the necessary kubernetes resources. Among all the resources created, it is useful to go into the details of the definition of the IPPools used to assign the localsid defined on the various nodes of the cluster. The first IPPool is the one needed to assign the BSID to each SRv6 policy created where we have specified the cidr cafe::/128 and the blockSize with value 122.

\begin{lstlisting}[language=yaml,label=lst:sr-policies-pool, caption=SR Policies IPPool]
- apiVersion: crd.projectcalico.org/v1
 kind: IPPool
 metadata:
   name: sr-policies-pool
 spec:
   blockSize: 122
   cidr: cafe::/118
   ipipMode: Never
   nodeSelector: '!all()'
   vxlanMode: Never
\end{lstlisting}
Then, for each cluster node we have decided to assign a defined subnet for the LocalSIDs, as shown below, in our three node environment we have three different IPPools
\begin{lstlisting}[language=yaml,label=lst:sr-localsids-pools, caption=LocalSIDs IPPool]
- apiVersion: crd.projectcalico.org/v1
 kind: IPPool
 metadata:
   name: sr-localsids-pool-master
 spec:
   cidr: fcff:0:0:00AA::/64
   ipipMode: Never
   nodeSelector: kubernetes.io/hostname == 'master'
   vxlanMode: Never
- apiVersion: crd.projectcalico.org/v1
 kind: IPPool
 metadata:
   name: sr-localsids-pool-node1
 spec:
   cidr: fcff:0:0:11AA::/64
   ipipMode: Never
   nodeSelector: kubernetes.io/hostname == 'node1'
   vxlanMode: Never
- apiVersion: crd.projectcalico.org/v1
 kind: IPPool
 metadata:
   name: sr-localsids-pool-node2
 spec:
   cidr: fcff:0:0:12AA::/64
   ipipMode: Never
   nodeSelector: kubernetes.io/hostname == 'node2'
   vxlanMode: Never

\end{lstlisting}
After that we can check that all the pods are running:
\begin{center}
\begin{verbatim}
    $ kubectl get pods -A
\end{verbatim}
\end{center}
\begin{figure}[ht]
    \centering
    \includegraphics[width=1\textwidth]{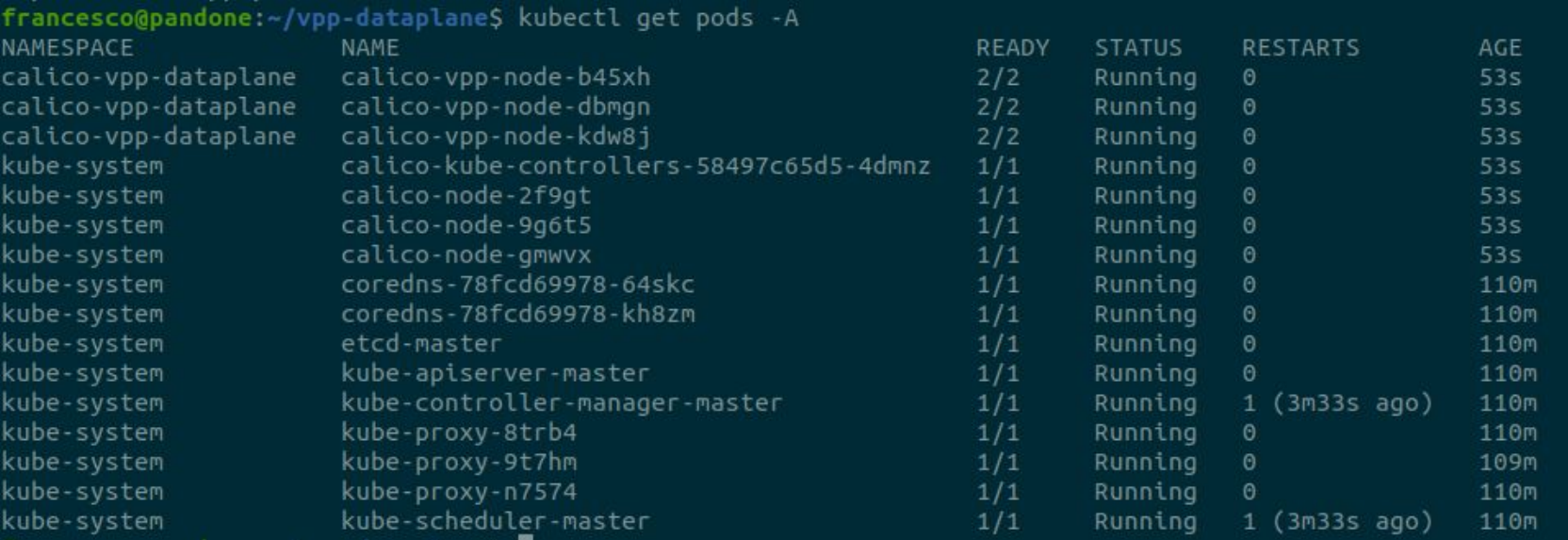}
    \caption{SRv6 basic scenario get pods}
    \label{fig:srv6-basic-get-pods-1}
\end{figure}

Since we rely on BGP to distribute all the information needed to create the SRv6 overlay, the pod subnets associated with a node will be available if at least one pod is deployed on the node.
In our case after the creation phase only the master node has associated pods in the workload subnet, therefore in order to obtain the complete overlay on the cluster we create pods distributed on the other nodes.  For instance, we can create a simple daemonset using a yaml definition as follows:
%

\begin{lstlisting}[language=yaml,label=lst:vpp-test-busybox, caption=Daemonset example]
apiVersion: apps/v1
kind: DaemonSet
metadata:
  name: vpp-test-busybox
spec:
  selector:
    matchLabels:
      name: vpp-test-busybox
  template:
    metadata:
      labels:
        name: vpp-test-busybox
    spec:
      tolerations:
        - key: node-role.kubernetes.io/control-plane
          operator: Exists
          effect: NoSchedule
        - key: node-role.kubernetes.io/master
          operator: Exists
          effect: NoSchedule
      containers:
        - name: vpp-test-busybox
          image: busybox
          command: ["sleep"]
          args: ["2d"]
\end{lstlisting}

After the creation of the daemonset we check the SRv6 configuration inside vpp by running the following command:
\begin{center}
\begin{verbatim}
    $ kubectl -n calico-vpp-dataplane exec -it <calico-vpp-node-podname> -c vpp -- vppctl

\end{verbatim}
\end{center}

from the vppctl prompt we can show the LocalSIDs configured on a node, in this example worker1, (can use the flag kubectl get pods -o wide in order to know which node the pod is on):
\begin{center}
\begin{verbatim}
    vpp# sh sr localsids
\end{verbatim}
\end{center}

\begin{figure}[ht]
    \centering
    \includegraphics[width=1\textwidth]{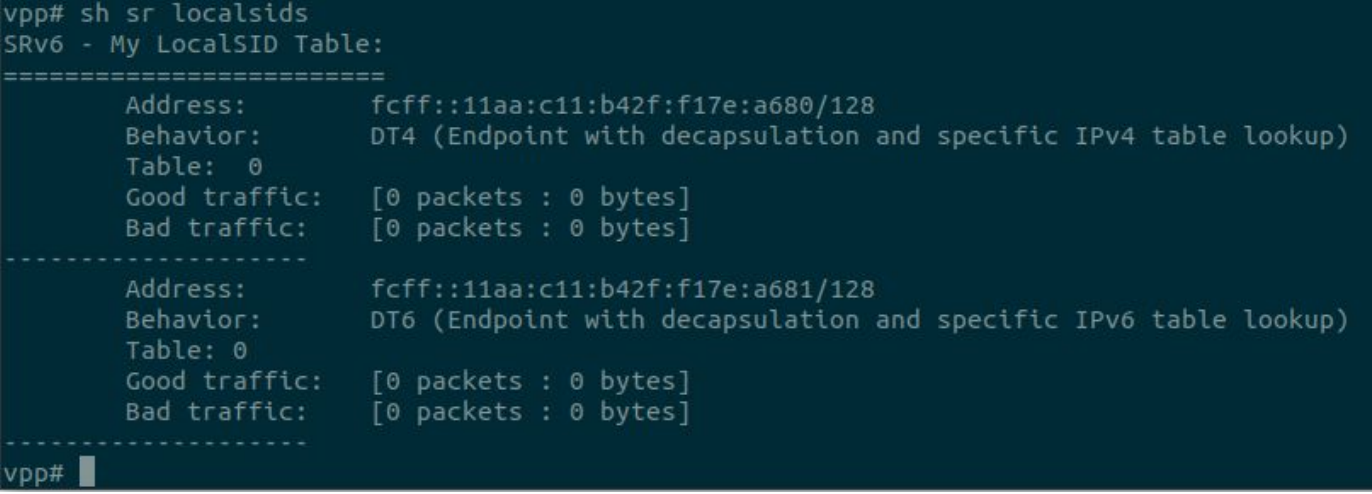}
    \caption{SRv6 basic scenario get localsids}
    \label{fig:srv6-basic-get-localsid-1}
\end{figure}

Show the SR policies and the steering-policies
\begin{center}
\begin{verbatim}
    vpp# sh sr policies
\end{verbatim}
\end{center}


\subsection{Test SRv6 connectivity between pods}
We use the pods created in the previous step to perform a connectivity test. 
In this example from the pod list we recognize that vpp-test-busybox-56csr is on node2 and vpp-test-busybox-bvnk9 in on node1

then from vpp-test-busybox-56csr we can test the ping command
\begin{center}
\begin{verbatim}
    $ kubectl exec -it vpp-test-busybox-56csr -- ping -c 4 172.16.166.128
\end{verbatim}
\end{center}

\begin{figure}[ht]
    \centering
    \includegraphics[width=1\textwidth]{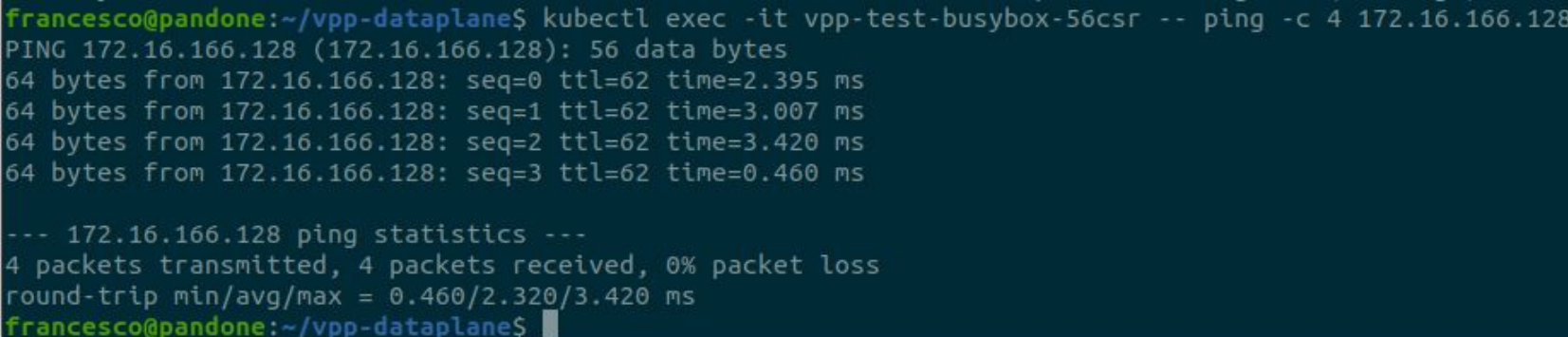}
    \caption{SRv6 basic scenario exec ping}
    \label{fig:srv6-basic-exec-ping}
\end{figure}

The vppctl also provides a counter of packets received for each localsid, so in another shell we can also check the ascending counter for node1's localsid.

\begin{figure}[ht]
    \centering
    \includegraphics[width=1\textwidth]{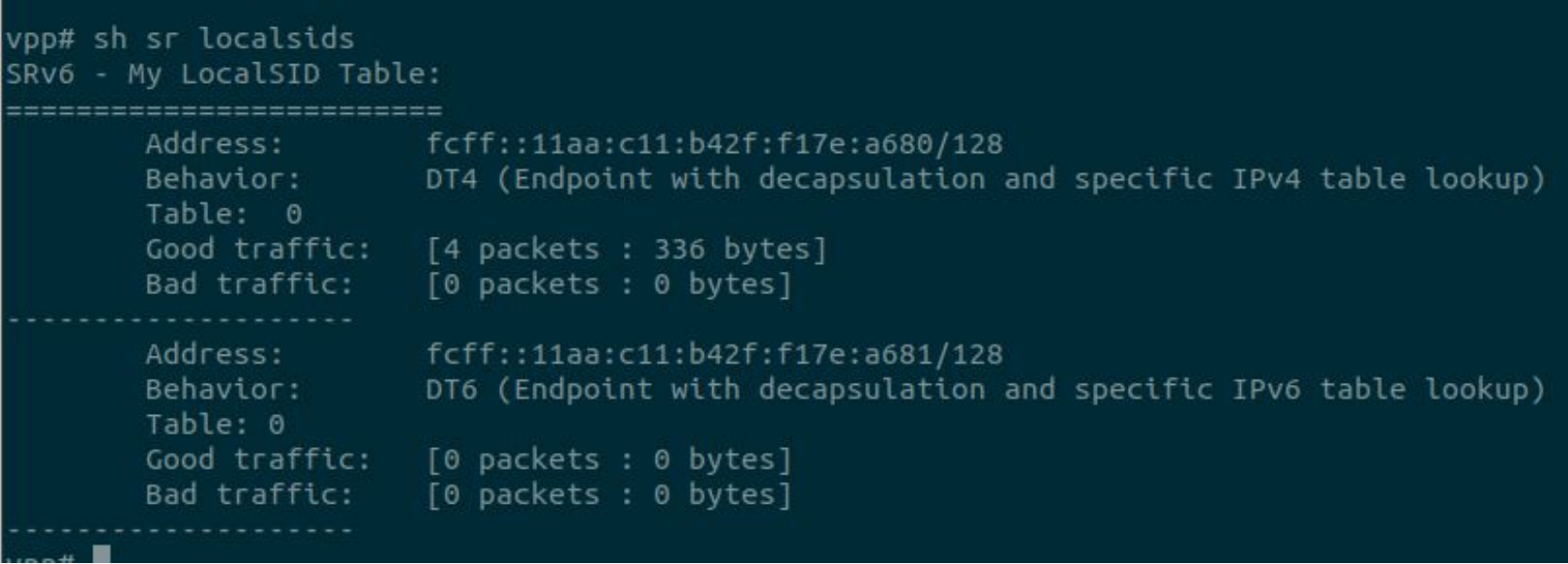}
    \caption{SRv6 basic scenario show localsids}
    \label{fig:srv6-basic-get-localsid-2}
\end{figure}

\section{SRv6 overlay with Traffic Engineering (SRv6-TE)}
\label{sec:srv6testbedte}
As a first step we need a scenario with a K8s cluster that can be configured with our calico-vpp cni. For this purpose we developed some test scenarios that can be used as a demo for a functional test.

\subsection{Environment Setup}
This tutorial is based on the repository that can be cloned from here \cite{walkthrough-repo}.
We assume that the same environment (host or VM) described for the basic testbed is used.

We want to deploy the topology depicted in the figure hereafter. The topology represents a Kubernetes cluster with one master and two workers, distributed on a network with 8 internal routers. We will use Vagrant to configure our virtualization environment (based on KVM-libvirt).
We will instantiate three VMs, one for the master and two for the workers. The internal network with 8 routers (and several other hosts) is emulated using the Mininet emulator and it is deployed in the virtualization hypervisor (host). 
Moreover, we include an emulated Mininet host, which is used to run the SRv6-PI (Policy Injector) \cite{SRv6-PI} based on goBGP.
The virtualization and networking environment is shown in the following figure. The VMs of the Master and of the workers are connected using veth pairs to the Mininet emulation environment. Both the Master and the workers are connected to Mininet emulated routers, in the same way as the Mininet emulated hosts.
To deploy the scenario, enter in the directory of the repository from command line and execute:
\begin{center}
\begin{verbatim}
    $ make start-scenario1
\end{verbatim}
\end{center}

This command takes care of setting up the virtualization environment using Vagrant and the networking emulation using Mininet. After the execution we have a Mininet network connected with 3 VMs that are part of our k8s cluster and a host with our SRv6-PI installed.

At this point we have this scenario with a Kubernetes cluster that needs to be configured with our CNI. In order to do this we access to the master node executing:
\begin{center}
\begin{verbatim}
    $  ssh vagrant@192.168.10.254
\end{verbatim}
\end{center}

(since this is a test environment, the default password is “vagrant”)

Inside the master node we have kubectl configured with admin access to the cluster. First we check that all nodes are correctly joined to our cluster by executing:
\begin{center}
\begin{verbatim}
    $ kubectl get nodes
\end{verbatim}
\end{center}

And we have an output like: 
\begin{figure}[ht]
    \centering
    \includegraphics[width=1\textwidth]{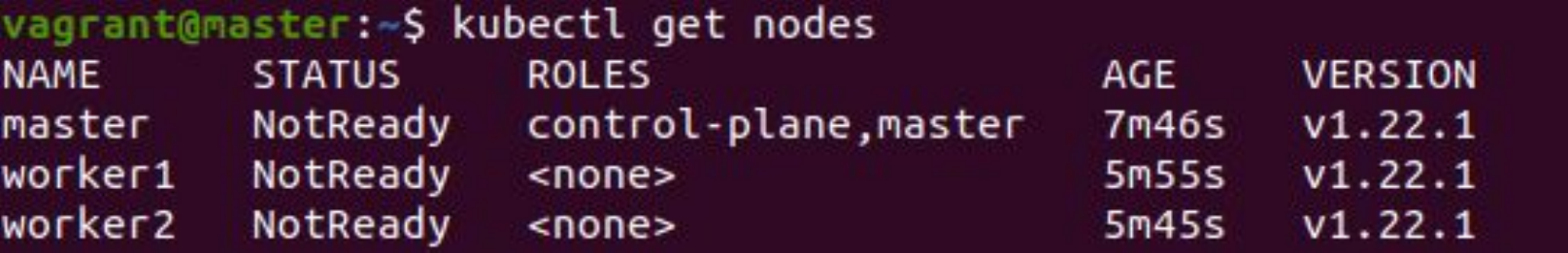}
    \caption{SRv6-TE setup environment get nodes}
    \label{fig:srv6-basic-get-localsid-2}
\end{figure}

For each of the following scenarios, the specific requirements must be installed before proceeding with the deployment, as documented in \cite{walkthrough-repo}.

\subsection{SRv6-TE overlay with BGP}
In this scenario we will deploy a cluster using the SRv6 overlay with Traffic Engineering (SRv6-TE) using the BGP protocol. In particular we demonstrate how with the use of our solution it is possible to inject SR policies leveraging the BGP protocol.

\subsubsection{Configure and deploy}
At this point we can complete the cluster configuration by installing / configuring our calico-vpp, we provide all the resource definition within a YAML file. 

First we check that the value of CALICOVPP\_SRV6\_TE is set to “bgp”

\begin{lstlisting}[language=yaml,label=lst:CALICOVPP_SRV6_TE, caption=Enable TE with BGP]
- name: CALICOVPP_SRV6_TE
  value: "bgp"
\end{lstlisting}

Then we can execute the apply command:

\begin{center}
\begin{verbatim}
   $ kubectl apply -f yaml/test-srv6.yaml
\end{verbatim}
\end{center}
\begin{figure}[ht]
    \centering
    \includegraphics[width=1\textwidth]{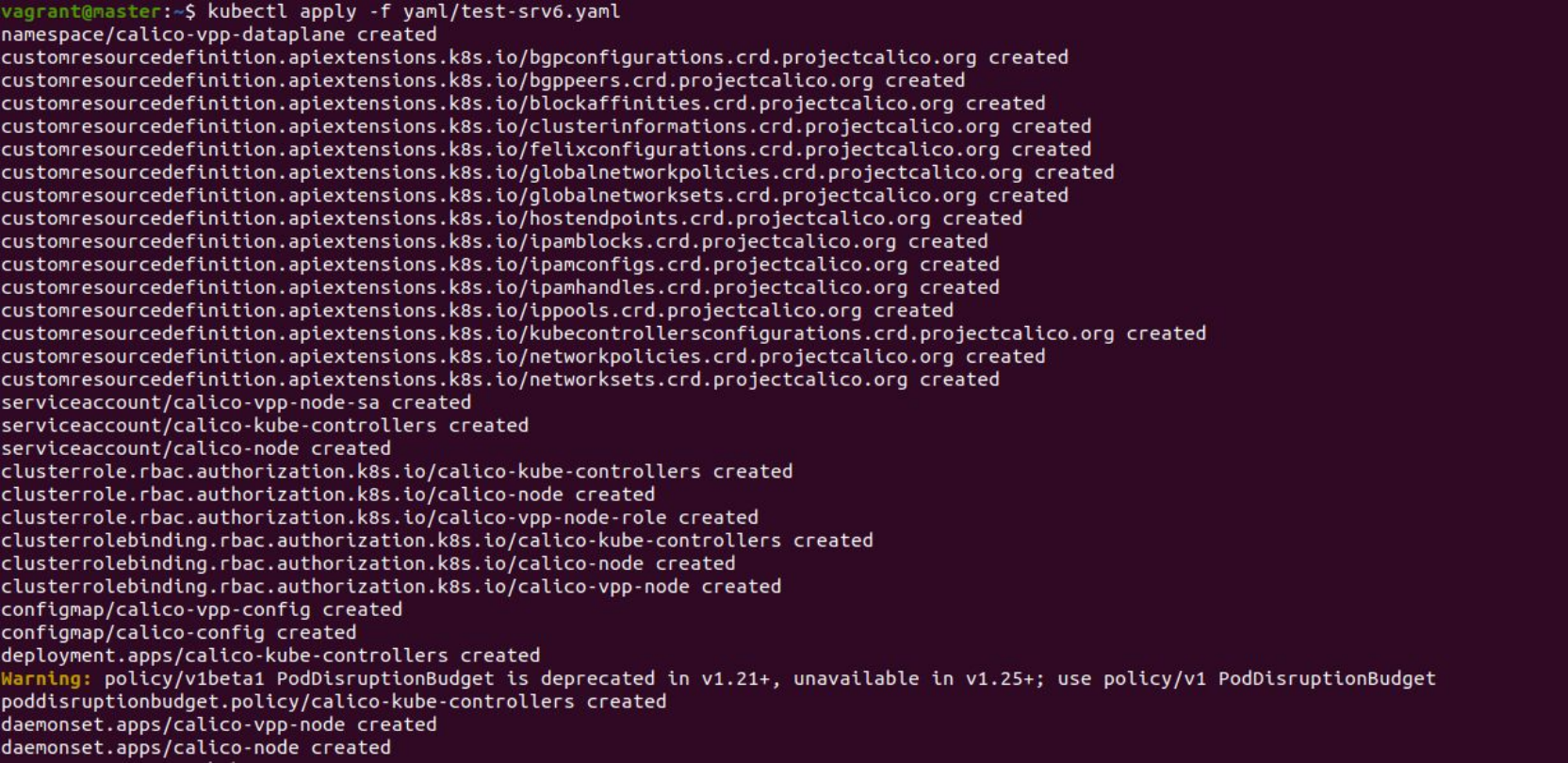}
    \caption{SRv6-TE deploy Calico VPP}
    \label{fig:srv6-te-kubectl-apply}
\end{figure}

Then we create all IPPools needed within the cluster
\begin{center}
\begin{verbatim}
    $ kubectl apply -f yaml/test-srv6-res.yaml
\end{verbatim}
\end{center}

After that we can check that all the pods are running:
\begin{center}
\begin{verbatim}
    $ kubectl get pods -A
\end{verbatim}
\end{center}

\begin{figure}[ht]
    \centering
    \includegraphics[width=1\textwidth]{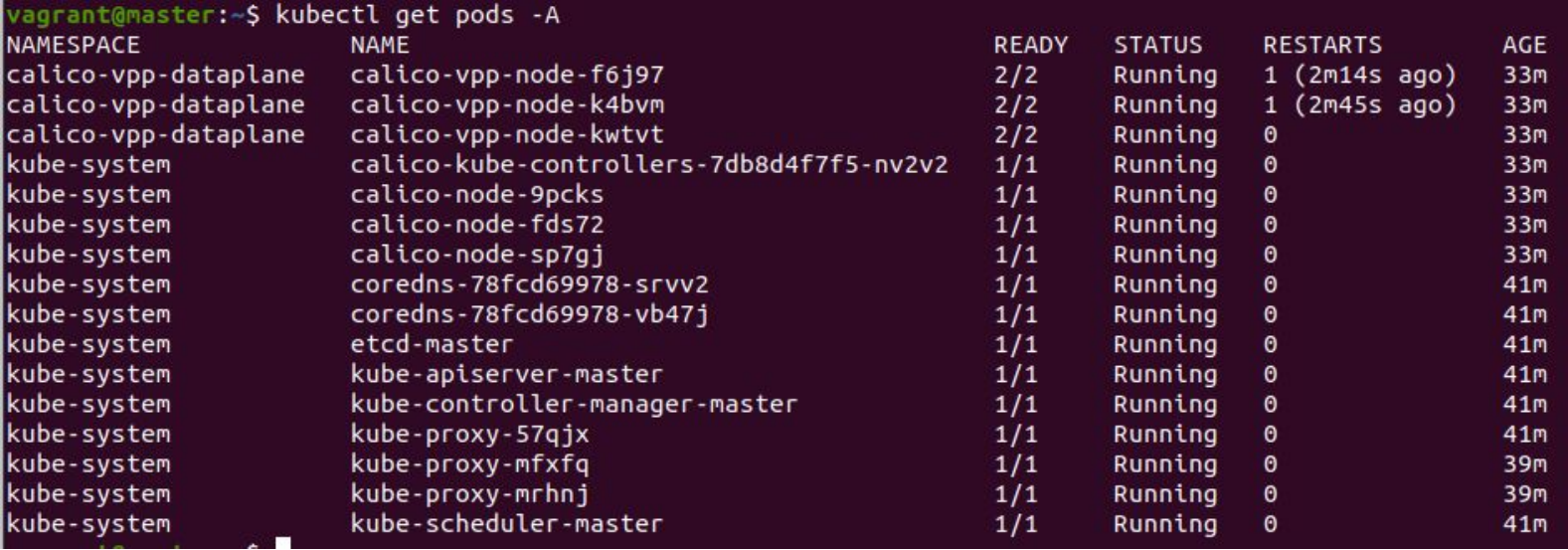}
    \caption{SRv6-TE BGP get pods}
    \label{fig:srv6-te-get-pods}
\end{figure}

We can check the SRv6 configuration within VPP by running the following command:
\begin{center}
\begin{verbatim}
    $ kubectl -n calico-vpp-dataplane exec -it <calico-vpp-node-podname> -c vpp -- vppctl
\end{verbatim}
\end{center}

from the vppctl prompt we can show the LocalSIDs configured on a node, in this example worker2, (can use the flag kubectl get pods -o wide in order to know which node the pod is on):
\begin{center}
\begin{verbatim}
    vpp# sh sr localsids
\end{verbatim}
\end{center}

\begin{figure}[ht]
    \centering
    \includegraphics[width=1\textwidth]{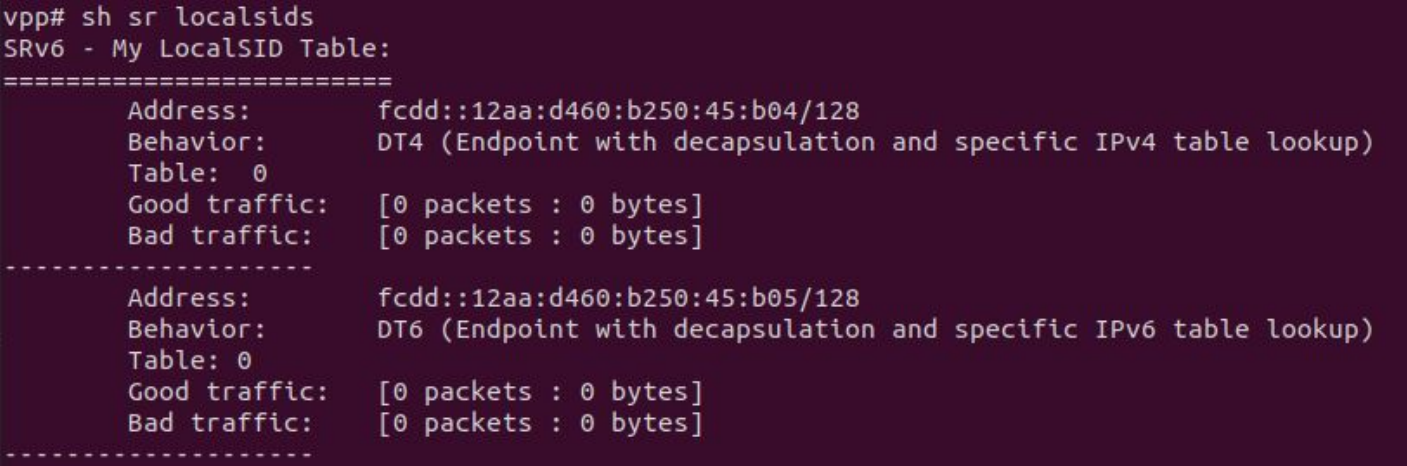}
    \caption{SRv6-TE BGP get localsids}
    \label{fig:srv6-te-bgp-get-localsid-1}
\end{figure}

Shows the SR policies, which in this case will be blank
\begin{center}
\begin{verbatim}
    vpp# sh sr policies
\end{verbatim}
\end{center}
\begin{figure}[ht]
    \centering
    \includegraphics[width=0.5\textwidth]{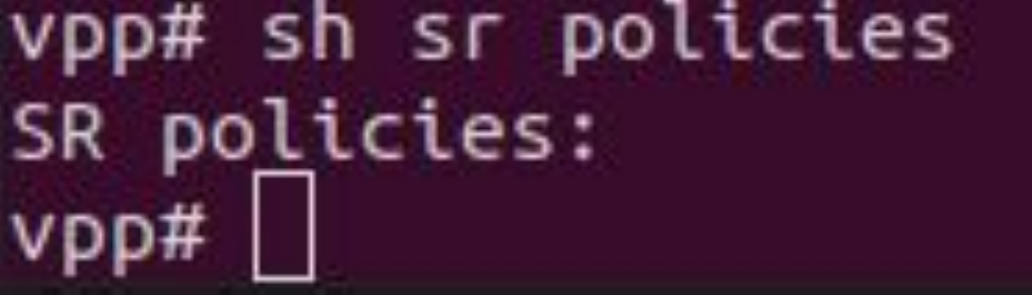}
    \caption{SRv6-TE BGP get policy empty}
    \label{fig:srv6-te-bgp-get-policies-empty}
\end{figure}

Shows the SR steering-policies, which in this case will be blank
\begin{center}
\begin{verbatim}
    vpp# sh sr steering-policies
\end{verbatim}
\end{center}

\begin{figure}[ht]
    \centering
    \includegraphics[width=0.5\textwidth]{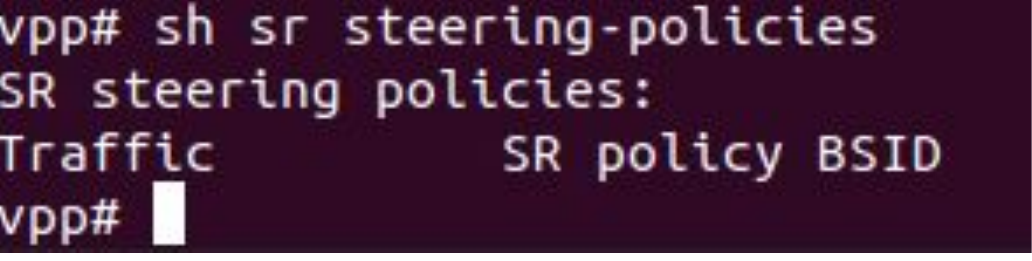}
    \caption{SRv6-TE BGP get steering-policies empty}
    \label{fig:srv6-te-bgp-get-steering-policies-empty}
\end{figure}
Shows the configured encapsulation source address i.e. for worker2
\begin{figure}[ht]
    \centering
    \includegraphics[width=0.5\textwidth]{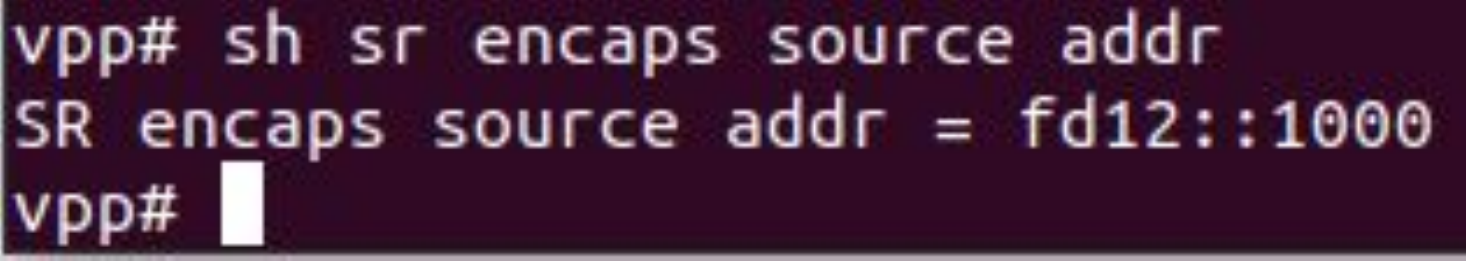}
    \caption{SRv6-TE BGP get source encap}
    \label{fig:srv6-te-bgp-get-encaps-sources-addr}
\end{figure}
As shown in the testbed figure, we have deployed a mininet host that has a goBGP instance that can be used as BGP peer within the k8s cluster. 
In order to use the SRv6-PI as an external BGP Peer, we need to register it as a Calico BGPPeer.
For this scenario we provide the YAML definition which can be used as follows:
\begin{center}
\begin{verbatim}
    $ kubectl apply -f yaml/bgp-peer.yaml
\end{verbatim}
\end{center}
And if we get the list of registered BGPeers, we should fetch something like the following:
In order to be able to inject the policies for IPv6 and IPv4 we need to register two instances for the two types of traffic. The content of bgp-peer.yaml:
\begin{lstlisting}[language=yaml,label=lst:bgp-peer, caption=Definition of BGPPeers]
apiVersion: crd.projectcalico.org/v1
kind: BGPPeer
metadata:
 name: my-peer-v4
spec:
 peerIP: 192.169.9.254
 asNumber: 64512
---
apiVersion: crd.projectcalico.org/v1
kind: BGPPeer
metadata:
 name: my-peer-v6
spec:
 peerIP: fd00:0:34::2
 asNumber: 64512
\end{lstlisting}

\subsubsection{Test SRv6 connectivity between pods}
In this step the workload pods are unable to communicate because there are no SRv6 policies configured. Let's  create a Daemonset to test the pods connectivity.
\begin{center}
\begin{verbatim}
    $ kubectl apply -f yaml/daemonset-test.yaml
\end{verbatim}
\end{center}

After creating the pods, we get something like this:
\begin{figure}[ht]
    \centering
    \includegraphics[width=1\textwidth]{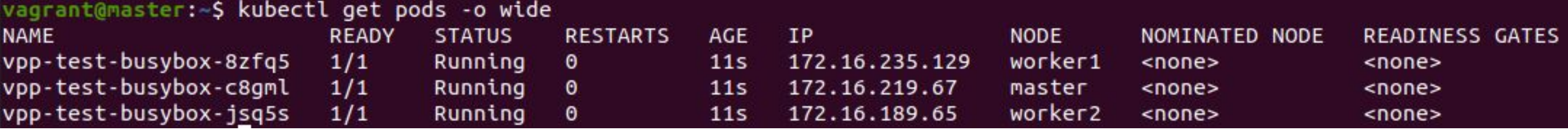}
    \caption{SRv6-TE BGP list daemonset pods}
    \label{fig:srv6-te-bgp-get-pods-daemonset}
\end{figure}

We verify that the ping between the worker2 and worker1 pods is not working
We retrieve the IPs of the pod running on worker1
\begin{figure}[ht]
    \centering
    \includegraphics[width=1\textwidth]{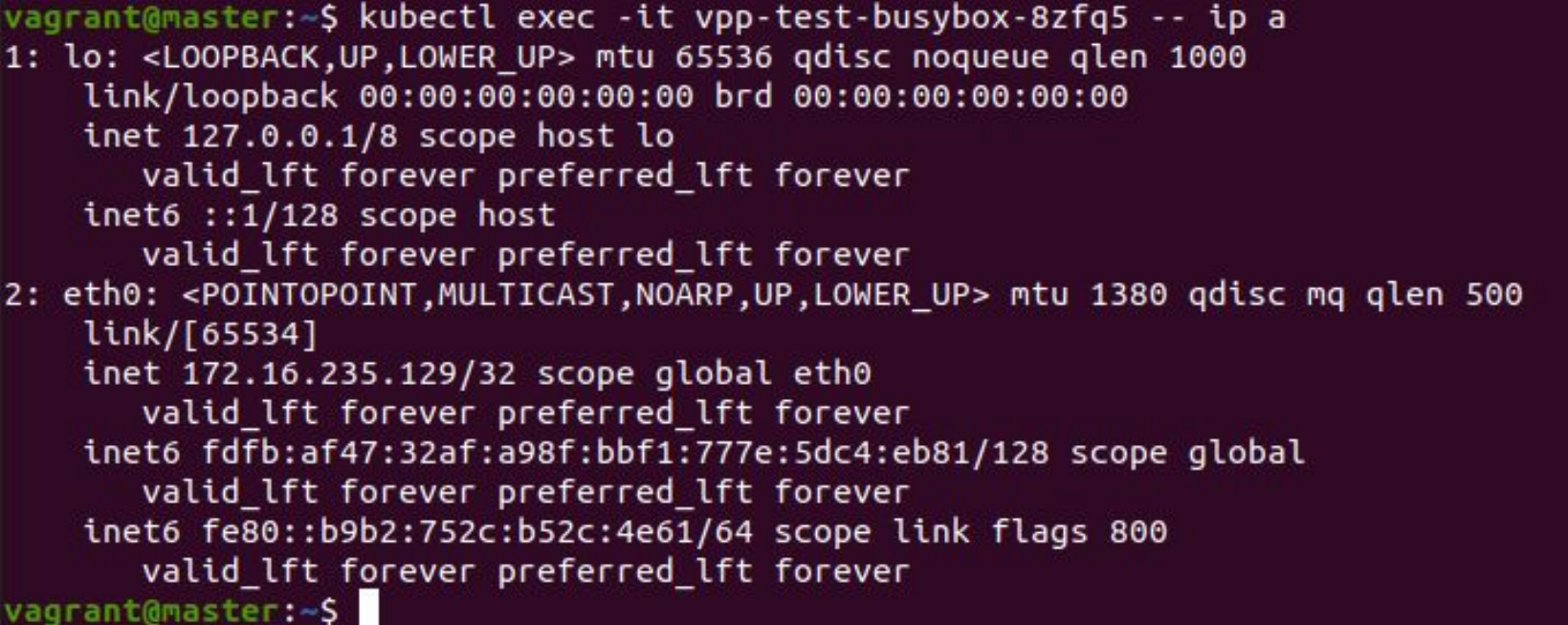}
    \caption{SRv6-TE BGP get pod IPs}
    \label{fig:srv6-te-bgp-get-pod-ip}
\end{figure}
We execute the ping command
\begin{figure}[ht]
    \centering
    \includegraphics[width=1\textwidth]{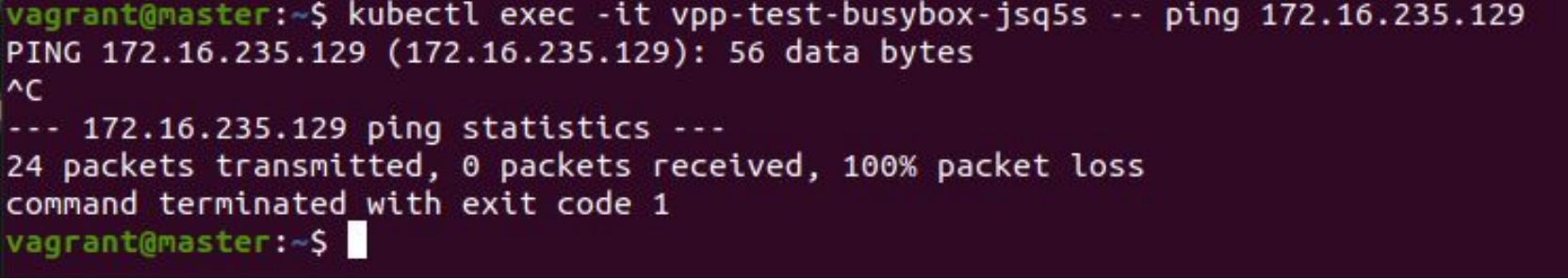}
    \caption{SRv6-TE BGP exec ping command}
    \label{fig:srv6-te-bgp-exec-pod-ping}
\end{figure}

\subsubsection{Test SRv6-TE edit policies}
We provide 6 yaml files with the policies needed for the communication between the cluster nodes, in particular for each node we need two policies for IPv6 and IPv4 traffic.
For instance, the two policies to reach worker2 are:
\begin{lstlisting}[language=yaml,label=lst:bgp-ipv4-traffic, caption=BGP Policies IPv4 traffic]
nlri:
 distinguisher: 4
 color: 94
 endpoint: fd12::1000
iswithdraw: false
age:
 seconds: 1638267871
 nanos: 992981348
sourceasn: 5600
family:
 afi: 2
 safi: 73
neighborip: fd12::1000
segmentlist:
 weight: 0
 segments:
 - sid: fcff:5::1
   behavior: 19
 - sid: fcff:7::1
   behavior: 19
 - sid: fcff:8::1
   behavior: 19
 - sid: fcdd::12aa:d460:b250:45:b04
   behavior: 19
bsid: cafe::4
priority: 0
nexthop: fd12::1000
\end{lstlisting}

\begin{lstlisting}[language=yaml,label=lst:bgp-iv6-traffic, caption=BGP Policies IPv6 traffic]
nlri:
 distinguisher: 4
 color: 94
 endpoint: fd12::1000
iswithdraw: false
age:
 seconds: 1638267871
 nanos: 992981348
sourceasn: 5600
family:
 afi: 2
 safi: 73
neighborip: fd12::1000
segmentlist:
 weight: 0
 segments:
 - sid: fcff:5::1
   behavior: 18
 - sid: fcff:7::1
   behavior: 18
 - sid: fcff:8::1
   behavior: 18
 - sid: fcdd::12aa:d460:b250:45:b05
   behavior: 18
bsid: cafe::5
priority: 0
nexthop: fd12::1000
\end{lstlisting}

At this point we open the shell with SRv6-PI, to do this we need to enter the namespace of the mininet host, running a utility command from the root directory of the repository
\begin{center}
\begin{verbatim}
    $ make enter-in name=gobgp
\end{verbatim}
\end{center}

We move to the SRv6-PI repository directory we cloned earlier and where we built the CLI.
We execute the following command to inject the policies for each file
\begin{center}
\begin{verbatim}
    ./SRv6-PI create -policyFile <policy_yaml_file>
\end{verbatim}
\end{center}

After the execution we can check the SRv6 configuration inside vpp by running the following command:
\begin{center}
\begin{verbatim}
    $ kubectl -n calico-vpp-dataplane exec -it <calico-vpp-node-podname> -c vpp -- vppctl
\end{verbatim}
\end{center}
\begin{figure}[ht]
    \centering
    \includegraphics[width=1\textwidth]{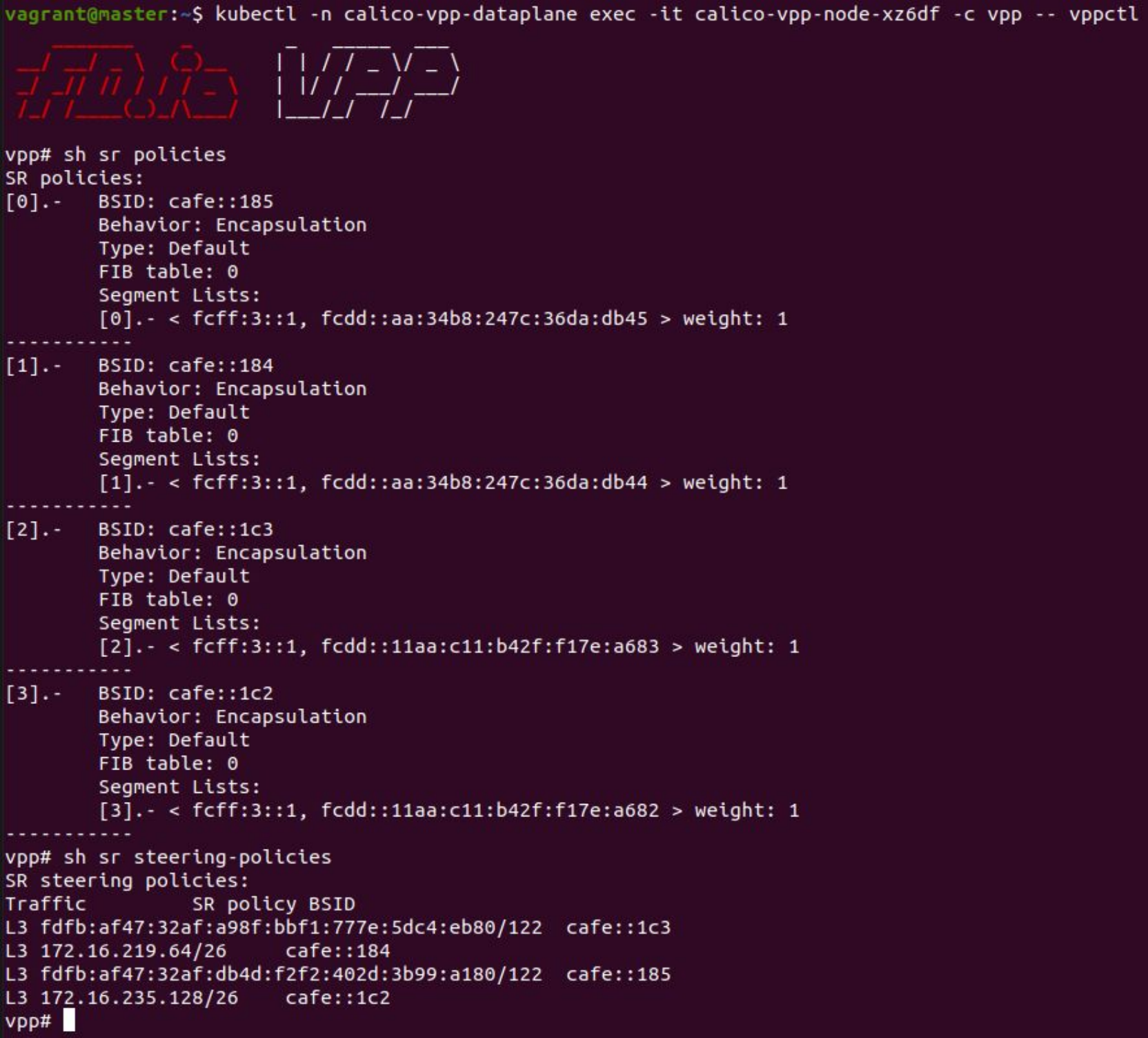}
    \caption{SRv6-TE BGP exec vppctl}
    \label{fig:srv6-te-bgp-vppctl-2}
\end{figure}
Then we can execute again the connectivity test between the pods, for IPv4
\begin{figure}[ht]
    \centering
    \includegraphics[width=1\textwidth]{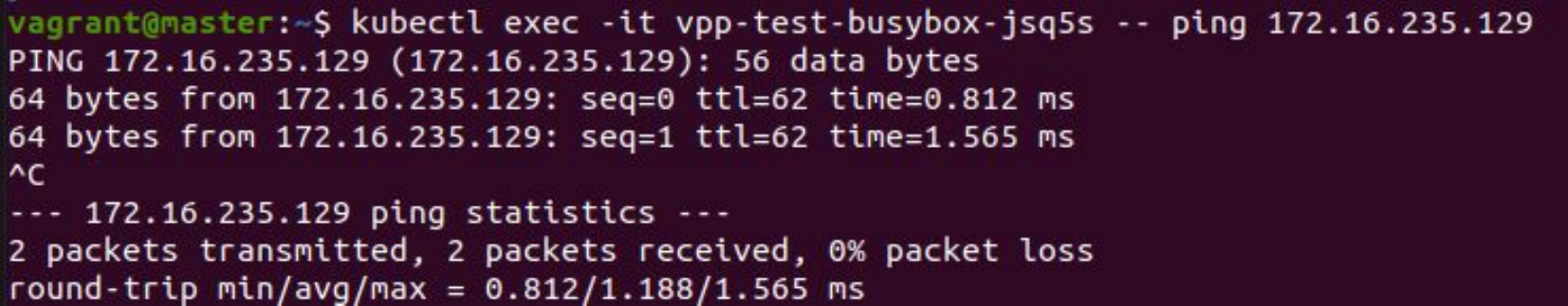}
    \caption{SRv6-TE BGP exec ping after TE update}
    \label{fig:srv6-te-bgp-ping-2}
\end{figure}
And for IPv6
\begin{figure}[ht]
    \centering
    \includegraphics[width=1\textwidth]{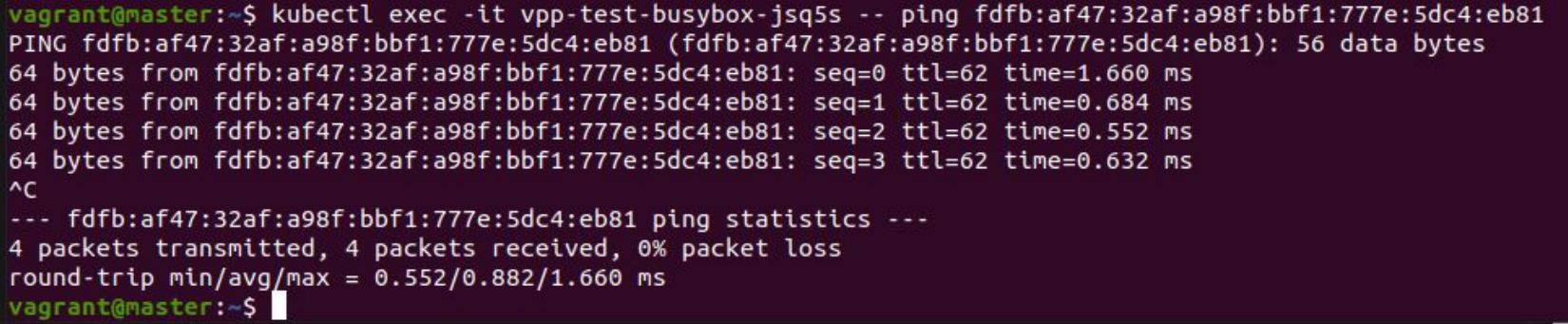}
    \caption{SRv6-TE BGP exec ping over IPv6 after TE update}
    \label{fig:srv6-te-bgp-ping-ipv6}
\end{figure}

\subsection{SRv6-TE overlay with Kubernetes control plane}
In this scenario we will deploy a cluster using the SRv6 overlay with Traffic Engineering (SRv6-TE) using the Kubernetes control plane. In particular we demonstrate how with the use of our solution it is possible to inject SR policies leveraging the ConfigMap resources.

\subsubsection{Configure and deploy}
At this point we can complete the cluster configuration installing/configuring our calico-vpp, we provide the definition of all the resources within a YAML file.
We first check that the value of CALICOVPP\_SRV6\_TE is set to “configmap”
\begin{lstlisting}[language=yaml,label=lst:CALICOVPP_SRV6_TE, caption=Enable TE with Kubernetes control plane]
- name: CALICOVPP_SRV6_TE
  value: "configmap"
\end{lstlisting}

Then we can execute the apply command:
\begin{center}
\begin{verbatim}
    $ kubectl apply -f yaml/test-srv6.yaml
\end{verbatim}
\end{center}

\begin{figure}[ht]
    \centering
    \includegraphics[width=1\textwidth]{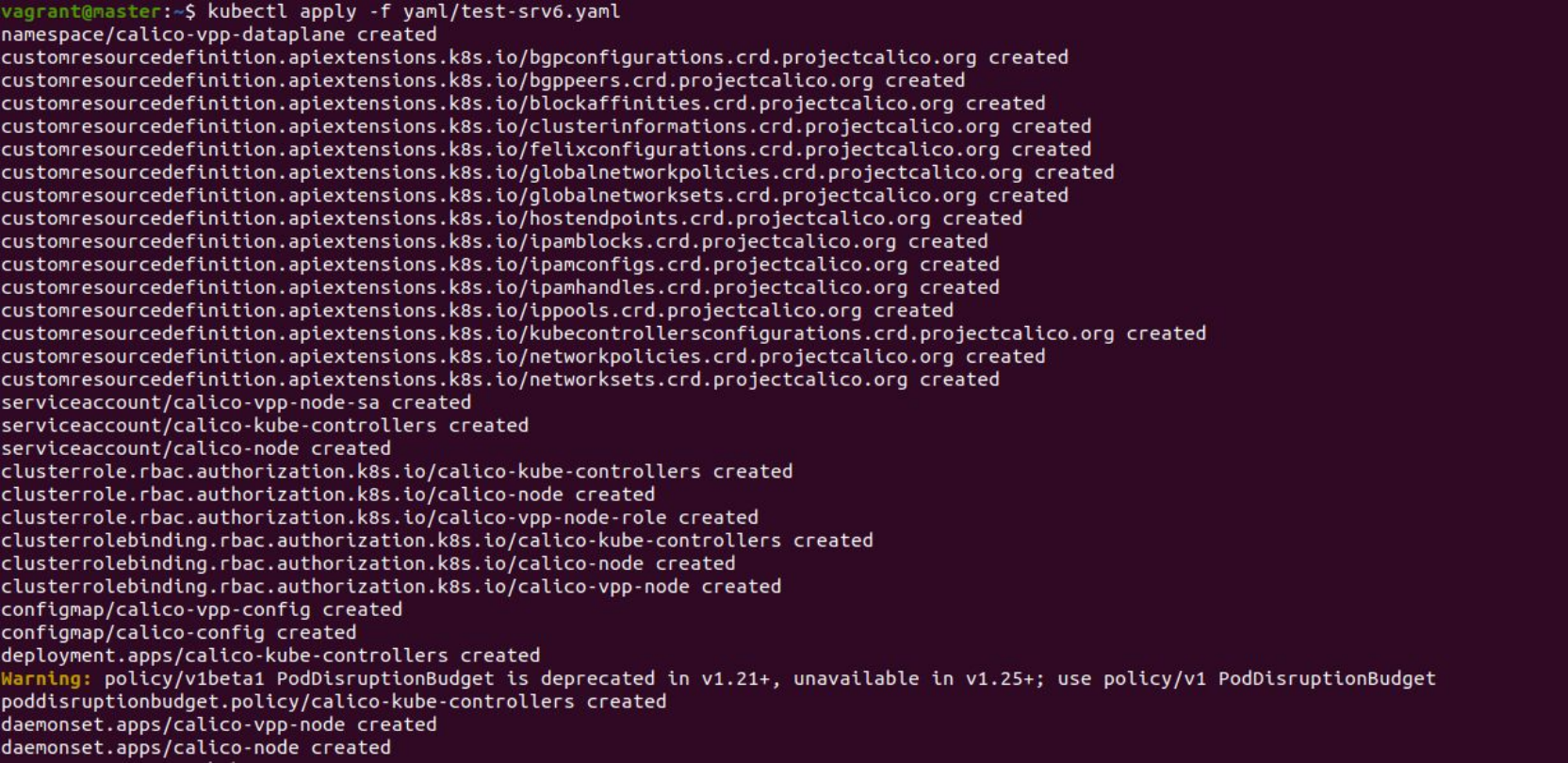}
    \caption{SRv6-TE ConfigMap deploy Calico VPP}
    \label{fig:srv6-te-kubectl-apply}
\end{figure}
Then we create all IPPools needed within the cluster
\begin{center}
\begin{verbatim}
    $ kubectl apply -f yaml/test-srv6-res.yaml
\end{verbatim}
\end{center}
Then we create the ConfigMap with our initial SRv6 configuration:
\begin{center}
\begin{verbatim}
    $ kubectl apply -f yaml/configMap.yaml
\end{verbatim}
\end{center}

The ConfigMaps named  srv6-config-<node\_name> will be created in the calico-vpp-dataplane namespace, so that they will be accessible from pods within the namespace, specifically our calico-vpp-node-xxxx. After that we can verify that all pods are running:
\begin{figure}[ht]
    \centering
    \includegraphics[width=1\textwidth]{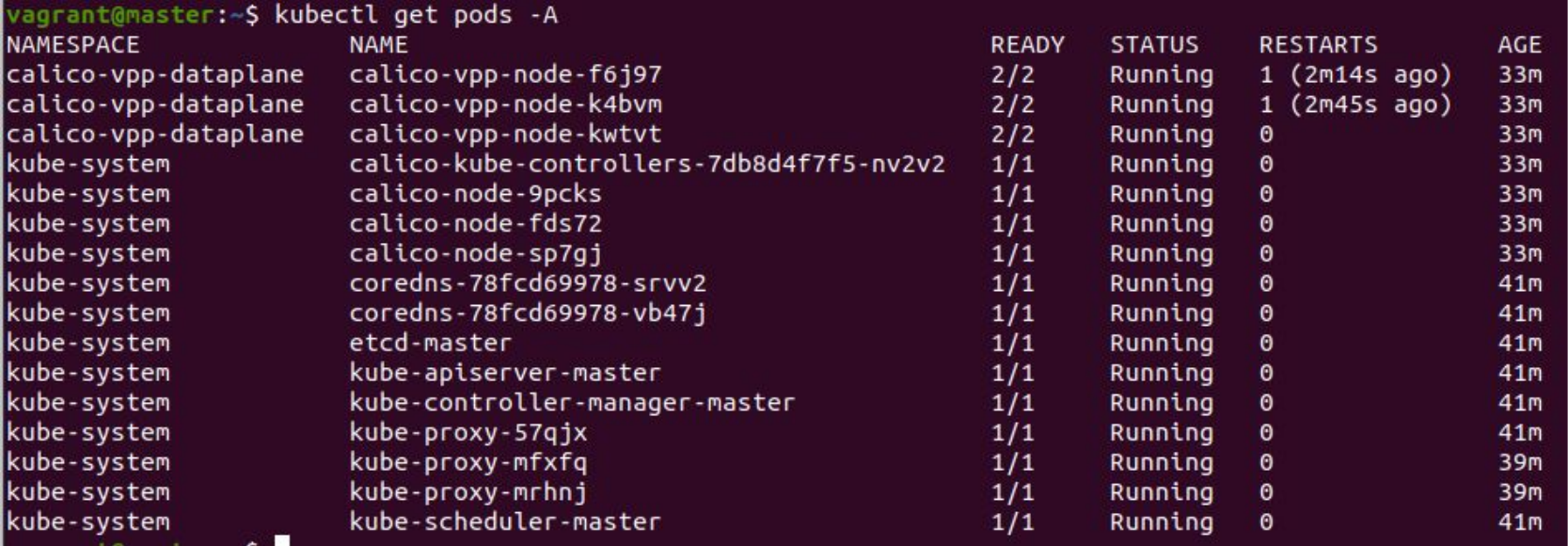}
    \caption{SRv6-TE K8s get pods}
    \label{fig:srv6-te-k8s-get-pods}
\end{figure}
We can check the SRv6 configuration inside vpp by running the following command:
\begin{center}
\begin{verbatim}
    $ kubectl -n calico-vpp-dataplane exec -it <calico-vpp-node-podname> -c vpp -- vppctl
\end{verbatim}
\end{center}
from the vppctl prompt we can show the LocalSIDs configured on a node, in this example worker1, (can use the flag kubectl get pods -o wide in order to know which node the pod is on):
\begin{center}
\begin{verbatim}
    vpp# sh sr localsids
\end{verbatim}
\end{center}

\begin{figure}[ht]
    \centering
    \includegraphics[width=1\textwidth]{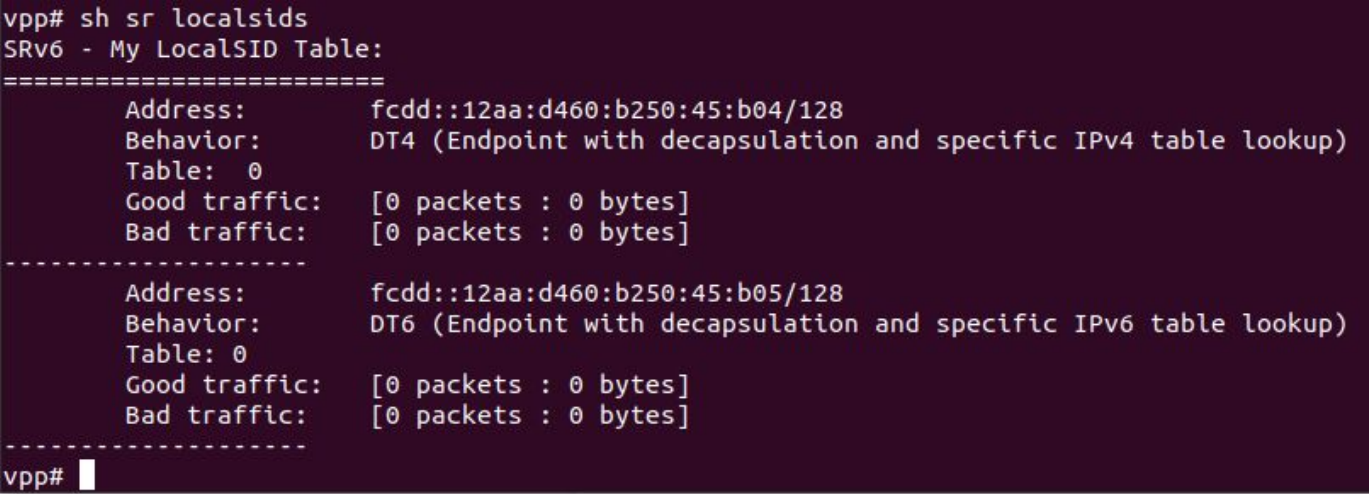}
    \caption{SRv6-TE K8s get localsids}
    \label{fig:srv6-te-k8s-get-localsid-1}
\end{figure}

Show the SR policies and the steering-policies
\begin{center}
\begin{verbatim}
    vpp# sh sr policies
\end{verbatim}
\end{center}

\begin{figure}[ht]
    \centering
    \includegraphics[width=1\textwidth]{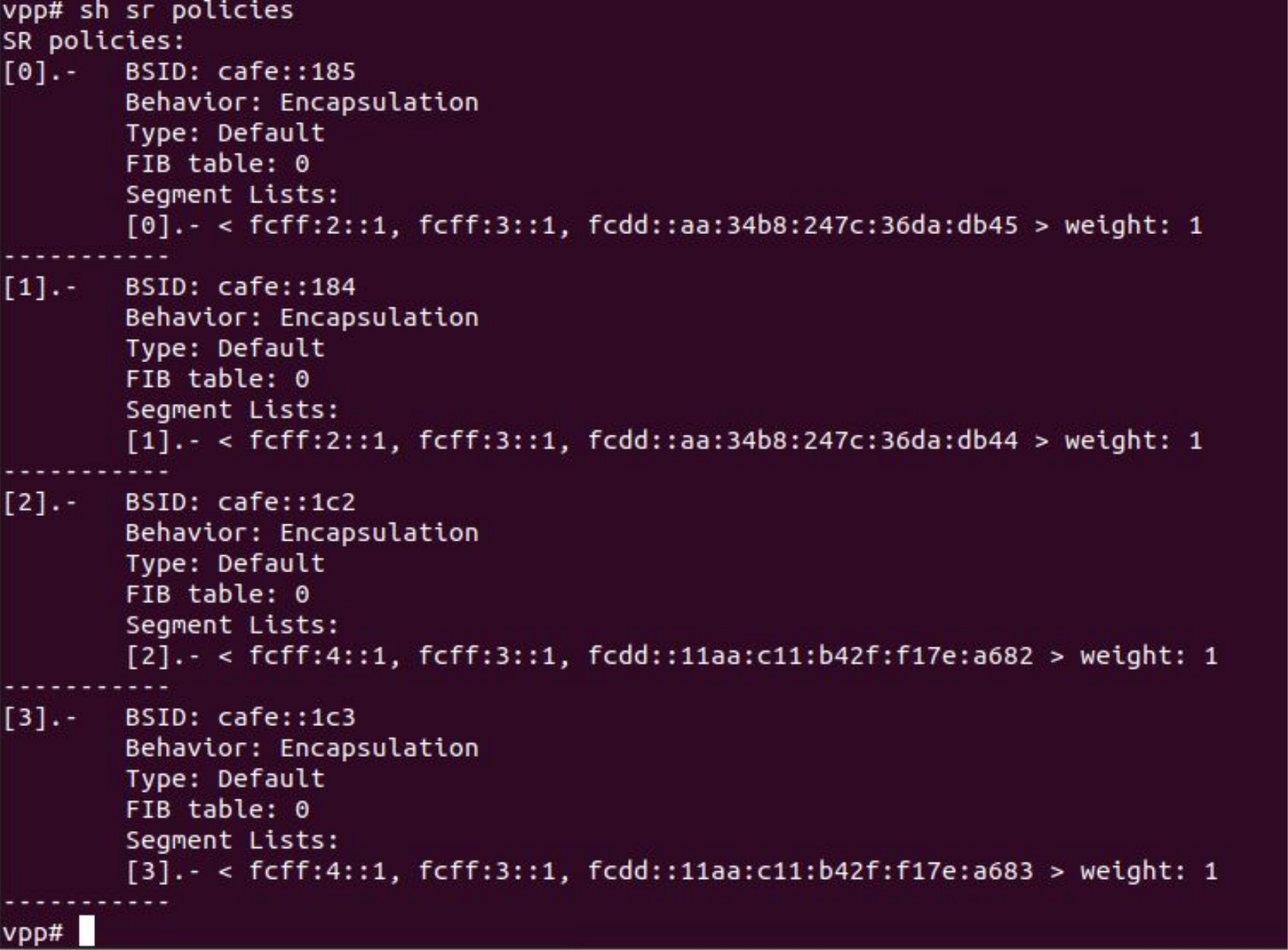}
    \caption{SRv6-TE K8s get policies}
    \label{fig:srv6-te-k8s-get-policies-1}
\end{figure}

show the SR steering-policies for worker2:
\begin{figure}[ht]
    \centering
    \includegraphics[width=1\textwidth]{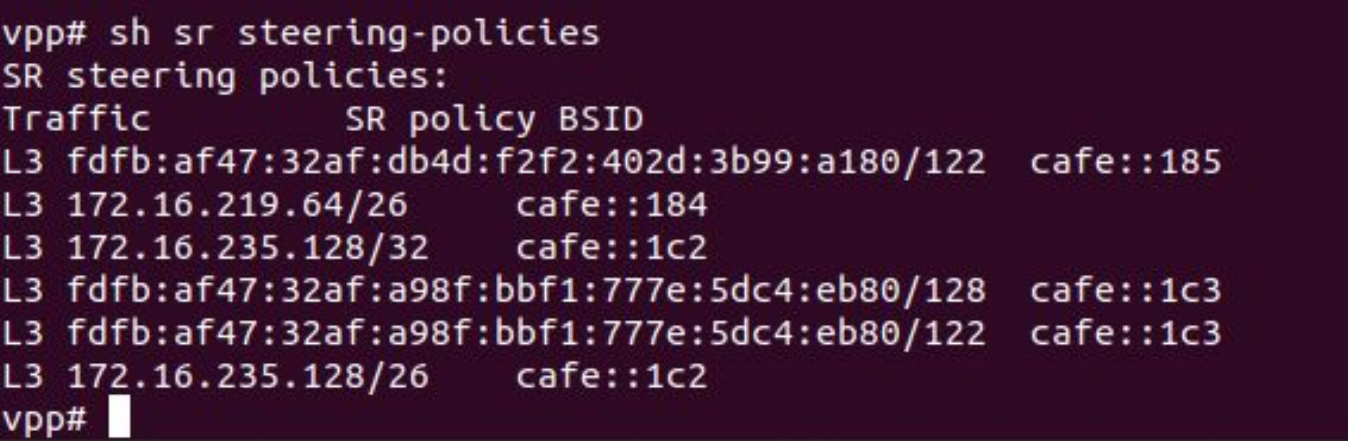}
    \caption{SRv6-TE K8s get steering-policies}
    \label{fig:srv6-te-k8s-get-steering-policies-1}
\end{figure}

Shows the configured encapsulation source address for worker2
\begin{figure}[ht]
    \centering
    \includegraphics[width=0.5\textwidth]{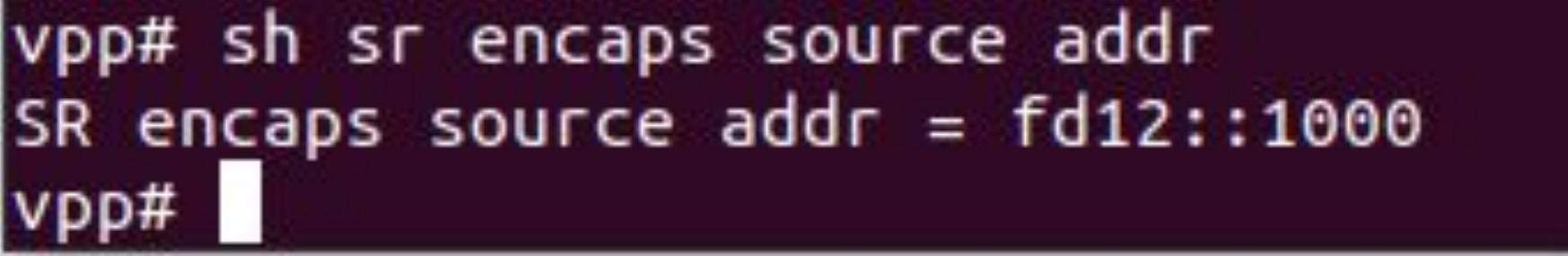}
    \caption{SRv6-TE K8s get source encap}
    \label{fig:srv6-te-k8s-get-encaps-source-addr-1}
\end{figure}

\subsubsection{Test SRv6 connectivity between pods}
\begin{center}
\begin{verbatim}
    $ kubectl apply -f yaml/daemonset-test.yaml
\end{verbatim}
\end{center}
\begin{figure}[ht]
    \centering
    \includegraphics[width=1\textwidth]{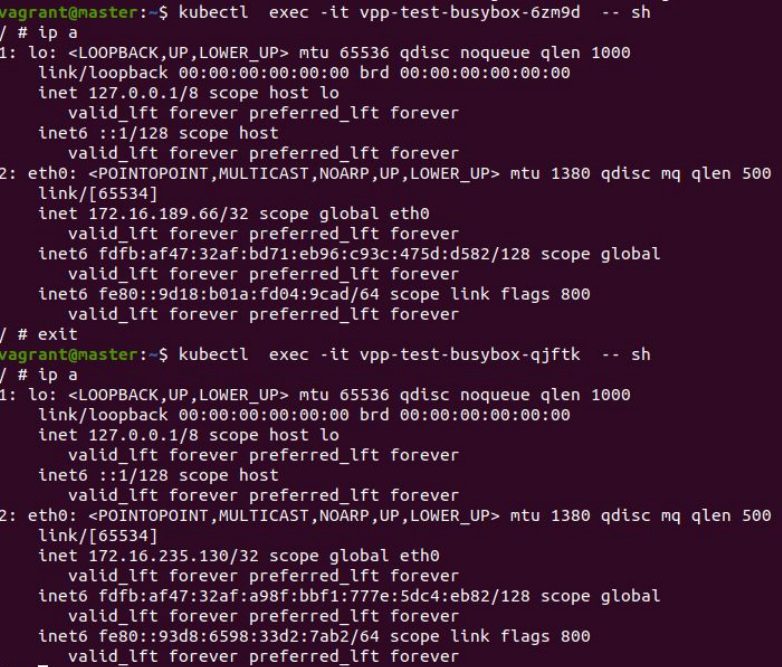}
    \caption{SRv6-TE K8s get pods IPs}
    \label{fig:srv6-te-k8s-pods-daemonset-show-ip-1}
\end{figure}

Since the pod vpp-test-busybox-6zm9d is running on worker2 node and vpp-test-busybox-qjftk pod on worker1 node,  we log into a shell and then execute the ping command
\begin{figure}[ht]
    \centering
    \includegraphics[width=1\textwidth]{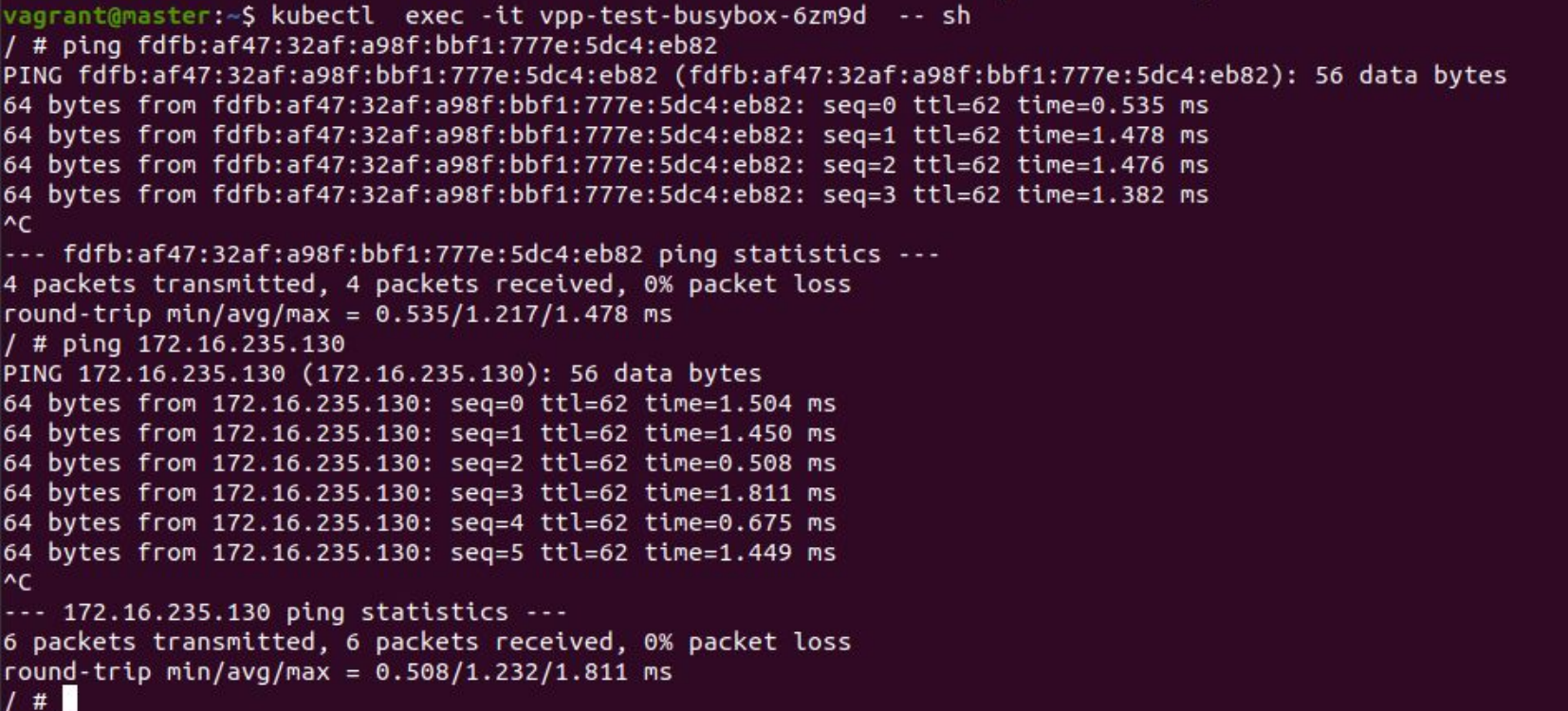}
    \caption{SRv6-TE K8s exec ping}
    \label{fig:srv6-te-k8s-pods-daemonset-ping-1}
\end{figure}

The vppctl also provides a counter of packets received for each localsid, so in another shell we can also check the ascending counter for worker1's localsid.
\begin{figure}[ht]
    \centering
    \includegraphics[width=1\textwidth]{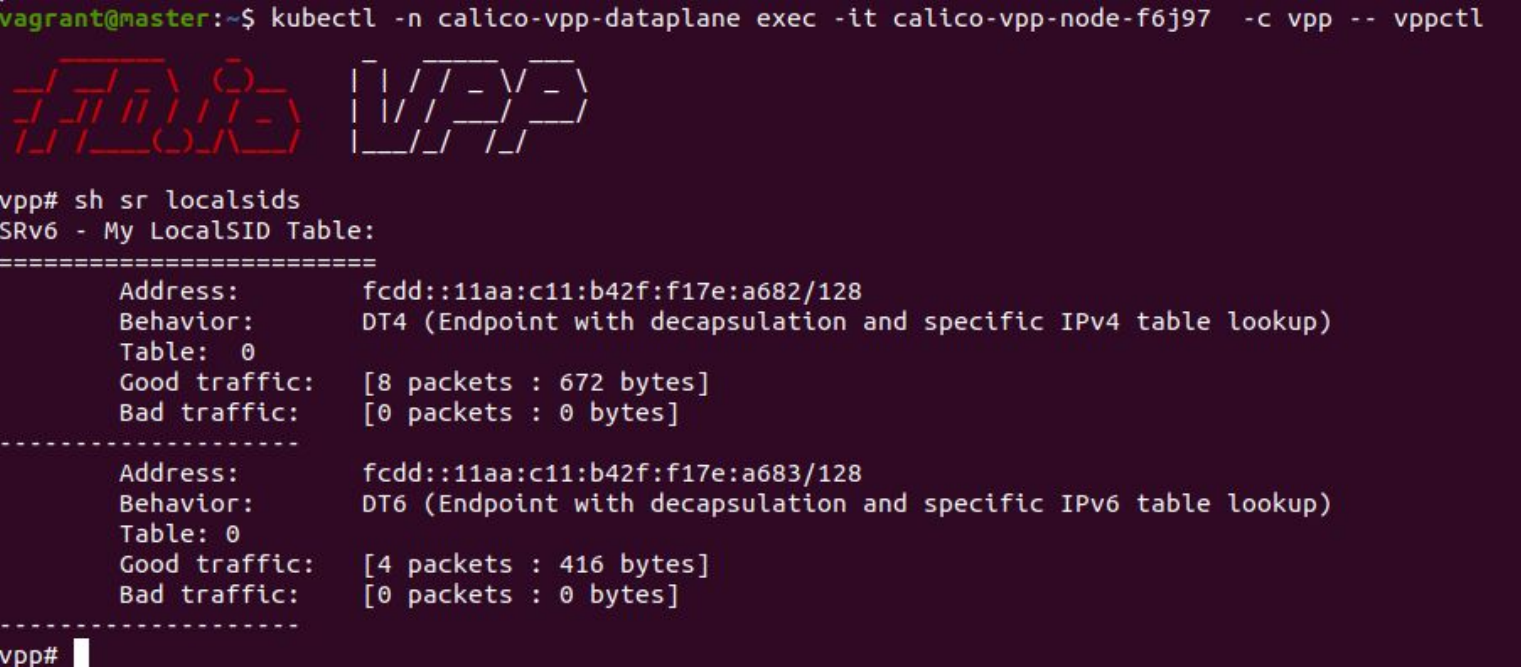}
    \caption{SRv6-TE K8s localsids counter}
    \label{fig:srv6-te-k8s-pods-show-localsid-counter-1}
\end{figure}

\subsubsection{Test SRv6-TE edit policies}
Let's take a look at the default configuration we defined in configMap.yaml, reported here for convenience:

\begin{lstlisting}[language=yaml,label=lst:bgp-iv6-traffic, caption=SRv6 Policies with ConfigMap]
apiVersion: v1
kind: ConfigMap
metadata:
 name: srv6-config-master
 namespace: calico-vpp-dataplane
data:
 srv6: |
   ---
   localsids:
     DT4: "fcdd::aa:34b8:247c:36da:db44"
     DT6: "fcdd::aa:34b8:247c:36da:db45"
   node: master
   policies:
     -
       bsid: "cafe::1c3"
       node: "fd11::1000"
       segment_list:
         - "fcff:3::1"
         - "fcdd::11aa:c11:b42f:f17e:a683"
       traffic: IPv6
     -
       bsid: "cafe::1c2"
       node: "fd11::1000"
       segment_list:
         - "fcff:3::1"
         - "fcdd::11aa:c11:b42f:f17e:a682"
       traffic: IPv4
     -
       bsid: "cafe::4"
       node: "fd12::1000"
       segment_list:
         - "fcff:6::1"
         - "fcff:8::1"
         - "fcdd::12aa:d460:b250:45:b04"
       traffic: IPv4
     -
       bsid: "cafe::5"
       node: "fd12::1000"
       segment_list:
         - "fcff:6::1"
         - "fcff:8::1"
         - "fcdd::12aa:d460:b250:45:b05"
       traffic: IPv6
---
apiVersion: v1
kind: ConfigMap
metadata:
 name: srv6-config-worker1
 namespace: calico-vpp-dataplane
data:
 srv6: |
   ---
   localsids:
     DT4: "fcdd::11aa:c11:b42f:f17e:a682"
     DT6: "fcdd::11aa:c11:b42f:f17e:a683"
   node: worker1
   policies:
     -
       bsid: "cafe::185"
       node: "fd10::1000"
       segment_list:
         - "fcff:3::1"
         - "fcdd::aa:34b8:247c:36da:db45"
       traffic: IPv6
     -
       bsid: "cafe::184"
       node: "fd10::1000"
       segment_list:
         - "fcff:3::1"
         - "fcdd::aa:34b8:247c:36da:db44"
       traffic: IPv4
     -
       bsid: "cafe::4"
       node: "fd12::1000"
       segment_list:
         - "fcff:6::1"
         - "fcff:8::1"
         - "fcdd::12aa:d460:b250:45:b04"
       traffic: IPv4
     -
       bsid: "cafe::5"
       node: "fd12::1000"
       segment_list:
         - "fcff:6::1"
         - "fcff:8::1"
         - "fcdd::12aa:d460:b250:45:b05"
       traffic: IPv6
---
apiVersion: v1
kind: ConfigMap
metadata:
 name: srv6-config-worker2
 namespace: calico-vpp-dataplane
data:
 srv6: |
   ---
   localsids:
     DT4: "fcdd::12aa:d460:b250:45:b04"
     DT6: "fcdd::12aa:d460:b250:45:b05"
   node: worker2
   policies:
     -
       bsid: "cafe::1c3"
       node: "fd11::1000"
       segment_list:
         - "fcff:4::1"
         - "fcff:3::1"
         - "fcdd::11aa:c11:b42f:f17e:a683"
       traffic: IPv6
     -
       bsid: "cafe::1c2"
       node: "fd11::1000"
       segment_list:
         - "fcff:4::1"
         - "fcff:3::1"
         - "fcdd::11aa:c11:b42f:f17e:a682"
       traffic: IPv4
     -
       bsid: "cafe::185"
       node: "fd10::1000"
       segment_list:
         - "fcff:2::1"
         - "fcff:3::1"
         - "fcdd::aa:34b8:247c:36da:db45"
       traffic: IPv6
     -
       bsid: "cafe::184"
       node: "fd10::1000"
       segment_list:
         - "fcff:2::1"
         - "fcff:3::1"
         - "fcdd::aa:34b8:247c:36da:db44"
       traffic: IPv4
\end{lstlisting}
For instance let’s take the policies defined for the worker2 to reach the worker1 using IPv6, specifically the one with BSID cafe::1c3.The policy is made up of three SIDs and, if we compare it with the information represented in Figure XX, it is clear that the packets from worker2 to worker1 must go through R4 and R3 before reaching worker1. If we want the packets from worker2 to worker1 to go through R7, R2, R3, we need to modify the policies in the file as follows:

\begin{lstlisting}[language=yaml,label=lst:bgp-iv6-traffic, caption=SRv6 Policies with ConfigMap modified]
- bsid: "cafe::1c3"
  node: "fd11::1000"
  segment_list:
    - "fcff:7::1"
    - "fcff:2::1"
    - "fcff:3::1"
    - "fcdd::11aa:c11:b42f:f17e:a683"
  traffic: IPv6
\end{lstlisting}
Then we just need to update the ConfigMap file by executing the command:
\begin{center}
\begin{verbatim}
    $ kubectl apply -f yaml/configMap.yaml
\end{verbatim}
\end{center}

After the execution, we can check the SR policies from the VPP cli for the node worker2 using the command
\begin{center}
\begin{verbatim}
    $ vpp# sh sr policies
\end{verbatim}
\end{center}

And we obtain an output similar to this:
\begin{figure}[ht]
    \centering
    \includegraphics[width=1\textwidth]{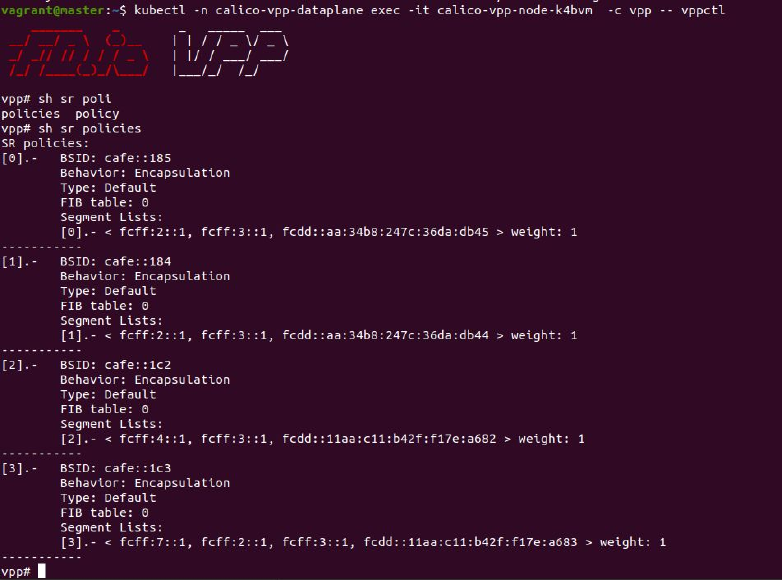}
    \caption{SRv6-TE K8s policied modified}
    \label{fig:srv6-te-k8s-policies-modified}
\end{figure}
After this check we can test again the connectivity between pods located respectively on worker1 and worker2.

}

\section{Performance Evaluation}
\label{sec:perf-eval}

In this section, we describe and discuss the results of the performance evaluation experiments that we performed using the knb (Kubernetes Network Benchmark) tool~\cite{kubernetes-network-benchmark}.

The knb tool generates traffic to evaluate the TCP and UDP throughput between pods in a Kubernetes cluster. We consider the communication between two pods (a client and a server) on different nodes, so that measurements include the encapsulation and decapsulation overheads (bytes and processing) associated with the overlay networking solutions. 
The knb tool also includes in its report the monitoring of RAM and CPU usage in the host for the client pod and for the server pod. Both TCP and UDP throughput measurements are taken by knb using the iperf3 tool~\cite{iperf3} that generates traffic and evaluates throughput.

Our goal for this performance evaluation is just to compare the proposed SRv6 overlay with the existing IP in IP overlay in Calico-VPP and show that the SRv6 overlay has comparable performance in the data plane, while offering a richer set of features. We are not interested in discussing the absolute performance (throughput) that can be achieved using the VPP data plane as it also depends on the hardware characteristics of the environment in which the Kubernetes cluster is deployed.

In the experiments, we use IPv4 for the pods networking, as the existing IP in IP overlay only supports IPv4. Hence, we will compare IPv4 in IPv4 (for the existing IP in IP overlay) with IPv4 in SRv6 (for the proposed SRv6 overlay). 

\subsection{Test environment and tools}

To execute the performance evaluation experiments, we used a PC with libvirt/KVM. We allocated three VMs using Vagrant, using the testbed setup shown in Fig.~\ref{fig:basic-testbed}. The PC and VMs specifications are reported in Table~\ref{tab:host-vm-specs}.

\begin{table}[ht]
\centering
\begin{tabularx}{\thesis{0.8}\paper{0.5}\textwidth}{|c|X|X|X|}
 \hline
  & CPU & RAM & Persistent disks \\
 \hline\hline
 Host & \makecell[l]{AMD Ryzen\texttrademark ~9 \\ 5900HX @3.3GHz} & \makecell[l]{32GB DDR4 \\ @3200 MHz} & 1x SSD (512 GB) \\
 \hline
 VM & 2x or 4x vCPU  & 4 GB & 128 GB taken from the SSD \\
 \hline
\end{tabularx}
\caption{Configuration for performance experiments}
\label{tab:host-vm-specs}
\end{table}

For the experiments, we have used a modified version of the \paper{knb}\thesis{\acrshort{knbLabel}} (Kubernetes Network Benchmark) tool~\cite{kubernetes-network-benchmark}. In particular, we extended the knb tool to support: i) the use of IPv6 in addition to IPv4 as pod networking; ii) the adaptation of the segment size/packet size to the \paper{MTU}\thesis{\acrshort{MTULabel}} (Maximum Transfer Unit) available in the overlays; iii) the use of service IPs instead of service names. Our fork of the knb Github repository is available at~\cite{kubernetes-network-benchmark-fork}.

\subsection{Experiments and results}
\label{sec:experiments}

\begin{figure}[ht]
    \centering
    \includegraphics[width=\paper{0.5}\thesis{1}\textwidth]{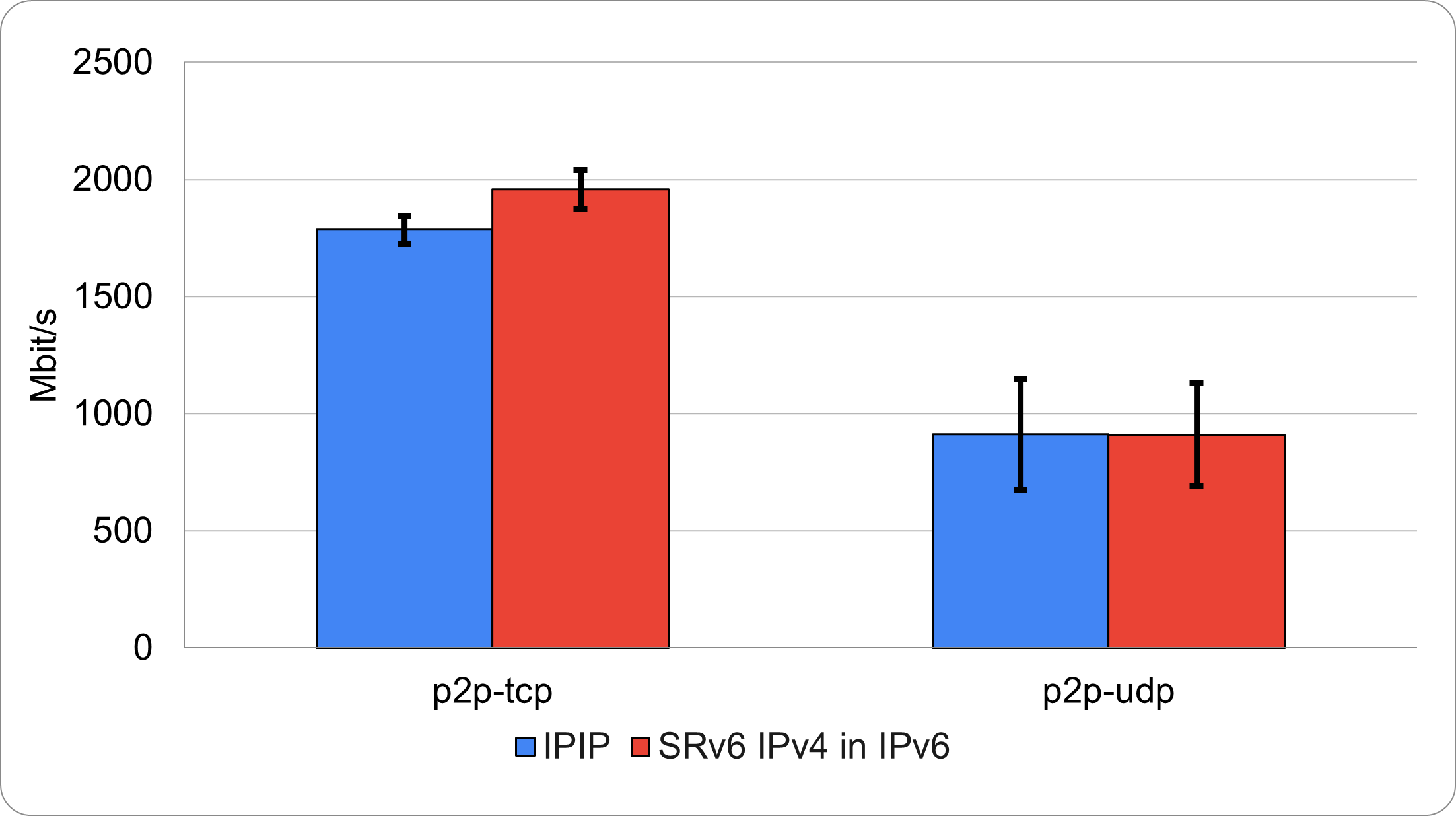}
    \caption{Pod to pod TCP and UDP Throughput \thesis{(2vCPU)}}
    \label{fig:bandwidth}
\end{figure}

Our experiments consist of evaluating the TCP and UDP throughput\thesis{ and monitoring the RAM and CPU usage in the client and server pods} for the IP in IP overlay (which represents the reference measurement) and for our proposed SRv6 overlay. We run a number of measurement runs with the modified knb tool, each run has a duration of 120 seconds.\thesis{ The tests were repeated in two different ways of communication between pods, specifically: simple pod communication to a pod between different nodes and pods to a service between different nodes.
\subsubsection{Tests Pod to Pod}}
The TCP and UDP throughput between a client pod and a server pod evaluated by the knb tool are reported in Fig.~\ref{fig:bandwidth}. The absolute values of the transfer rate are in the order of 2 Gb/s for TCP and 1 Gb/s for UDP. These values refer to the software-based processing of packets inside the testbed depicted in Fig.~\ref{fig:basic-testbed}, including the encapsulation and decapsulation operations performed by the IP in IP or SRv6 overlays that we want to assess. Actually, we are not interested in these absolute values, but in the relative comparison of performance of IP in IP and SRv6 overlay. We note that the performance of the new proposed SRv6 overlay is comparable with the performance or the existing IP in IP overlay. In particular, the SRv6 overlay slightly improves the performance with respect to the IP in IP overlay for the TCP throughput while it shows the same performance for the UDP throughput. In the reported experiments we have performed 10 runs of each measurements and considered the average of the results, the error bars in the figure represent the 95\% confidence interval. 

\thesis{The knb tool used for the measurement experiments also reports the RAM and CPU utilization for the pods client side and the server side, we have collected these measurements in Fig.~\ref{fig:ram_2cpu} and Fig.~\ref{fig:cpu_2cpu}.

\begin{figure}[ht]
    \centering
    \includegraphics[width=0.8\textwidth]{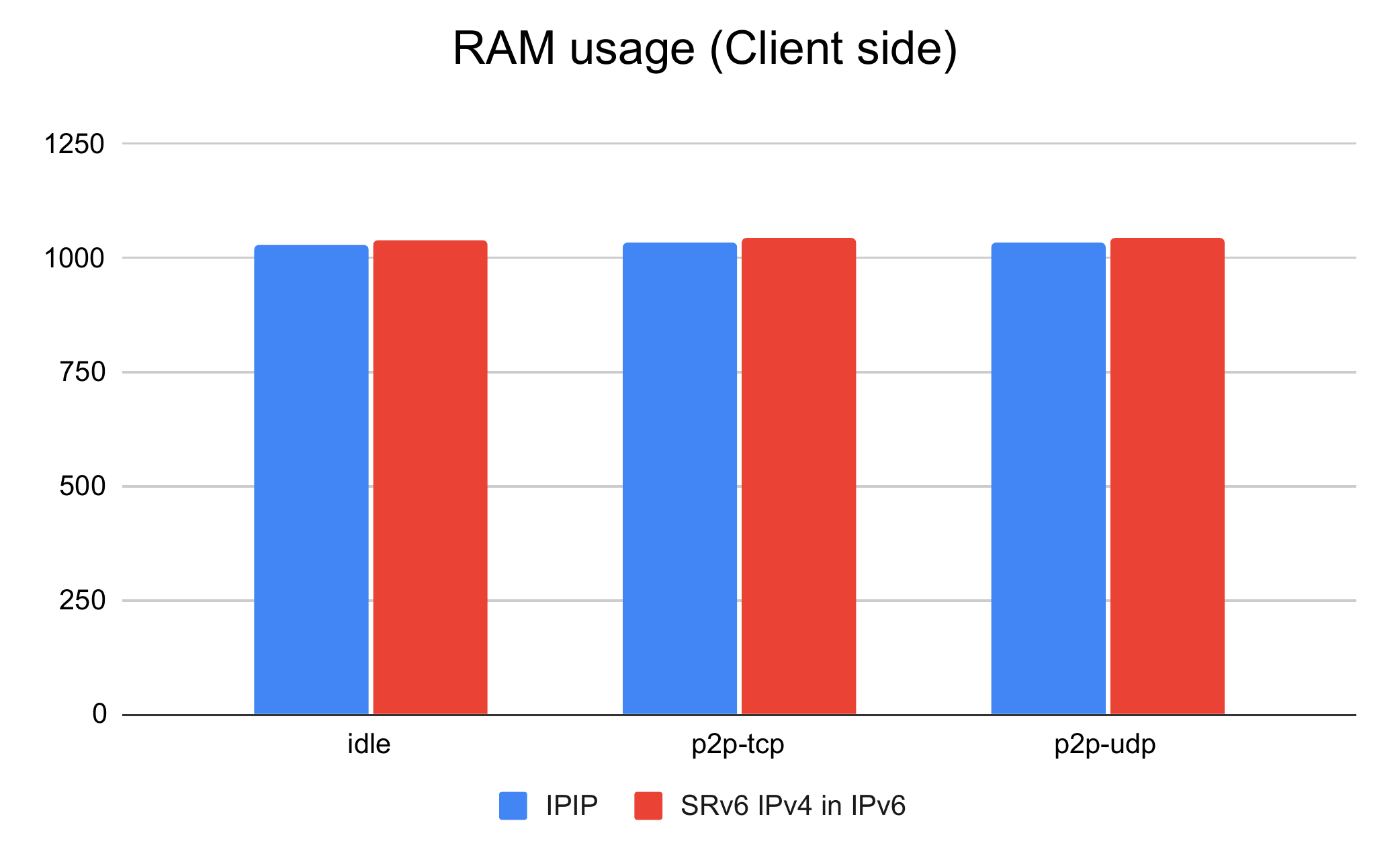}
    \includegraphics[width=0.8\textwidth]{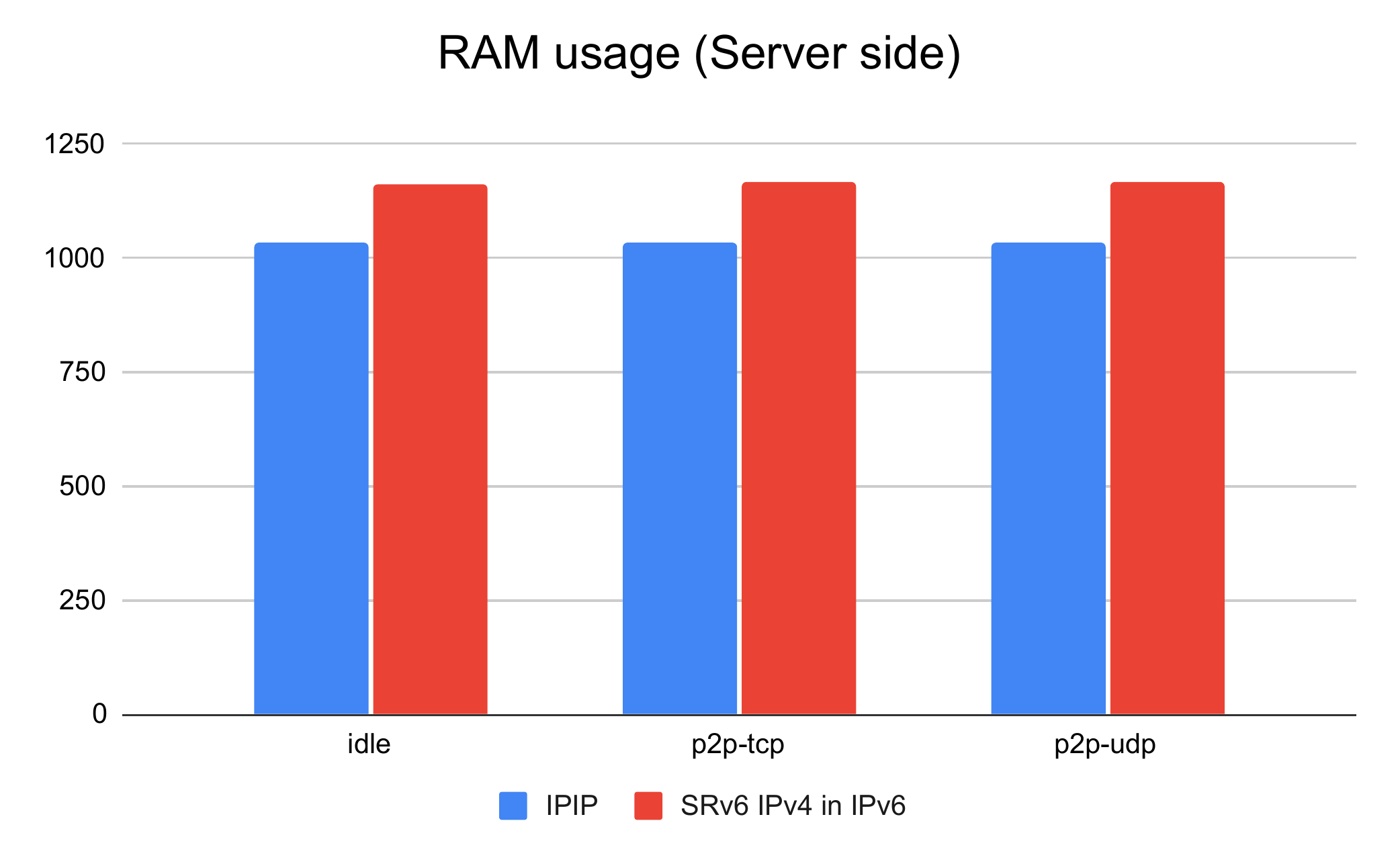}
    \caption{RAM usage client and server side (2vCPU)}
    \label{fig:ram_2cpu}
\end{figure}

Evaluation of TCP and UDP throughput is performed using the iperf3 tool. The TCP throughput evaluation with iperf3 is straightforward as the TCP client starts retrieving the data on the TCP connection and the TCP flow control mechanism will automatically adapt the transfer rate to the maximum available capacity.
On the other hand, the evaluation of the maximum throughput with UDP is more complex as it requires generating UDP packets at increasing rates and evaluating if the packets are received or not by the correspondent node. The process of generating UDP packets in iperf3 is CPU intensive so that often the measured throughput corresponds to the CPU bottleneck for the packet generation. In our testing we started with the default allocation of 2 vCPU to the VMs and we were not able to obtain stable throughput results for both the IP in IP and the SRv6 overlay for UDP. The CPU utilization in the client and server side was measured at 100\% during these early experiments Fig.~\ref{fig:cpu_2cpu}. Then we moved to an allocation of 4vCPU to the VMs and we achieved stable UDP throughput measurements reported in Fig.~\ref{fig:bandwidth_4cpu}.

\begin{figure}[ht]
    \centering
    \includegraphics[width=0.8\textwidth]{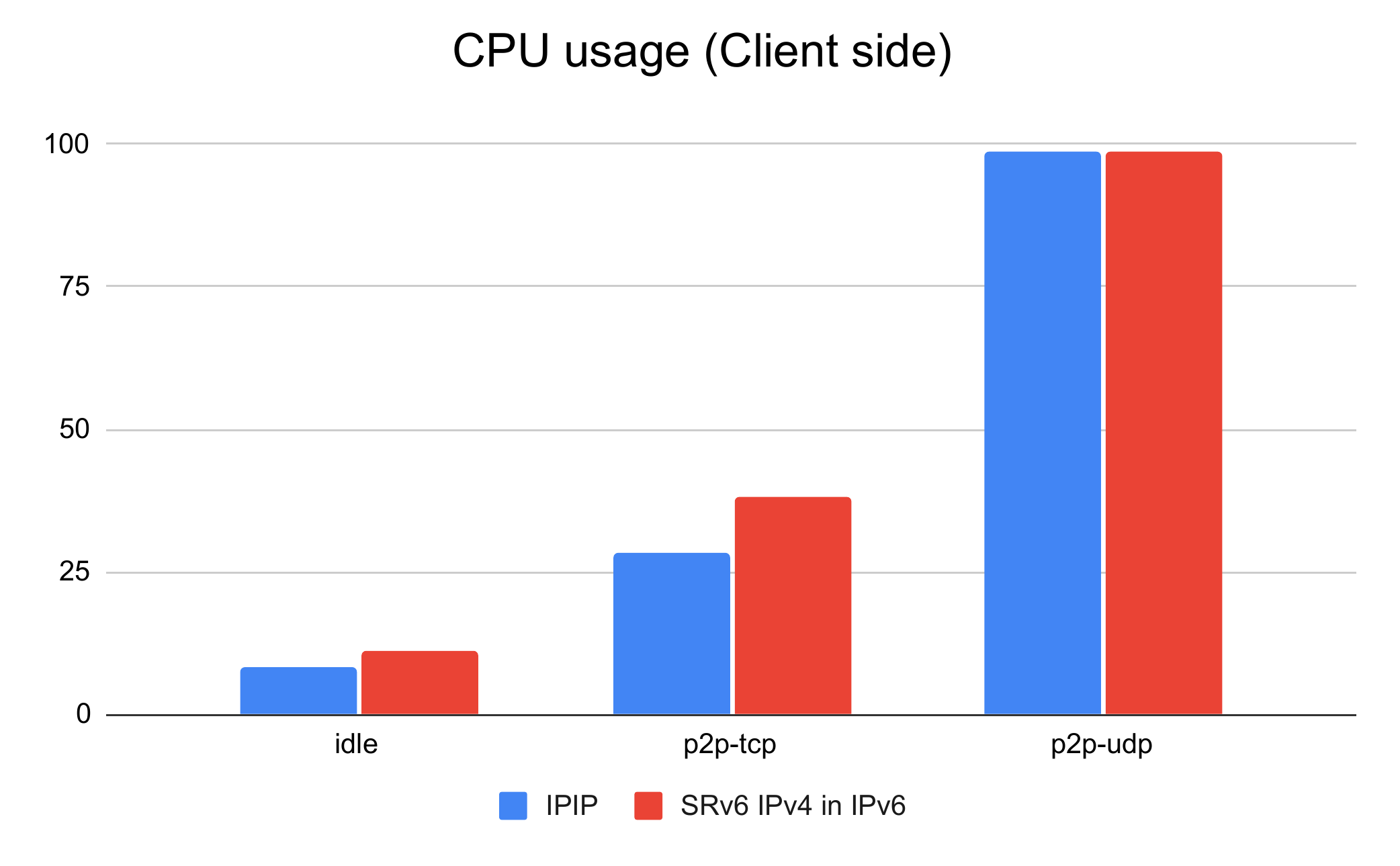}
    \includegraphics[width=0.8\textwidth]{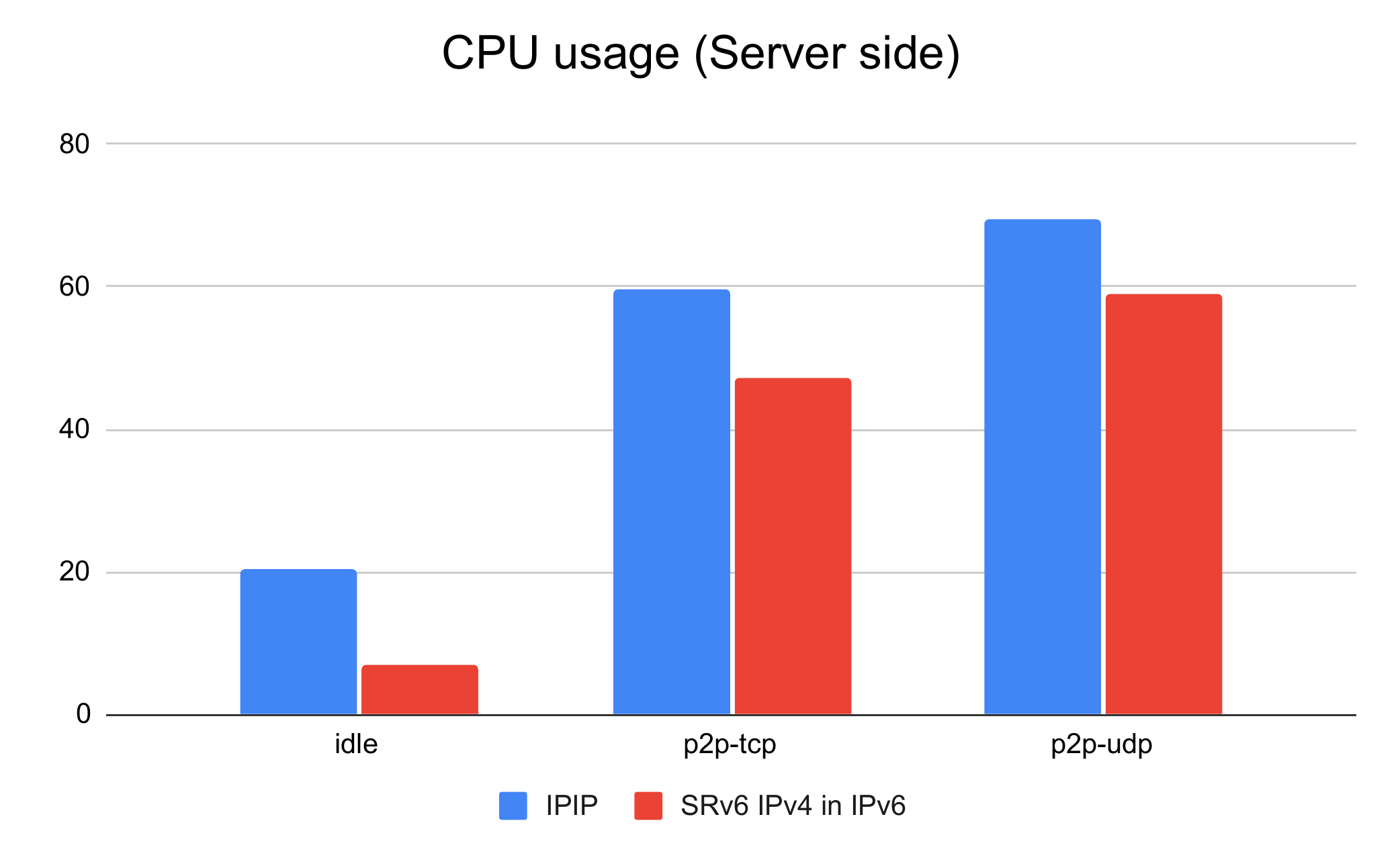}
    \caption{CPU usage client and server side (2vCPU)}
    \label{fig:cpu_2cpu}
\end{figure}

}

\thesis{ 
\begin{figure}[ht]
    \centering
    \includegraphics[width=1\textwidth]{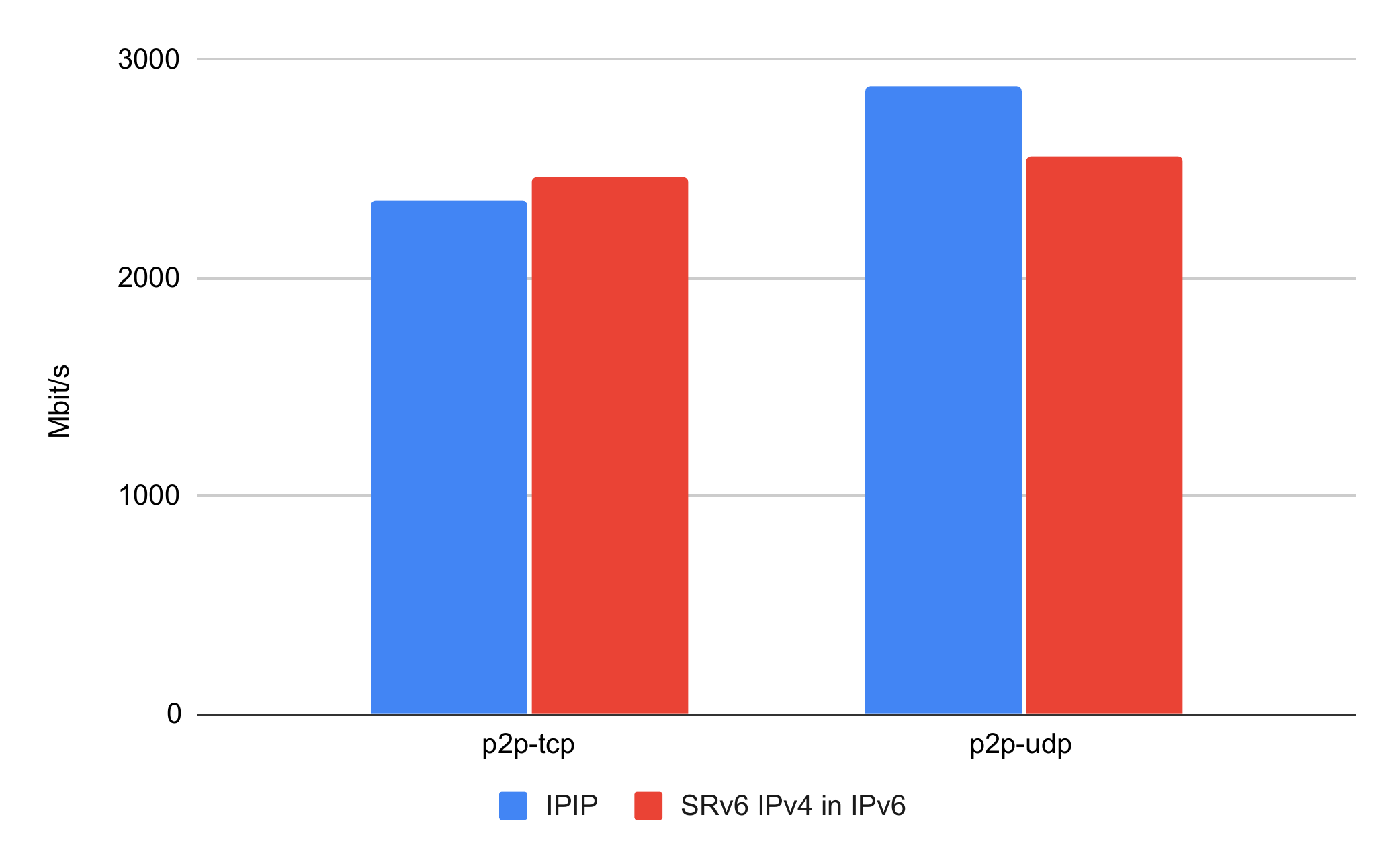}
    \caption{Pod to pod TCP and UDP Throughput (4vCPU)}
    \label{fig:bandwidth_4cpu}
\end{figure}
The CPU utilization in the client and server side was in the order of 60\% as shown in Fig.~\ref{fig:cpu_4cpu}. \\
\begin{figure}[ht]
    \centering
    \includegraphics[width=0.8\textwidth]{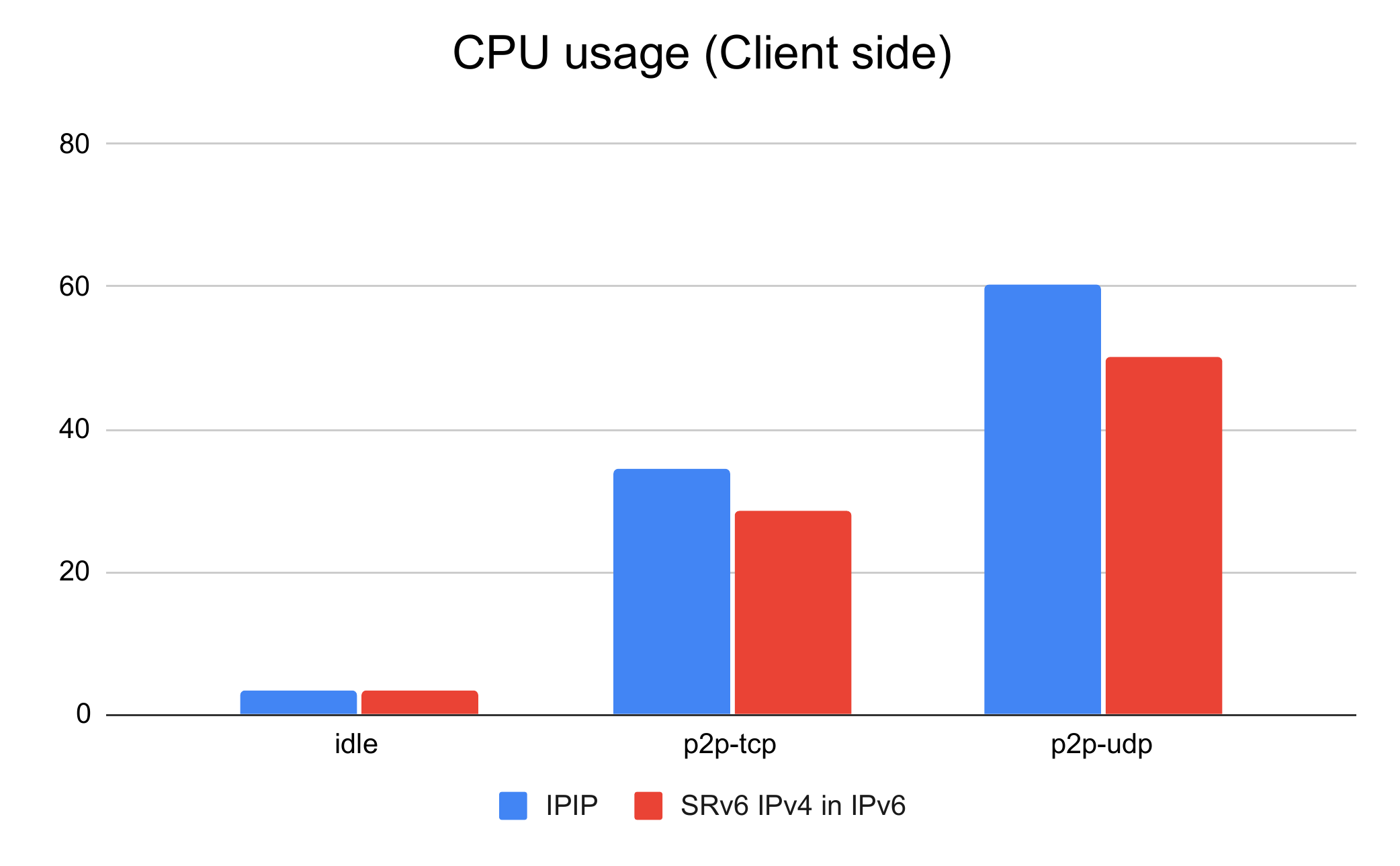}
    \includegraphics[width=0.8\textwidth]{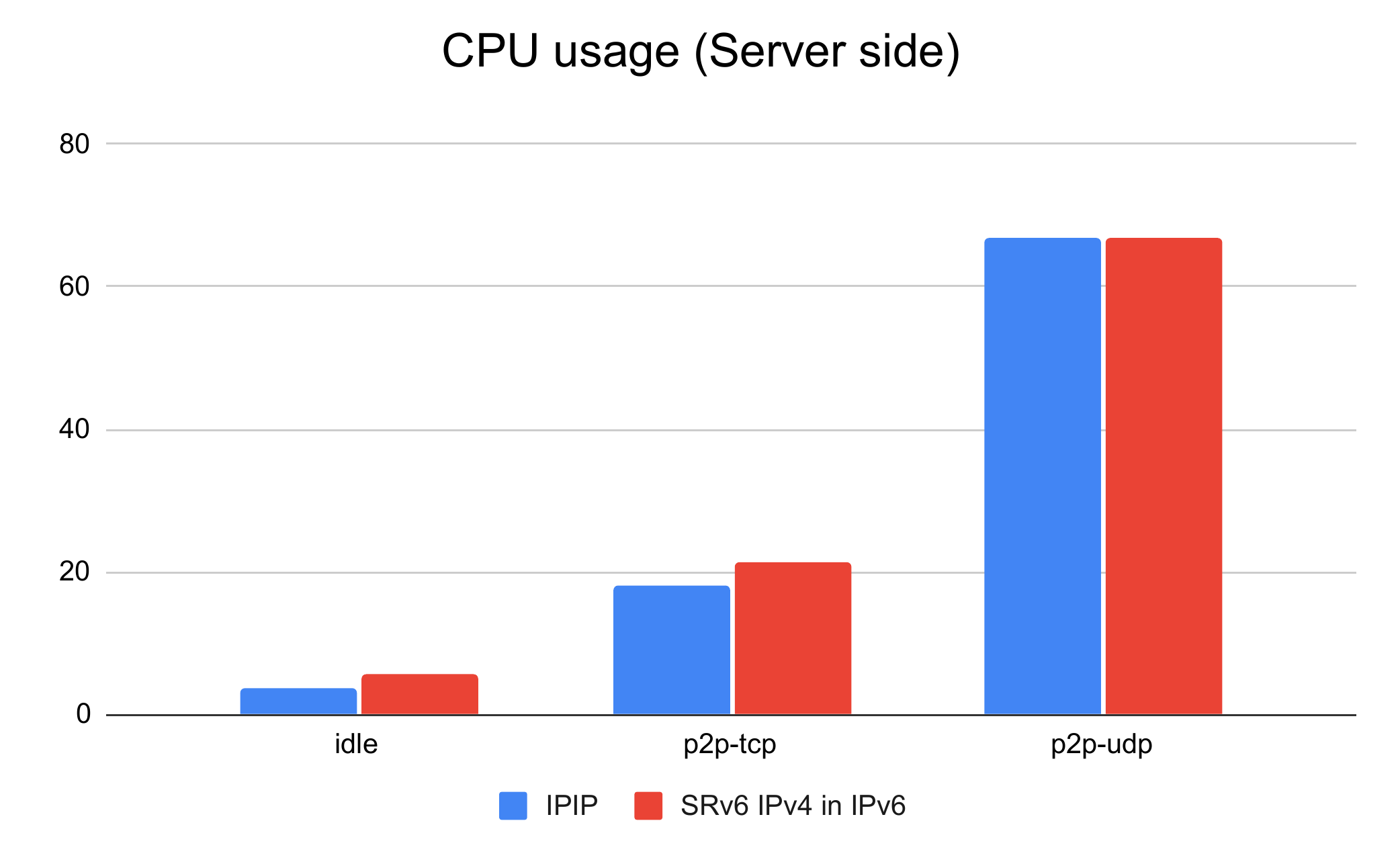}
    \caption{CPU usage client and server side (4vCPU)}
    \label{fig:cpu_4cpu}
\end{figure}


}

\thesis{

\subsubsection{Tests Pod to Services}
The evaluation of the TCP and UDP throughput is made using the iperf3 tool as the previous configuration, but in this case the address used as endpoint of the iperf server is a service address that act as a loadbalancer for the pod that are part of the service. For the sake of the experiment the service expose just one pod. In this way we are able to collect the data of the experiment. 
The absolute values of the transfer rate are in the order of 2 Gb/s for TCP and 1 Gb/s for UDP as reported in Fig.~\ref{fig:bandwidth_p2s}. As observed in the previous tests the process of generating UDP packets in iperf3 is CPU intensive so that often the measured throughput corresponds to the CPU bottleneck for the packet generation.  
Then we moved to an allocation of 4vCPU to the VMs and we achieved stable UDP throughput measurements reported in Fig.~\ref{fig:bandwidth_4cpu_p2s}.
As for the previous scenario the CPU utilization in the client and server side was in the order of 60\% as shown in Fig.~\ref{fig:cpu_4cpu_p2s}.
Regarding the consumption of RAM, both for IPnIP and SRv6 there are no particular differences, indeed the values are very similar to each other.



\begin{figure}[ht]
    \centering
    \includegraphics[width=1\textwidth]{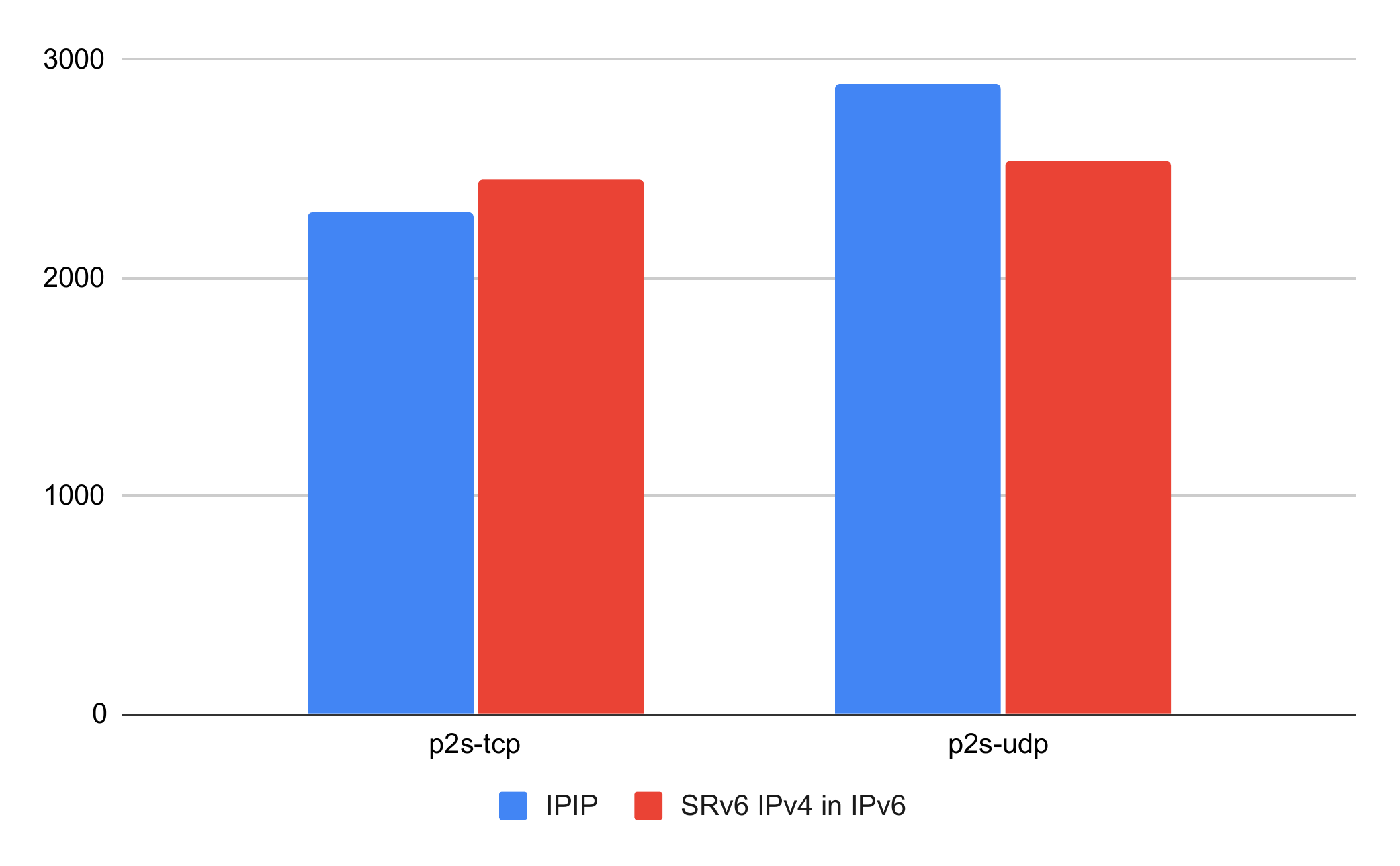}
    \caption{Pod to pod TCP and UDP Throughput (4vCPU)}
    \label{fig:bandwidth_4cpu_p2s}
\end{figure}

\begin{figure}[ht]
    \centering
    \includegraphics[width=0.8\textwidth]{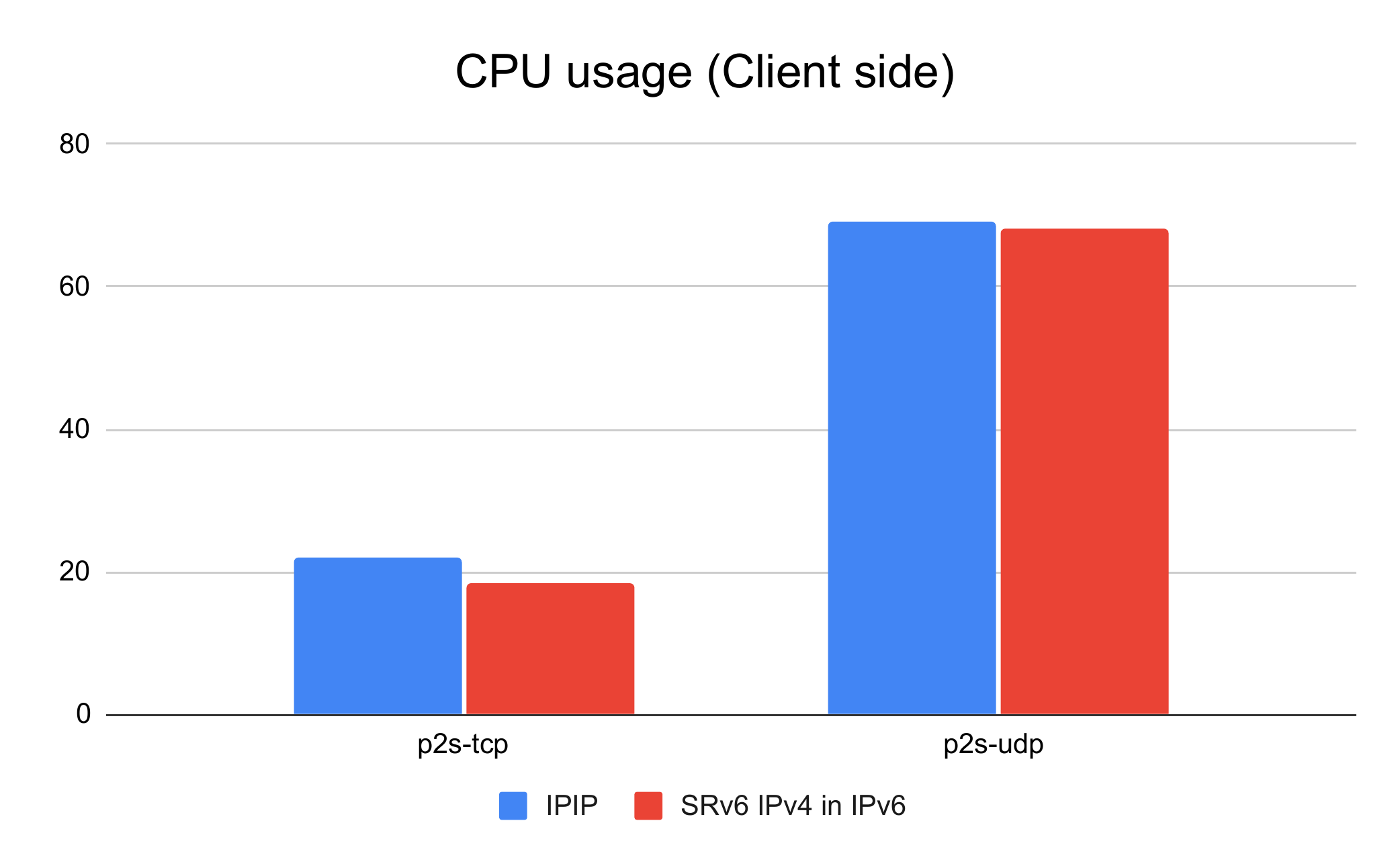}
    \includegraphics[width=0.8\textwidth]{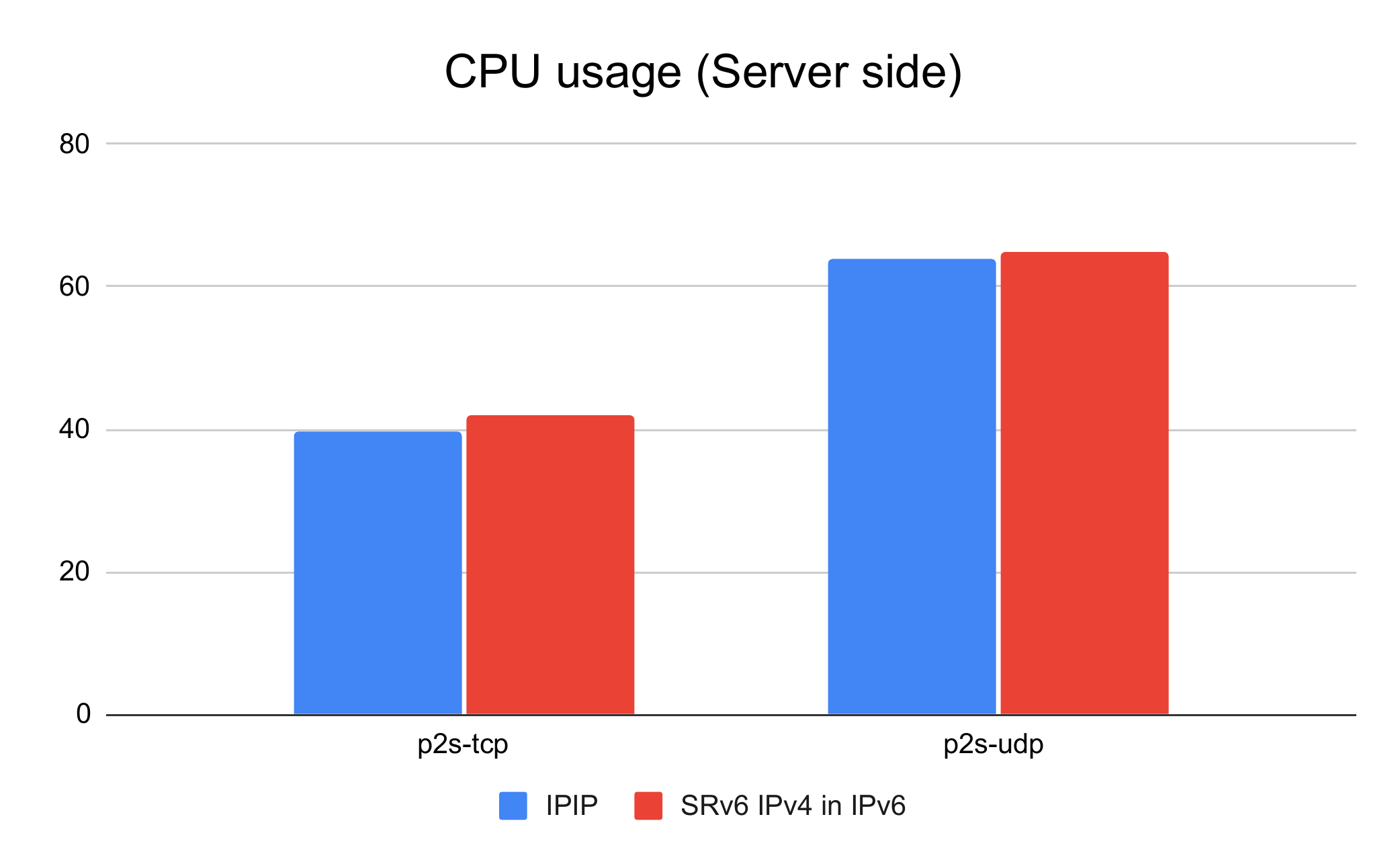}
    \caption{CPU usage client and server side (4vCPU)}
    \label{fig:cpu_4cpu_p2s}
\end{figure}

\begin{figure}[ht]
    \centering
    \includegraphics[width=0.8\textwidth]{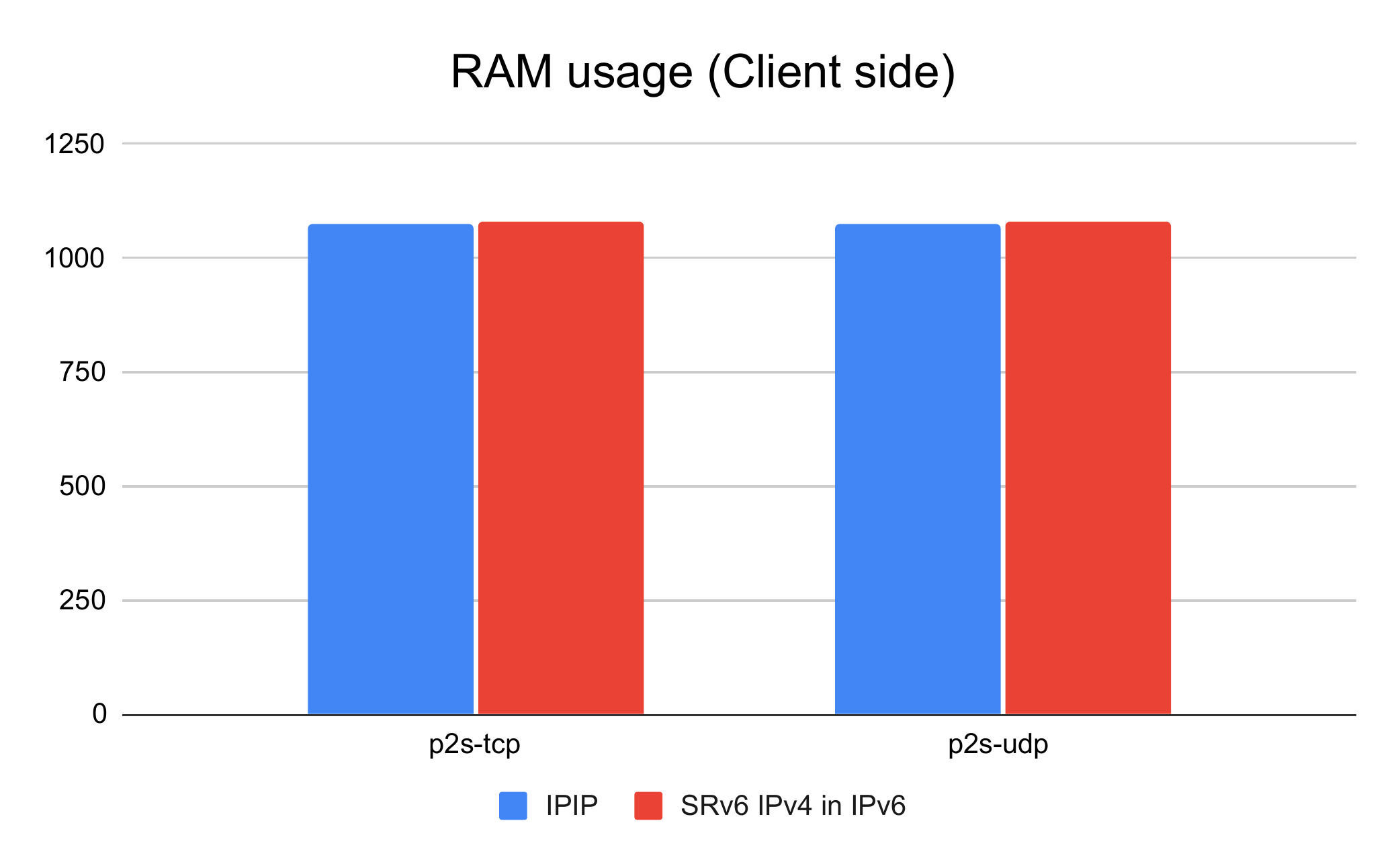}
    \includegraphics[width=0.8\textwidth]{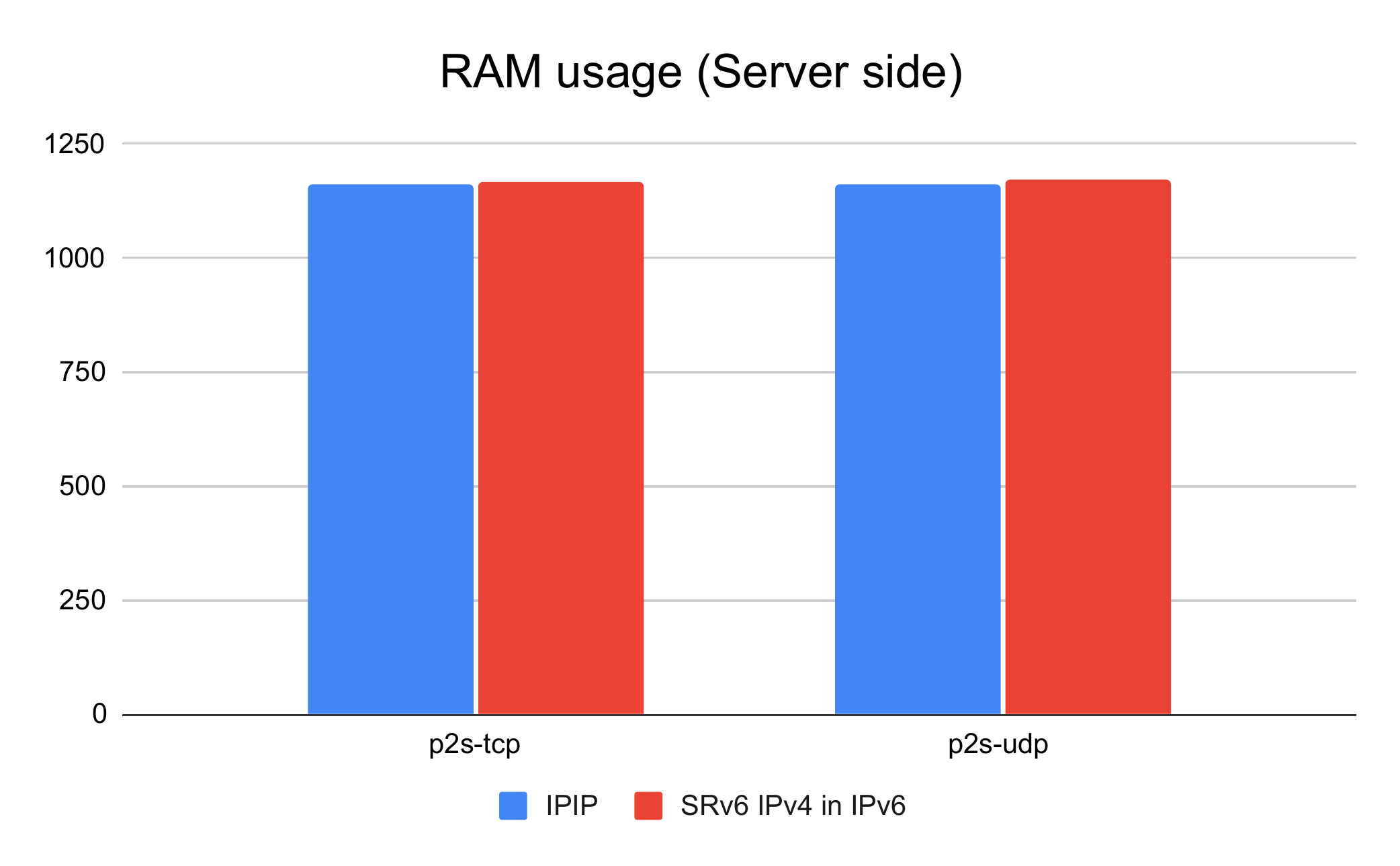}
    \caption{RAM usage client and server side (4vCPU)}
    \label{fig:ram_4cpu_p2s}
\end{figure}

}



\paper{
\section{Future work}
\subsection{Additional scenarios enabled by SRv6}
\label{sec:srv6-additional-scenarios}

The use of SRv6 as an overlay networking mechanism for Kubernetes opens up the possibility of supporting further features in addition to Traffic Engineering, which has been the main focus of this work. The most important one for Telco operators is the extra cluster communications, i.e. the possibility to easily interconnect pods running in the Kubernetes cluster with nodes on external networks. Complex multi-tenant scenarios are of interest for Telco operators and the recent RFC (BGP Overlay Services Based on SRv6 \cite{rfc9252}) describes how to support these service scenarios using SRv6. An example of these services is Ethernet VPN (EVPN) \cite{rfc8365}. Conceptually, our proposed approach based on BGP can support these scenarios, as external networks can be advertised using BGP updates message. However, more work is needed to extend our current implementation to support all the scenarios described in \cite{rfc9252}.  

\subsection{Extending Kubernetes models}
Our proposed solution starts from the assumption that the specific networking configurations are only done inside the networking plugin (Calico-VPP in our case) and are not exposed to the generic Kubernetes cluster configuration. Thanks to this approach, it is possible to decouple the concerns related to networking from the cluster definition at the application level, and it is possible to introduce the advanced networking features offered by SRv6 using the specific plugin (Calico-VPP) that we have designed and implemented.

In a longer-term perspective, we envisage that other Kubernetes networking plugins will integrate SRv6 capabilities. For example, Cilium \cite{cilium} has recently introduced SRv6 features \cite{bernier-cilium-srv6}. For this reason, an interesting and challenging future work is the extension of Kubernetes models to consider features that can be mapped into SRv6 capabilities. The extended models will be used to automatically derive the configuration of networking plugins capable of supporting SRv6. In this approach, the overall Kubernetes cluster definition at application level should also include networking aspects, likely in terms of high-level requirements.  
}

\paper{
\section{Conclusions}
\label{sec:conclusions}
}

\paper{
The paper has shown how to design and implement a networking plugin for Kubernetes based on SRv6, capable of taking advantage of the features offered by this powerful networking technology. In particular, we have demonstrated the Traffic Engineering capabilities of the proposed SRv6 overlay, which can help optimize the utilization of the transport networks. In our solution, we have extended an existing networking plugin (Calico-VPP) and its overlay solution based on IP-in-IP tunneling, implementing our IP-in-SRv6 tunneling. The basic configuration of a Kubernetes cluster and its interaction with the CNI plugin is not changed, our current solution is completely transparent for the Kubernetes users. The administrator of the Kubernetes cluster can configure the SRv6 overlay with minimal changes with respect to the configuration of the IP-in-IP overlay. We have achieved a dynamic and automatic configuration of the nodes, capable of supporting the advanced features (i.e. Traffic Engineering) offered by the SRv6 overlay. 

With respect to our first set of research questions, we have shown that it is possible to add advanced networking features without disrupting the current CNI interface. In our solution this is hidden from the regular configuration of the Kubernetes cluster and it is done inside the proposed CNI plugin. This is consistent with the Kubernetes model, which is based on the separation of the networking concerns and it has the advantage that the existing clusters and workloads can be supported in a relatively easy way. On the other hand, we observe that this approach has the disadvantage that the new advanced networking features are only available inside a given plugin, while it would be desirable to have a generalization of the models, so that multiple networking plugins could offer a comparable set of features like Traffic Engineering.

Coming to the second set of research questions, related to the control mechanisms needed to deal with advanced networking features, we have designed, implemented, and demonstrated a dynamic and automatic solution that supports Traffic Engineering features in the SRv6 overlay. We have analyzed two approaches: one is based on the extension of routing protocols (BGP) and the other one is based on Kubernetes control plane. The solution based on the BGP extension has been merged in the Calico-VPP project mainstream, the one based on Kubernetes control plane is still in an evaluation phase. We believe that both approaches have merit and their applicability scenarios and suggest that further work continues in both directions. We have found that the extension of BGP protocol is more complex from the design and development point of view, but offers the advantage of easing the interaction with remote nodes that do not belong to the Kubernetes cluster. On the other hand, the use of Kubernetes control plane makes the introduction of new features easier, but it can be applied to a homogeneous environment in which all nodes to be controlled belong to the Kubernetes cluster.


}

\thesis{
This thesis work has shown how to design and implement an extension of a networking plugin for Kubernetes based on SRv6, capable of taking advantage of the features offered by this powerful networking technology. In particular, I have demonstrated the Traffic Engineering capabilities of the proposed SRv6 overlay, which can help optimizing the utilization of the transport networks. In this solution, I have extended an existing networking plugin (Calico-VPP) and its overlay solution based on IP-in-IP tunneling, implementing a custom IP-in-SRv6 tunneling. The basic configuration of a Kubernetes cluster and its interaction with the CNI plugin is not changed, this current solution is completely transparent for the Kubernetes users. The administrator of the Kubernetes cluster can configure the SRv6 overlay with minimal changes with respect to the configuration of the IP-in-IP overlay. I have achieved a dynamic and automatic configuration of the nodes, capable of supporting the advanced features (i.e. Traffic Engineering) offered by the SRv6 overlay. 

With respect to the first set of research questions, I have shown that it is possible to add advanced networking features with no disruption of the current CNI interface. In this solution this is hidden to the regular configuration of the Kubernetes cluster and it is done inside the proposed CNI plugin. This is consistent with the Kubernetes model that is based on the separation of the networking concerns and it has the advantage that the existing clusters and workloads can be supported in a relatively easy way. On the other hand, I verified that this approach has the disadvantage that the new advanced networking features are only available inside a given plugin, while it would be desirable to have a generalization of the models, so that multiple networking plugins could offer a comparable set of features like Traffic Engineering.

Coming to the second set of research questions, related to the control mechanisms to be used to deal with the advanced networking features, I have designed, implemented and demonstrated a dynamic and automatic solution that supports Traffic Engineering features in the SRv6 overlay. I have analyzed two approaches: one is based on the extension of routing protocols (BGP) and the other one is based on Kubernetes control plane. The solution based on the BGP extension has been merged in the Calico-VPP project mainstream, the one based on Kubernetes control plane is still in an evaluation phase. To enable the BGP-based implementation, it was necessary to contribute to the goBGP open source project \cite{go-bgp} to support the advertising of the candidate path of a Segment Routing (SR) Policy using the BGP SAFI (Subsequent Address Family Identifiers) defined in [38] (BGP SAFI 73) specifically, the support for the TypeB segment defined in [38].
In order to demonstrate the feasibility of the solution with BGP, I have implemented a BGP peer (SRv6-PI) capable of injecting the policies, the SRv6-PI implementation is open source and available at \cite{SRv6-PI}.
I believe that both approaches have their merits and applicability scenarios, therefore I suggest that further work continues in both directions. I have found that the extension of the BGP protocol is more complex from a design and development point of view, but offers the advantage of facilitating interaction with remote nodes that do not belong to the Kubernetes cluster. On the other hand, the use of the Kubernetes control plane facilitates the introduction of new features, but it can be applied to a homogeneous environment where all the nodes to be controlled belong to the Kubernetes cluster.

Finally, during the research activity, I found the need to have a tool to perform automated performance tests on the Kubernetes network plugins considering different types and traffic flows. In particular in the development of this tool I found it necessary to support IPv6 and to be able to use the traffic generators most suitable for the type of protocol. To address these requirements, I have contributed to the development of the Kites tool, which has been released as open source at \cite{kites}

} 

\appendices

\ifCLASSOPTIONcaptionsoff
  \newpage
\fi



\bibliographystyle{IEEEtran}
\bibliography{references}

\end{document}